\documentclass[fleqn,usenatbib,onecolumn]{mnras}
\usepackage{comment}
\usepackage{array}
\usepackage{longtable}
\usepackage{hyperref}
\usepackage{enumitem} 
\usepackage{threeparttable}
\usepackage{graphicx}  
\usepackage{booktabs}  
\usepackage{subcaption}  
\usepackage{multirow}
\usepackage{amsmath}
\usepackage{pifont}
\usepackage{xcolor}
\usepackage{booktabs}
\usepackage{multirow}
\usepackage{threeparttable}
\usepackage[font=small,labelfont=bf]{caption}

\usepackage{newtxtext,newtxmath}

\usepackage[T1]{fontenc}

\DeclareRobustCommand{\VAN}[3]{#2}
\let\VANthebibliography\thebibliography
\def\thebibliography{\DeclareRobustCommand{\VAN}[3]{##3}\VANthebibliography}

\usepackage{graphicx}	
\usepackage{amsmath}	

\usepackage[]{lineno}
\title[]{A Novel Pipeline for the Identification of New Gamma-Ray Blazars from the 4FGL-Xiang Catalog Based on Multi-wavelength Flux Distributions}

\author[Y.C. Xiang et al.]{
	Yunchuan Xiang,$^{1,2}$\thanks{E-mail: xiang\_yunchuan@yeah.net}
	Yingzhi Ye,$^{1}$
	Peng Feng,$^{1}$
	Hui Li,$^{1}$ 
	Xiankai Pang,$^{1}$
	Xiaofei Lan,$^{1}$
	\newauthor
	Qingquan Jiang,$^{1}$
	and Ningfeng Zhang$^{1}$\\
	$^{1}$School of Physics and Astronomy, China West Normal University, Nanchong 637009, China\\
	$^{2}$Department of Astronomy, Yunnan University; Key Laboratory of Astroparticle Physics of Yunnan Province, Kunming 650091, People's Republic of China
}

\date{accepted 2026-01-19}

\pubyear{Monthly Notices of the Royal Astronomical Society}

\begin{document}
\label{firstpage}
\pagerange{\pageref{firstpage}--\pageref{lastpage}}
\maketitle

\begin{abstract}

The identification and classification of Fermi blazars are core topics in high-energy astrophysics. To enable precise spatial cross-identification, we constructed two high-precision catalogs: the updated 4FGL-Xiang-DR2 (DR2) and a supplementary version of the fifth edition of Roma-BZCAT (\texttt{5BZCAT\_err}). We then developed and applied a novel four-step analytical pipeline combining cross-matching with the statistical analysis of multi-band flux distributions to identify new Fermi blazars. 
The analytical pipeline has yielded several key results in the systematic comparison of BZBs and BZQs. We found that among single statistical metrics, kurtosis is the most powerful discriminator (MAD~$>$~1.64). At the overall distribution level, the 1.4~GHz, 843~MHz, 5~GHz, 0.1--2.4~keV, and 0.3--10~keV bands show significant divergence (JSD~$>$~0.3). Building on these findings, our proposed ``$\mathrm{Box\text{-}Cox+TND}$'' model successfully fits the observed flux distributions between BZBs and BZQs. 
Applying this entire pipeline, we successfully identified 17 new blazars. The validity of these associations is strongly supported by our multi-wavelength flux model, which confirms that 15 of the 17 candidates are statistically consistent with the known blazar population, falling within the $2\sigma$ confidence interval. Although the two remaining sources exhibit some statistical deviation in the gamma-ray band, their strong consistency in other wavebands, coupled with high spatial association probabilities, leads us to conclude that their associations are also reliable and should not be readily excluded.
\end{abstract}

\begin{keywords}
galaxies: active -- BL Lacertae objects: general -- quasars: general -- gamma-rays: galaxies -- galaxies:
statistics.
\end{keywords}



\section{Introduction}  \label{sec1} 
Blazars are a special class of active galactic nuclei (AGNs) whose powerful relativistic jets point toward our line of sight and produce broadband emission spanning the electromagnetic spectrum, from radio to high-energy gamma rays
 \citep{Urry1995,Ghisellini1998,Abdo2010c,bottcher2019}. Their typical observational properties encompass exceptionally high luminosities, pronounced and rapid variability, intense and highly variable polarization, apparent superluminal motion observed in the radio regime, non-thermal continuum spectra dominated by nuclear emission, and strong $\gamma$-ray emission  \citep{Angel1980,Wills1992,Abdo2010b}. This makes them ideal natural laboratories for studying extreme physical processes in the Universe, such as jet formation and the acceleration mechanisms of cosmic particles, and high-energy emission mechanisms \citep{Fossati1998,Aharonian2006,rieger2007fermi,blandford2019relativistic,Tavecchio2021,Liodakis2022}.

Based on the characteristics of their optical spectra, blazars are traditionally divided into two main subclasses: Flat-Spectrum Radio Quasars (FSRQs) and BL Lacertae objects (BL Lacs). This classification scheme is often referred to as the ``blazar dichotomy''. 
The criterion for distinguishing between these two classes is the equivalent width (EW) of the emission lines in their spectra. FSRQs exhibit prominent broad emission lines, with an equivalent width typically greater than 5~\AA~\citep{Urry1995}. Their spectra contain a significant radiative component from the accretion disk, and they generally have high Eddington ratios; for example, the ratio of the broad line region luminosity to the Eddington luminosity is about $L_{BLR}/L_{Edd} \sim 5 \times 10^{-4}$~\citep{Ghisellini2011}.
In contrast, BL Lacs display weaker spectral features, often lacking prominent emission lines (\(\text{EW} < 5\)~\AA) \citep{Urry1995}, with jet emission dominating their radiation and minimal contribution from the accretion disk  \citep{Sbarrato2012}.

Although a general consensus exists within the astronomical community regarding the differences between BZBs and BZQs\footnote{Throughout this paper, we follow the 5BZCAT nomenclature \citep{Massaro2009}: BZQ denotes FSRQ and BZB denotes BL Lac; hereafter we use BZQ $\boldsymbol{\equiv}$ FSRQ and BZB $\boldsymbol{\equiv}$ BL Lac consistently.}, several key questions remain unresolved. Specifically, the statistical characteristics of their multi-band flux distributions, and to what extent these distributions truly differ, have not yet been clearly established. Furthermore, how to comprehensively and accurately quantify such differences, and ultimately, what physical mechanisms drive them, remain open questions requiring systematic investigation.

As one of the most fundamental observable physical quantities, flux is not only closely related to key parameters such as luminosity, magnitude, and redshift, but is also strongly linked to the radiation mechanisms, jet formation processes, and beaming effects of blazars \citep{oke1968energy,Blandford1979,Abdo2010c,Chen2016,Foschini2024,Xie2024}. This makes flux an ideal physical probe, whose statistical behavior can effectively reveal the underlying physical processes of blazars.  
Moreover, the extensive availability of flux measurements across diverse observational catalogs offers a crucial data foundation for conducting large-sample statistical analyses of blazar flux distributions \citep{NED2019}. 
Conducting a systematic statistical analysis of the multi-band flux distributions of blazars carries the following significant scientific implications:
\begin{enumerate}[label=(\alph*)]
	\item To provide key observational evidence for systematically revealing the fundamental differences between BZBs and BZQs in their multi-band energy output and radiation mechanisms. 
	
	\item By precisely quantifying the differences in these flux distributions, this study will provide key observational constraints on the radiation mechanisms and particle acceleration processes of these two subclasses, and offer new supporting evidence for the blazar dichotomy. 
	
	\item To explore novel classification criteria based on the statistical distributions of multi-band flux, thereby establishing a more robust classification paradigm that overcomes the limitations of traditional methods reliant on single, variable optical spectral observations.
	
\end{enumerate}
Therefore, our core motivation is to quantitatively characterize and compare the multi-band flux distribution features of BZBs and BZQs. We aim to answer the following questions: 
\begin{enumerate}
	\item What are the characteristics of the flux distributions for BZBs and BZQs in different energy bands?
	\item Are the distributional characteristics of BZBs and BZQs significantly different, and how can we comprehensively quantify the degree of this difference?
	\item If such differences exist, what physical mechanisms or factors are responsible for them?
	\item If such differences exist, can we construct a unified distribution model for the flux of BZBs and BZQs that can serve as a new criterion for identifying new candidates of these subclasses?
\end{enumerate}

To address the above scientific questions, we developed a systematic four-step analytical pipeline consisting of the following: (1) retrieving positional uncertainties and multiwavelength fluxes from the fifth-edition Roma-BZCAT (5BZCAT) and constructing the derivative catalog 5BZCAT\_err using the compiled positional-uncertainty metadata (see Section~\ref{sec2.2}), in preparation for subsequent cross-identification analyses; (2) calculating spatial association probabilities with the second 4FGL-Xiang release (DR2); (3) quantifying and identifying the differences in the flux distributions; and (4) modeling the multi-band fluxes with the $\mathrm{Box\text{-}Cox+TND}$ model by applying a Box-Cox transformation followed by a truncated-normal fit to the transformed data (see Section~\ref{sec2.6.2}).

In this study, we have chosen the Roma-BZCAT as our primary candidate catalog, utilizing its sample for the statistical analysis of multi-band fluxes.
The Roma-BZCAT is one of the most comprehensive blazar catalogs, initially compiled and continuously updated by \cite{Massaro2009}. This catalog identifies blazars through optical spectroscopy and integrates multi-wavelength observational data. The latest version,  5BZCAT\footnote{\url{https://www.ssdc.asi.it/bzcat/}}  published in 2015, includes 3,561 blazars, comprising 1,151 BL Lacs (5BZB), 1,909 FSRQs (5BZQ), and 227 uncertain blazar candidates (5BZU). 
Additionally, a new category, denoted as 5BZG, was introduced in the 5BZCAT catalog.  
This category encompasses sources that are typically reported as BL Lac objects but exhibit a spectral energy distribution (SED) characterized by significant dominance of emission from the host galaxy at optical wavelengths. The 5BZG category contains a total of 274 sources \citep{Massaro2015}. 
Its well-defined classification and extensive multi-wavelength data provide a solid foundation for investigating the differences in flux distributions and for performing statistical modeling.


The Fermi Large Area Telescope (Fermi-LAT) is a $\gamma$-ray telescope with a wide field of view of approximately 2.4 sr and is capable of completing an all-sky survey roughly every three hours, thereby continuously detecting GeV $\gamma$-ray photons that may originate from blazars\footnote{https://fermi.gsfc.nasa.gov/ssc/observations/types/allsky/?utm\_source} \citep{Atwood2009}.	
 Benefiting from improvements in event reconstruction and instrument response functions (such as Pass 8)\footnote{https://fermi.gsfc.nasa.gov/ssc/data/analysis/documentation/Cicerone/Cicerone\_LAT\_IRFs/IRF\_overview.html}, the Fermi-LAT exhibits excellent localization capability in the GeV energy range. For example, at 20 GeV, its point spread function (PSF) has a 68\% containment radius of approximately 0.1°. 
Compared with the Energetic Gamma Ray Experiment Telescope (EGRET) \citep{thompson1993}, a similar space-based instrument, the Fermi-LAT offers a broader energy range, from about 20 MeV to 1 TeV. This enables it to probe the inverse-Compton (IC) peak in blazar SEDs and to monitor their pronounced variability on timescales from minutes to years \citep[e.g.,][]{Bottcher2013,pandey2022,penil2024}. While typical approaches to detect new sources is a stacking of all available data, blazar variability has also been used to detect faint sources that only reach the detection threshold of Fermi-LAT over a short (week to month-long) amount of time \citep[e.g.,][]{arsioli2018,kreter2020}.

In this study, 
to validate the performance of our entire analytical pipeline in identifying new gamma-ray blazars, we selected a target source catalog that has not been cross-identified: the ``4FGL-Xiang'' catalog (DR1), constructed based on Fermi-LAT observational data. This catalog was first published by \cite{Xiang2024}, who developed an all-sky automated algorithm based on Fermipy to efficiently search for new sources using multi-threaded parallel computing. After processing 15.41 years of Fermi-LAT all-sky survey data, this work ultimately identified 1379 new gamma-ray sources compared to the latest 4FGL-DR4 
\citep{Abdollahi2022,Ballet2023} with a significance exceeding 4$\sigma$, providing an excellent test sample for validating the identification capability of our pipeline.

Another motivation for performing a spatial cross-identification between the 4FGL-Xiang and 5BZCAT catalogs stems from our systematic survey of source classifications in 4FGL-DR4\footnote{\url{https://fermi.gsfc.nasa.gov/ssc/data/access/lat/14yr\_catalog/}}. We found that AGNs \footnote{The AGN subclasses include agn, bcu, bll, css, fsrq, nlsy1, rdg, sey, and ssrq, with detailed explanations available at \url{https://heasarc.gsfc.nasa.gov/W3Browse/fermi/fermilpsc.html}.} account for 56.3\% of the total $\gamma$-ray sources in the entire catalog. Among these AGNs, blazars constitute  97.09\%, indicating that blazars are among the most prevalent $\gamma$-ray sources in the all-sky survey.  
This strongly motivated us to investigate the potential associations between  5BZCAT and 4FGL-Xiang.
Identifying and confirming more blazar candidates, along with analyzing the statistical properties of BZBs and BZQs based on existing blazar samples, is of great significance for studying the classification, evolution, jet dynamics, non-thermal radiation mechanisms, and contributions of blazars to the cosmic high-energy background \citep{Abdo2010c,Volonteri2011,Massaro2012,Ackermann2015,Ajello2015,Baring2017}. 

In Section \ref{sec2}, we provide a detailed description of our data analysis methods and results, while Section \ref{sec3} presents a discussion and Section \ref{sec4.0} presents a summary of our findings.

\section{Data Analysis} \label{sec2}

\subsection{Construction of the 4FGL-Xiang-DR2 Including Elliptical Positional Uncertainties} \label{sec2.1}


Prior to conducting data analysis, we reviewed the limitations of DR1 \citep{Xiang2024}, which primarily include the limited degrees of freedom in defining the region of interest (ROI) and the absence of elliptical positional uncertainty. These issues are discussed in detail below, along with the corresponding improvement measures.

During the Fermi-LAT likelihood analysis, if the number of model degrees of freedom is large relative to the effective information of the data, the optimal spectral and spatial parameters of all sources within the ROI become poorly constrained. This destabilizes the Fermipy maximum-likelihood optimization and makes it difficult for the optimizer to converge, thereby causing the binned-likelihood analysis to fail \citep{Xiang2024}. To solve this problem, a common remedy is to restrict the number of free parameters in the ROI model to a reasonable range. However, this in turn reduces the accuracy with which Fermipy can determine the spectral and spatial properties of both the target source and the background emission within the ROI. 
In a binned likelihood analysis, \cite{Xiang2024} set the ROI as a circular area with a 5-degree radius centered on the target position\footnote{The target position refers to the fixed reference coordinates at the center of each of the 72 square observation regions shown in Figure 1 of \citet{Xiang2024}.} , and freed the spectral parameters of all sources within this area to investigate residual background radiation within a larger 30° × 30° square region.

By analyzing the position and significance level of the surrounding residual emission, \cite{Xiang2024} identified 1,379 new $\gamma$-ray sources.
For sources detected beyond the 5-degree radius, the limited degrees of freedom in the spectral parameters of surrounding background sources, combined with the broad point-spread function (PSF) of the low-energy band, increased the probability of contamination from background noise during fitting, potentially reducing the accuracy of the significance levels of these new sources. To address this, we  referred to a precise background radiation simulation scheme, centered on the target source, as recommended by the Fermi Science Support  Center\footnote{\url{https://fermi.gsfc.nasa.gov/ssc/data/access/lat/BackgroundModels.html}}, to re-evaluate the significance levels and related parameters of the 1,379 new $\gamma$-ray sources.
DR1 lacks key elliptical positional uncertainty information. 
The localization accuracy of a source is primarily determined by the spatial distribution of a limited number of high-energy photons. Due to statistical fluctuations in the arrival directions of these photons, the resulting positional uncertainty often deviates from an ideal circular shape. In the early stages of Fermi data analysis, the circular positional uncertainty was adopted as a rough approximation and applied in the Fermi-LAT Bright Source List \citep[BSL;][]{Abdo2009fermi}. 
This is because the bright $\gamma$-ray sources included in the BSL typically have high Test Statistic (TS) values and a sufficient number of high-energy photons, which make the asymmetry in the positional uncertainty distribution relatively minor. As a result, the circular approximation has only a limited impact on the analysis outcomes in such cases.

Based on observational data, the Fermi team constructed a likelihood function to characterize the probability distribution of point source positions. This function can be used to estimate the positional uncertainty of a source, and its width is proportional to the width of the PSF. 
The positional uncertainty of a source is typically represented by an elliptical distribution that defines the region of uncertainty on the sky. An error ellipse is uniquely defined by five parameters: the right ascension and declination of its center, the lengths of the semi-major and semi-minor axes, and the orientation angle of the ellipse \citep{Abdo2010a}. 
At the 95\% confidence level, there is an approximate relationship between the semi-major and semi-minor axes of the elliptical positional uncertainty (denoted as $\sigma_{x,95}$ and $\sigma_{y,95}$, respectively) and the corresponding circular positional uncertainty ($r_{95}$) \citep{Abdo2010b}: 
$r_{95} = \sqrt{\sigma_{x,95} \times \sigma_{y,95}}$. 
Compared to circular positional uncertainty, which has three degrees of freedom , elliptical positional uncertainty involves five degrees of freedom and can therefore more accurately represent the true distribution of observational errors at the same confidence level. Since the release of the 1FGL, elliptical positional uncertainty has been recognized as an important parameter and consistently included in subsequent Fermi-LAT source catalogs 
\citep[e.g.,][]{Abdo2010a,Nolan2012,Acero2015,Abdollahi2020-4FGL}.

\begin{figure*}
	\centering
	\includegraphics[width=1.05\textwidth]{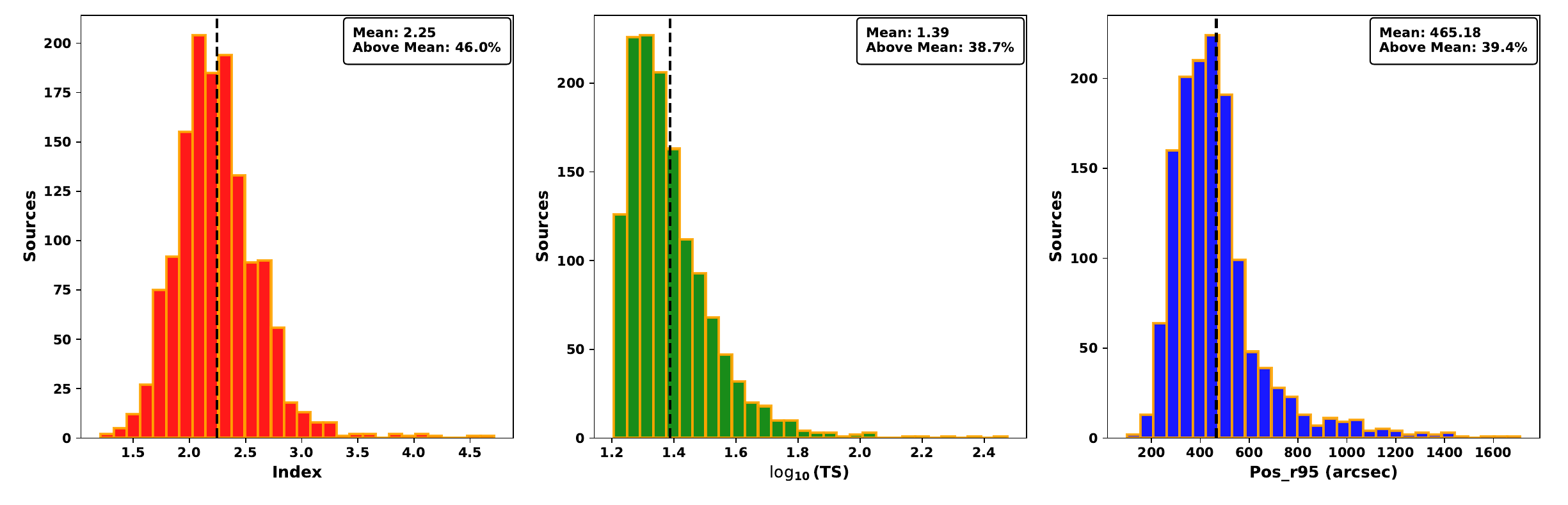}  
	\caption{
		Distributions of spectral index (left), $\rm log_{10}(TS)$ values  (middle), and $r_{95}$ positional uncertainties (right) for sources in the DR1. In all panels, the vertical black dashed line marks the mean value.
	} 
	\label{fig:index_ts_r95}
\end{figure*}

As shown in Figures 2 and 3 of \citet{Abdo2010a}, $\gamma$-ray sources with higher TS values and harder spectra tend to have smaller 95\% confidence-level positional uncertainties. This is primarily due to the narrower PSF at higher energies. This trend indicates that bright  $\gamma$-ray sources with harder spectra and a greater number of high-energy photons generally achieve better localization accuracy compared to sources with softer spectra and fewer high-energy photons.
We subsequently investigated the distributions of TS values, spectral indices, and $r_{95}$  for sources in the DR1, as illustrated in Figure \ref{fig:index_ts_r95}. 
The left panel shows that the average spectral index is 2.25, with more than 46\% of the sources exhibiting values above this average, indicating that the majority of sources have relatively soft spectra. The middle panel reveals a mean TS value of 26.3  ($\rm log_{10}(TS)=1.39$)  , with approximately 68.7\% of the sources falling below this level, suggesting that most DR1 sources are weak  $\gamma$-ray emitters. The right panel shows that the average $r_{95}$ positional uncertainty is $465.18^{\prime\prime}$ (0.13$^{\circ}$).
In Section \ref{sec2.2}, we obtained the average length of the semi-major axis of the 95\% confidence-level elliptical positional uncertainties for the 5BZCAT\_err sample, which is approximately $4^{\prime\prime}$. 
In comparison, the positional uncertainties in DR1 are nearly two orders of magnitude larger than those in 5BZCAT\_err, further emphasizing the lower localization accuracy of Fermi-LAT sources.

The above analysis indicates that most sources in DR1 are faint  $\gamma$-ray sources, generally exhibiting relatively soft spectral characteristics. In addition, their elliptical positional uncertainties are significantly larger than those of the sources in the 5BZCAT\_err. Given the directional asymmetry of positional uncertainties, approximating them using circular error models may increase the risk of mismatches in spatial cross-identification.
The use of a high-precision elliptical positional uncertainty model is essential for improving the accuracy and reliability of  subsequent spatial cross-identification.



To address the aforementioned limitations, in this analysis, we employed the Fermi-LAT data analysis toolkit \textbf{Fermipy}\footnote{https://fermipy.readthedocs.io/en/latest/} \citep{wood2017fermipy} (version 1.2.0) to investigate $\gamma$-ray sources at specified positions. Referring to the standard Binned Likelihood Tutorial provided by the Fermi Science Support Center \footnote{\url{https://fermi.gsfc.nasa.gov/ssc/data/analysis/scitools/binned_likelihood_tutorial.html}}, we performed a binned likelihood analysis for each $\gamma$-ray source in DR1, centered on its coordinates. The analysis utilized the Pass 8 photon dataset (P8R3\_V3), with the instrument response function (IRF) set to P8R3\_SOURCE\_V3. Photon events were filtered using the criteria \textbf{evtype = 3} and \textbf{evclass = 128}. The energy binning was configured with 10 bins per decade (\textbf{binsperdec = 10}), and the spatial resolution was set to 0°.1 (\textbf{binsz = 0°.1}), and the photon energy range spanned 100 MeV to 1 TeV. 
For DR1, the time range spanned August 4, 2008 (MET 239557427) to December 29, 2023 (MET 725575162).
For this analysis (DR2), the selected time range extended from August 4, 2008 (MET 239557417) to August 29, 2024 (MET 745981270). Thus, DR2 uses 8 months of additional data compared with DR1.
To minimize contamination from Earth-limb atmospheric cosmic rays, the maximum zenith angle was restricted to 90° (\textbf{zmax = 90°}). 
The input model files for the likelihood analysis included all sources within a 15° radius of the center of each analysis region from both the 4FGL-DR4 and DR1. For sources within a 6° radius of each ROI, the normalization factor and spectral index were set as free parameters  during the likelihood fitting. 
Additionally, the normalization parameters for the two diffuse background templates, namely the Galactic diffuse emission (\texttt{gll\_iem\_v07.fits}) and the extragalactic isotropic diffuse emission (\texttt{iso\_P8R3\_SOURCE\_V3\_v1.txt})\footnote{\url{https://fermi.gsfc.nasa.gov/ssc/data/access/lat/BackgroundModels.html}}, were also allowed to vary freely in the likelihood fitting.

\begin{figure*}
	\centering
	\includegraphics[width=\textwidth,height=0.6\textheight,keepaspectratio]{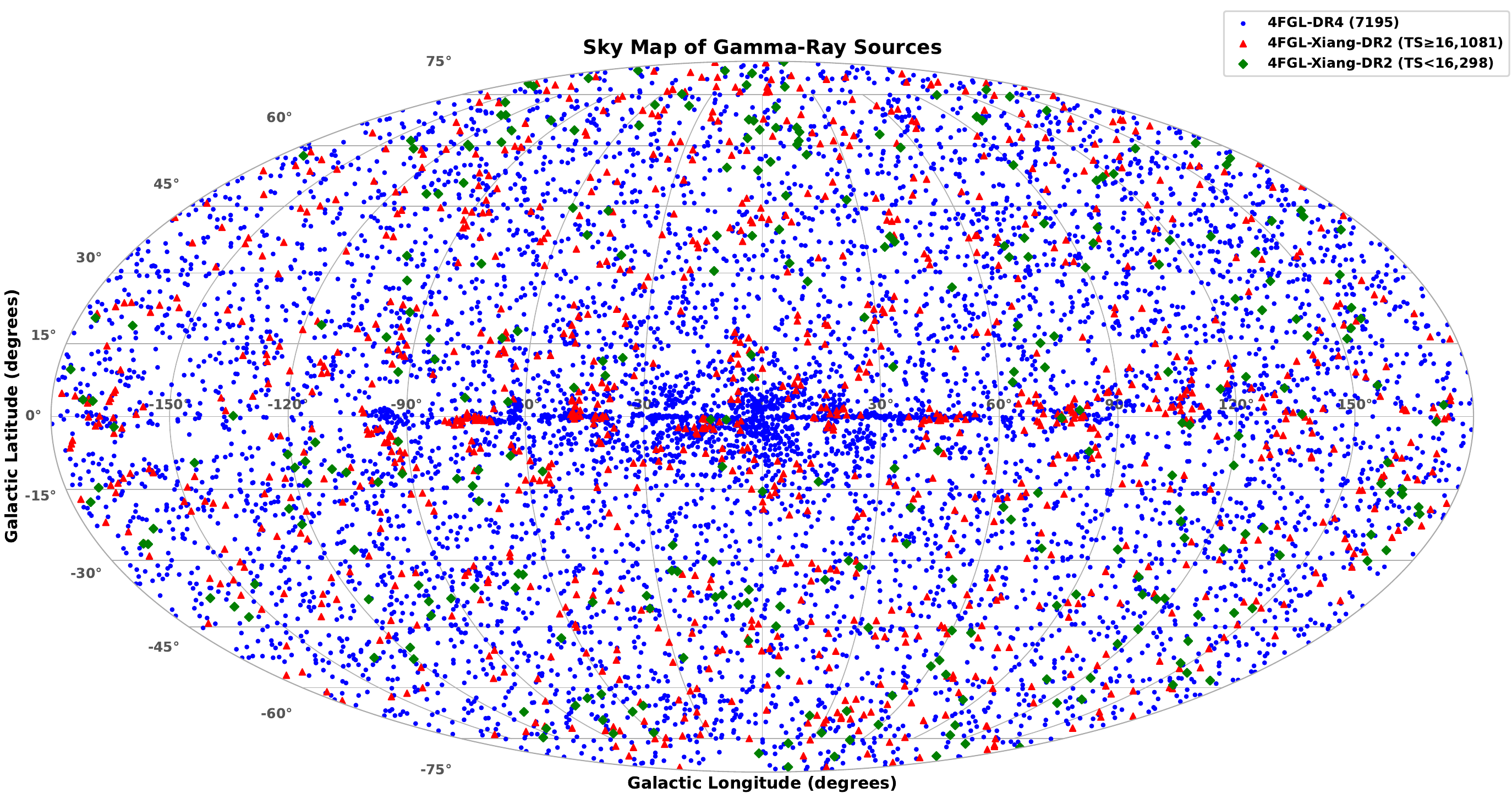}
	\caption{Sky map showing the spatial distribution of $\gamma$-ray sources from 4FGL-DR4 (blue dots, 7,195 sources) and sources in DR2 with TS $\geq 16$ (red triangles, 1,081 sources) and $< 16$ (green squares, 298 sources).}
	\label{fig:spatial_distribution}
\end{figure*}

We first employed the \textbf{GTAnalysis} module to load the configuration file containing the aforementioned parameters, to perform the initial setup and optimization procedures. 
Prior to conducting the binned likelihood fit, we invoked the \textbf{optimize()} function, which employs an iterative strategy to automatically optimize the fit for all sources within the ROI. 
Running this method early in the analysis typically ensures that all parameters are close to the global likelihood maximum, thereby improving the efficiency of subsequent fits and accelerating convergence. 
Next, we performed the likelihood fit using the \textbf{fit()} function, specifying \textbf{NEWMINUIT} as the likelihood optimizer (\textbf{optimizer=`NEWMINUIT'}) and setting \textbf{min\_fit\_quality=3} to ensure that the fit results for each source reached an optimal state.

We then used the \textbf{gta.localize()} function to refine the position of each target source by determining its best-fit coordinates and associated elliptical positional uncertainty. By enabling the \texttt{update=True} option, the optimized coordinates were automatically substituted for the original ones, thereby ensuring positional accuracy.
After obtaining the best-fit positions, we re-ran the \textbf{fit()} function to confirm that all relevant parameters were optimized at the updated locations.
Based on the above procedure, we compiled the fit results for 1,379 $\gamma$-ray sources and incorporated them into a new catalog, 4FGL-Xiang-DR2 (DR2).
The complete catalog was saved in FITS format and deposited in the China-VO Paper Data Repository (see Section~\ref{sec4}).

The parameter information for DR2 is presented in Readme.txt in Section \ref{sec4}. Compared to DR1, this version provides more comprehensive and detailed source parameters, including elliptical positional uncertainties, photon fluxes, and energy fluxes across different energy bands,  which were not included in DR1. 
Subsequently, we compared the significance levels of the 1,379 sources between DR2 and DR1.
The results show that 298 faint GeV sources, which had an average significance level of 4.5$\sigma$ in DR1, exhibit reduced significance levels in DR2, falling below the 4$\sigma$ threshold.

\subsection{Exploring the Decline in Detection Significance of 298 DR1 Sources}\label{sec3.1}

To investigate the causes of the TS value decline for the 298 sources, we examined their significance levels above 500\,MeV (Sig$_{0.5}$) in DR1. The results show that their Sig$_{0.5}$ values range from 4.01 to 5.91\,$\sigma$, with 93.29\% of the sources falling below the 5\,$\sigma$ threshold. This suggests that the majority of these objects are intrinsically weak $\gamma$-ray emitters. 
Moreover, the average photon count above 500 MeV ($N_{\mathrm{p}}$) for these sources was only 115, significantly lower than the DR1 sample average of 238.
These results indicate that these faint sources originally contained relatively few high-energy photons, making their TS values—already near the 4$\sigma$ detection threshold—susceptible to fluctuations due to variations in spectral parameters and background noise \citep{Xiang2024}. 

In this analysis, we referred to the target-centered strategy recommended by the Fermi Science Support Center and expanded the ROI radius from 5$^\circ$ to 6$^\circ$, allowing more surrounding sources and diffuse components to be included and modeled with increased degrees of freedom. 
This strategy, combined with refined background, spectral, and spatial models in DR2, may have contributed to a more realistic representation of the $\gamma$-ray environment.
In the likelihood analysis process, some photons previously attributed to the target sources might have been reassigned to nearby sources or background, potentially leading to a reduction in TS values for 298 faint GeV sources.

In \textit{Fermi}-LAT data analysis, it is common for the significance levels of certain sources to decrease with the accumulation of data and the updating of catalog parameters. This phenomenon is substantiated by a comparative analysis of the average significance level (\texttt{Signif\_Avg}) values for 6,599 sources across the 4FGL-DR3 (\texttt{gll\_psc\_v31.fit}) and 4FGL-DR4 (\texttt{gll\_psc\_v35.fit}) catalogs.
As shown in Table~\ref{tab1}, 1,931 sources in 4FGL-DR4 exhibit the lower \texttt{Signif\_Avg} values compared to their corresponding values in 4FGL-DR3, with a \texttt{Ratio\_Sig} (defined as the ratio of \texttt{Signif\_Avg\_v35} to \texttt{Signif\_Avg\_v31}) less than 1. This finding suggests that, with the accumulation of observational data and updates to the spectral and spatial model parameters, approximately 30\% of the sources exhibit a decline in significance level. 
\begin{table}
	\centering
	\caption{Comparison of \texttt{Signif\_Avg} and \texttt{Energy\_Flux100} between 4FGL-DR3 and 4FGL-DR4}
	\resizebox{\textwidth}{!}{%
		\begin{tabular}{lllllllll}
			\toprule
			Source\_Name &  RAJ2000 &  DEJ2000 &  Signif\_Avg\_v31 & Energy\_Flux100\_v31 &  Signif\_Avg\_v35 & Energy\_Flux100\_v35 &  Ratio\_Sig &  Ratio\_Eflux \\
			\midrule
			4FGL J0000.3-7355 & 0.098 & -73.922 & 7.460 & 1.58E-12 & 8.493 & 1.74E-12 & 1.138 & 1.097 \\
			4FGL J0000.5+0743 & 0.138 & 7.727 & 5.272 & 1.12E-12 & 5.681 & 1.93E-12 & 1.078 & 1.721 \\
			4FGL J0000.7+2530 & 0.188 & 25.515 & 4.177 & 8.57E-13 & 4.197 & 8.05E-13 & 1.005 & 0.939 \\
			... &... &... &... &... &... &... &...&...  \\
			\bottomrule
		\end{tabular}%
	}
	\label{tab1}
	\begin{tablenotes}
		\small
		\item \textbf{{Note.}} The complete dataset can be found in the file \texttt{Comparison\_v31\_v35.xlsx}, as referenced in Section~\ref{sec4}. The keywords in this table are uniformly presented in the format ``Keyword: Unit: Definition'', and the same format is applied to all subsequent tables. \\
		RAJ2000 : deg : Right ascension \\
		DEJ2000 : deg : Declination \\
		Signif\_Avg\_v31 : \(\sigma\) : Source significance in \(\sigma\) units over the 100 MeV to 1 TeV band in 4FGL-DR3 \\
		Signif\_Avg\_v35 : \(\sigma\) : Source significance in \(\sigma\) units over the 100 MeV to 1 TeV band in 4FGL-DR4 \\
		Ratio\_Sig : ... : Ratio of Signif\_Avg\_v35 to Signif\_Avg\_v31 \\
		Energy\_Flux100\_v31 : erg cm\(^{-2}\) s\(^{-1}\) : Energy flux from 100 MeV to 100 GeV obtained by spectral fitting in 4FGL-DR3 \\
		Energy\_Flux100\_v35 : erg cm\(^{-2}\) s\(^{-1}\) : Energy flux from 100 MeV to 100 GeV obtained by spectral fitting in 4FGL-DR4 \\
		Ratio\_Eflux : ... : Ratio of Energy\_Flux100\_v35 to Energy\_Flux100\_v31\\
	\end{tablenotes}
\end{table}

This finding strongly supports the conclusion that decreases in significance levels are a common phenomenon in \textit{Fermi}-LAT data analysis.
However, unlike 4FGL-DR4, which retained 322 sources with \texttt{Signif\_Avg} values below 4$\sigma$, we excluded such low-significance sources to enhance the statistical reliability of the sample and minimize background confusion in subsequent spectral and temporal variability analyses.
As a result, 1,082 sources were ultimately included in DR2. 
Additionally, we plotted the spatial distribution of sources from DR2 and 4FGL-DR4 in Figure~\ref{fig:spatial_distribution}. The results suggest that, except for a relatively higher density of sources along the Galactic plane, DR2 sources are approximately uniformly distributed across the entire sky.

\subsection{Construction of the Roma-BZCAT Including Elliptical Positional Uncertainties} \label{sec2.2}

Currently, 5BZCAT contains 3,561 sources across four types—5BZQ, 5BZB, 5BZU, and 5BZG—and provides preliminary classifications along with multi-wavelength flux data  \citep{Massaro2015}. However, 5BZCAT does not include positional uncertainty information, which presents significant challenges for probabilistic spatial cross-matching analyses.
The NED\footnote{\url{https://ned.ipac.caltech.edu/}} is an online astronomical database maintained by NASA in collaboration with the Infrared Processing and Analysis Center (IPAC) at the California Institute of Technology.   
It specializes in compiling comprehensive information on extragalactic objects and is continuously updated with the latest astronomical observations and research findings to ensure the timeliness and reliability of its data.

To obtain a catalog of blazar candidates with complete positional uncertainty information, we integrated the data from the 5BZCAT and the NED. 
First, we automatically obtained the positional uncertainties for the vast majority of sources, and for the few sources that were missing positional errors, we used a positional cross-identification method to complete their missing positional information.
Finally, we constructed a new catalog named 5BZCAT\_err, which contains the complete positional information for 3,561 sources, for subsequent cross-matching analysis. The detailed construction process for this catalog is described in Appendix \ref{appendixA}.
In the cross-matching analysis described in Section \ref{sec2.3}, 5BZCAT\_err served as the blazar candidate catalog for DR2 in subsequent analysis.

\begin{table*}

	\centering
	\renewcommand{\arraystretch}{1.2}
	\setlength{\tabcolsep}{5pt}
	\small
	
	\begin{threeparttable}
		\caption{Cross-Matching Results Between DR2 and 5BZCAT\_err}
		\label{tab2}
		\begin{tabular}{lllllllll}
			\hline\hline
			DR2\_name & DR2\_RA & DR2\_DEC & 5BZ\_name & 5BZ\_RA & 5BZ\_DEC & Sep &
			p$\rm _{any}$ & p$\rm _{i}$\tnote{a}\\
			\hline
			J1216.2+3451 & 183.999 & 34.800   & 5BZQ J1215+3448 & 183.982   & 34.804     & 53.762  & 99.59\% & 1 \\
			J1219.9+0200 & 185.069 & 2.029    & 5BZQ J1220+0203 & 185.050   & 2.062      & 136.678 & 99.70\% & 1 \\
			J1342.3-2901 & 205.637 & -28.970  & 5BZQ J1342-2900 & 205.564   & -29.012    & 274.991 & 98.92\% & 1 \\
			J1012.0+2313 & 153.074 & 23.243   & 5BZQ J1012+2312 & 153.068   & 23.204     & 141.115 & 99.72\% & 1 \\
			J0756.6-1534 & 119.212 & -15.682  & 5BZQ J0756-1542 & 119.211   & -15.702    & 70.995  & 99.87\% & 1 \\
			J1411.6+3407 & 212.910 & 34.070   & 5BZB J1411+3404 & 212.919   & 34.073     & 29.595  & 99.88\% & 1 \\
			J1221.3+0825 & 185.377 & 8.384    & 5BZG J1221+0821 & 185.384   & 8.362      & 80.453  & 99.64\% & 1 \\
			J1004.4+3750 & 151.173 & 37.857   & 5BZB J1004+3752 & 151.187   & 37.870     & 58.542  & 99.92\% & 1 \\
			J1207.0-1746 & 181.811 & -17.766  & 5BZB J1207-1746 & 181.798   & -17.768    & 44.797  & 99.98\% & 1 \\
			J0643.4+4215 & 100.882 & 42.210   & 5BZG J0643+4214 & 100.862   & 42.239     & 117.116 & 99.85\% & 1 \\
			J0742.9+3021 & 115.652 & 30.390   & 5BZB J0742+3018 & 115.679   & 30.310     & 299.750 & 99.45\% & 1 \\
			J0250.4-2130 & 42.546  & -21.479  & 5BZB J0250-2129 & 42.579    & -21.495    & 122.862 & 99.93\% & 1 \\
			J1243.9+4046 & 191.004 & 40.730   & 5BZQ J1243+4043 & 190.982   & 40.733     & 59.848  & 99.75\% & 1 \\
			J1336.8+0040 & 204.300 & 0.440    & 5BZB J1337+0035 & 204.381   & 0.591      & 617.766 & 88.40\% & 1 \\
			J1043.7+5323 & 161.024 & 53.376   & 5BZQ J1044+5322 & 161.044   & 53.372     & 46.200  & 99.97\% & 1 \\
			J1531.7+0845 & 232.879 & 8.867    & 5BZG J1531+0852 & 232.899   & 8.864      & 71.387  & 99.67\% & 1 \\
			J0205.4+1441 & 31.298  & 14.740   & 5BZQ J0205+1444 & 31.305    & 14.742     & 23.726  & 99.89\% & 1 \\
			\hline
		\end{tabular}
		
		\begin{tablenotes}
			\small
		    \item[a] To avoid ambiguity, we clarify that because most p$\rm _{i}$ values are at the $99.9\%$ level, they are approximately $1$ after rounding. \\			
			DR2\_Name : ... : Source name in DR2 \\
			DR2\_RA : deg : Right ascension \\
			DR2\_DEC : deg : Declination \\
			5BZ\_Name : ... : Matched source name from the 5BZCAT\_err \\
			5BZ\_RA : deg : Right ascension of the matched 5BZCAT\_err counterpart \\
			5BZ\_DEC : deg : Declination of the matched 5BZCAT\_err  counterpart \\
			Sep : arcsec : Angular separation between the source positions in DR2 and 5BZCAT\_err \\
			$\rm p_{any}$ : ... : Any-match probability \\
			$\rm p_{i}$ : ... : Individual relative match probability \\
		\end{tablenotes}
	\end{threeparttable}
	
\end{table*}

\subsection{Cross-Matching Analysis Between DR2 and 5BZCAT\_err}
\label{sec2.3}

\textbf{nway} is a numerical analysis software based on Bayesian statistics, specifically designed for multi-wavelength source cross-matching \citep{Salvato2018}. 	
It effectively incorporates positional uncertainties, magnitude information (or other priors) to compute matching probabilities, thereby improving the accuracy of cross-identification. 
In multi-wavelength cross-matching analysis, magnitude serves as an important prior in the computation of matching probabilities.
Incorporating magnitude information can effectively reduce mismatches that may arise from relying solely on positional data, which is particularly important in cases of high source density or relatively large observational uncertainties  \citep{Salvato2018}. 
\cite{Ackermann2015} examined the distribution characteristics of blazars in the Roma-BZCAT, comparing those detected and undetected by Fermi-LAT across three wavebands: radio flux density at 1.4 GHz, optical R-band magnitude, and X-ray flux in the 0.1–2.4 keV waveband.
The results indicate that $\gamma$-ray-bright blazars are, on average, brighter in all three bands, consistent with earlier studies \citep{Ackermann2011,Lister2011}.
This observed correlation supports the use of magnitude as a prior in cross-matching analyses.

We performed a probabilistic cross-matching between DR2 and 5BZCAT\_err using \textbf{nway}. To improve accuracy, we adopted an elliptical error model and incorporated optical magnitude information as a prior distribution. 
We identified high-confidence candidates by applying stringent probabilistic thresholds, requiring the match probabilities ($\rm p_{\text{any}}$ and $\rm p_i$) to each exceed 85\%. This process yielded 17 matched pairs.
These results were consistently verified using the cross-matching tool in  \textbf{TOPCAT} and independently validated through visual inspection of their spatial correlations in \textbf{DS9}, confirming a high degree of association for the matched pairs. A detailed description of the methodology and verification process can be found in Appendix \ref{appendixB}.

\subsection{Analysis of the GeV Emission Characteristics of 17 New Blazar Candidates}  \label{sec2.4}
In Section~\ref{sec2.3}, we identified 17 newly discovered  GeV blazar candidates. 
In this section, we systematically analyze the fundamental emission properties of these 17 sources using the \textbf{Fermipy} package.

\subsubsection{$\gamma$-Ray Spatial Distribution Analysis}  \label{sec2.4.1}

In this study, we conducted a spatial extension analysis of the 17 previously identified sources using the \textbf{extension()} function to evaluate their spatial distribution characteristics. For each source, we calculated the spatial extension significance ($\rm TS_{\mathrm{ext}}$) and adopted the criterion proposed by  \citet{Lande2012}, which defines a source as significantly extended if $\rm TS_{\mathrm{ext}} > 16$. The extension significance is defined as:
$\rm TS_{\mathrm{ext}} = 2 \log \left( \frac{L_{\mathrm{ext}}}{L_{\mathrm{ps}}} \right)$, 
where $\rm L_{\mathrm{ext}}$ and $\rm L_{\mathrm{ps}}$ denote the likelihood values for the extended and point source models, respectively. 

In our analysis, we applied two spatial extension templates—the two-dimensional Gaussian (2D-Gaussian) and the uniform disk (Disk) models—for source fitting. 
The radius ($\sigma$) of the spatial models was set to range from $0.1^\circ$ to $3.0^\circ$ with a step size of $0.01^\circ$. 
The threshold for spatial extension significance was set to $\rm TS_{\text{ext}} = 16$ (corresponding to \textbf{Tsqrt\_ts\_threshold = 4.0}), and the \texttt{update} parameter was set to \texttt{True} to ensure that the best-fit spatial model adopted for subsequent analyses.  
By comparing the fitting results from the two spatial models, we found that the $\rm TS_{\text{ext}}$ values derived from both the 2D-Gaussian ($\rm TS_{\text{ext, Gauss}}$) and Disk ($\rm TS_{\text{ext, Disk}}$) models were below the threshold of 16 for all 17 sources (see Table~\ref{tab3}), indicating that none of the sources exhibit significant spatial extension.
Consequently, the point source model was adopted as the best-fit model for all 17 sources in subsequent analyses.

\subsubsection{Spectral Analysis} \label{sec2.4.2}

To investigate the spectral curvature properties of the sources, we calculated the widely used spectral curvature index, $\rm TS_{cur}$ , defined as
$\rm TS_{\text{cur}} = 2 \left( \log L_{\text{curved spectrum}} - \log L_{\text{power law}} \right)$. 
According to \citet{Nolan2012}, a source is considered to exhibit significant spectral curvature if $\rm TS_{cur} > 16$. 
In this study, we employed the \textbf{curvature()} function from \textbf{Fermipy} to compute the $\rm TS_{cur}$ values for each target source under two commonly used spectral models: the Log-Parabola (LP) model and the Power-Law with Super-Exponential Cutoff (PLEC) model.

\begin{figure*}
	\centering
	\begin{subfigure}[b]{0.33\textwidth}
		\centering
		\includegraphics[width=\linewidth]{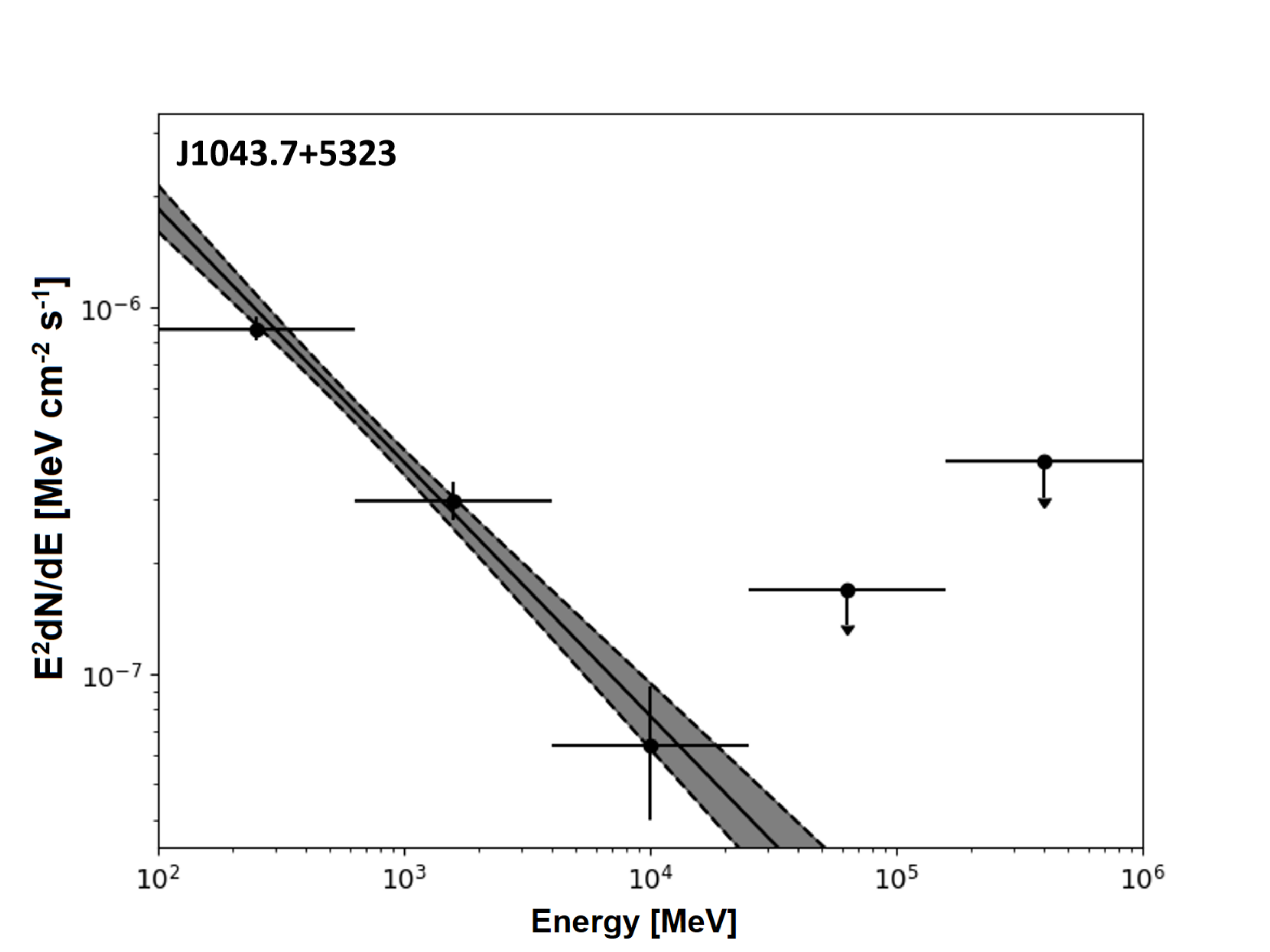}
		\caption{}
		\label{fig:subA}
	\end{subfigure}\hfill
	\begin{subfigure}[b]{0.33\textwidth}
		\centering
		\includegraphics[width=\linewidth]{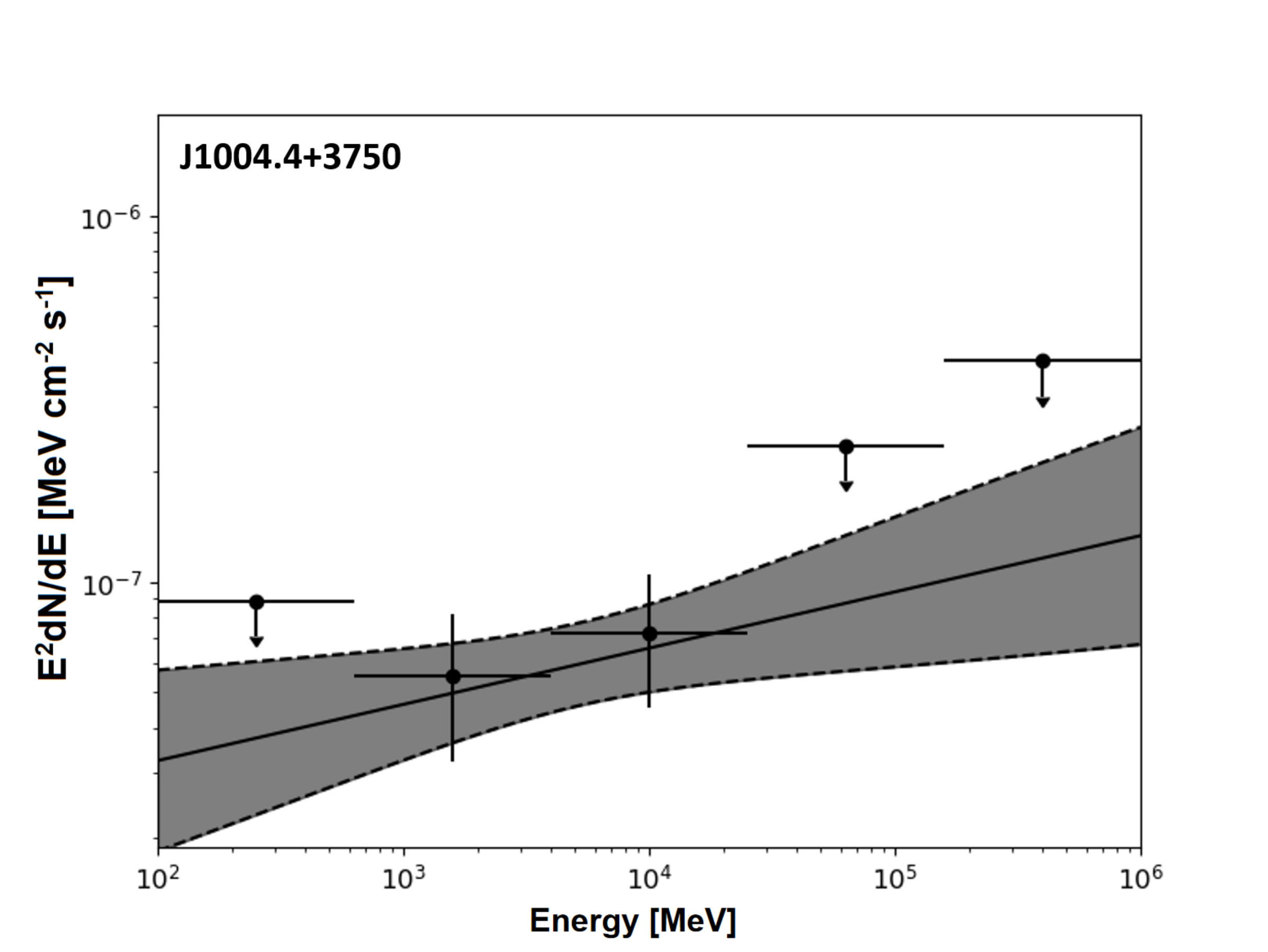}
		\caption{}
		\label{fig:subB}
	\end{subfigure}\hfill
	\begin{subfigure}[b]{0.33\textwidth}
		\centering
		\includegraphics[width=\linewidth]{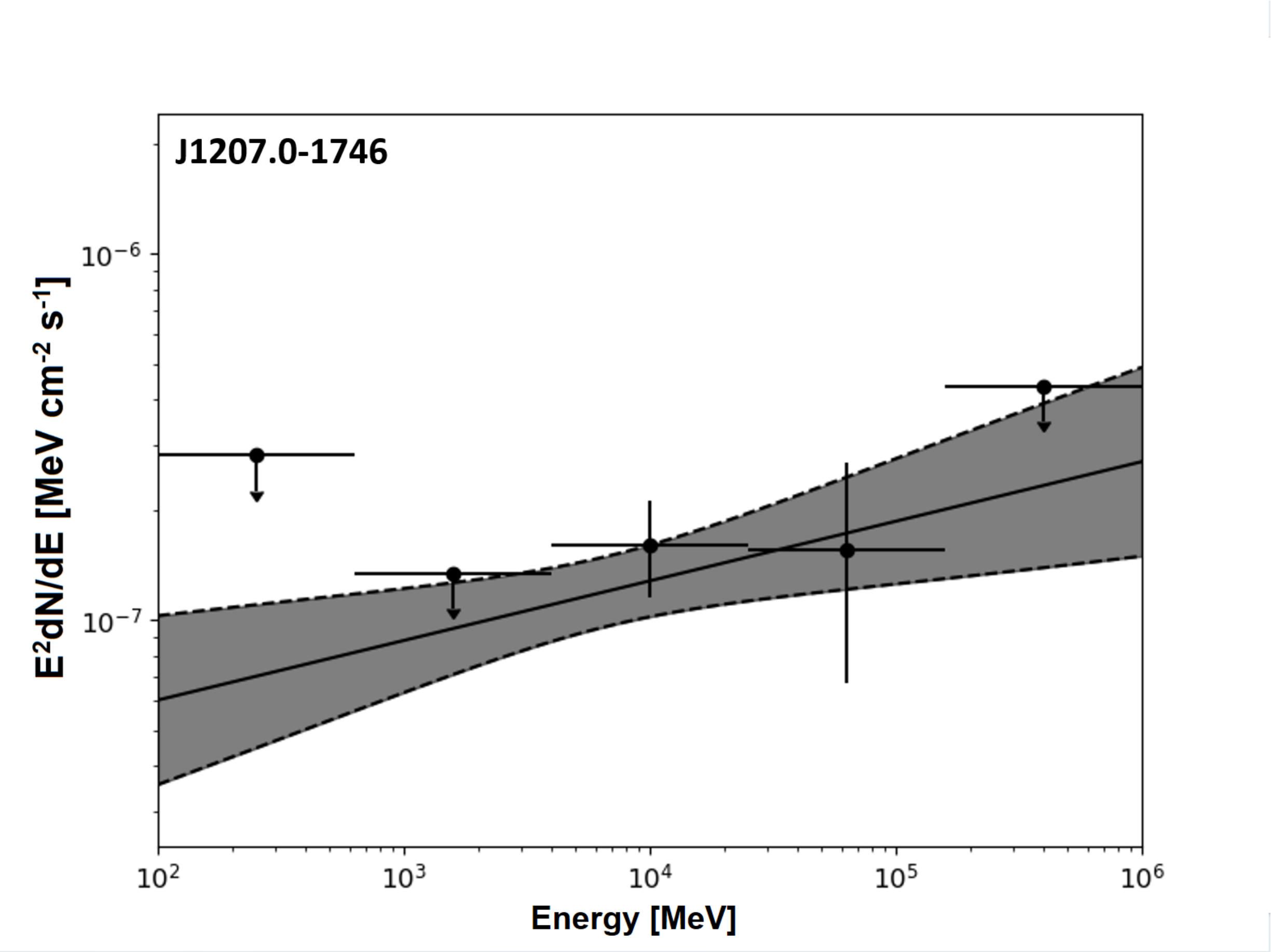}
		\caption{}
		\label{fig:subC}
	\end{subfigure}
	
	\caption{SEDs of three representative GeV sources: (a) J1043.7+5323, (b) J1004.4+3750, and (c) J1207.0-1746. The black data points represent the binned photon flux measurements with 1$\sigma$ error bars, where data points with TS values less than 4 are replaced by the 95\% confidence level upper limits. The black lines indicate the best-fit PL models, and the shaded regions represent the 1$\sigma$ uncertainties of the fits.}
	\label{fig:sed_images}
\end{figure*}

We performed spectral fitting above 100 MeV using both spectral models and calculated the corresponding $\rm TS_{cur}$ values to evaluate whether they exceeded the threshold of 16. The results show that, for all 17 sources, both $\rm TS_{cur, LP}$ and $\rm TS_{cur, PLEC}$ values are below 16 (see Table~\ref{tab3}), indicating that none of the sources exhibit significant spectral curvature.
Consequently, the power-law (PL) model was adopted as the best-fit spectral model for all subsequent analyses. The SEDs of three representative GeV blazars—(a) J1043.7+5323, (b) J1004.4+3750, and (c) J1207.0-1746—are shown in Figure~\ref{fig:sed_images}. 
All spectral data are compiled into the \texttt{SED17.tar.xz} archive, which is available in Section~\ref{sec4}.

\begin{figure*}
	\centering
	\begin{subfigure}[b]{0.7\textwidth}
		\centering
		\includegraphics[width=\textwidth]{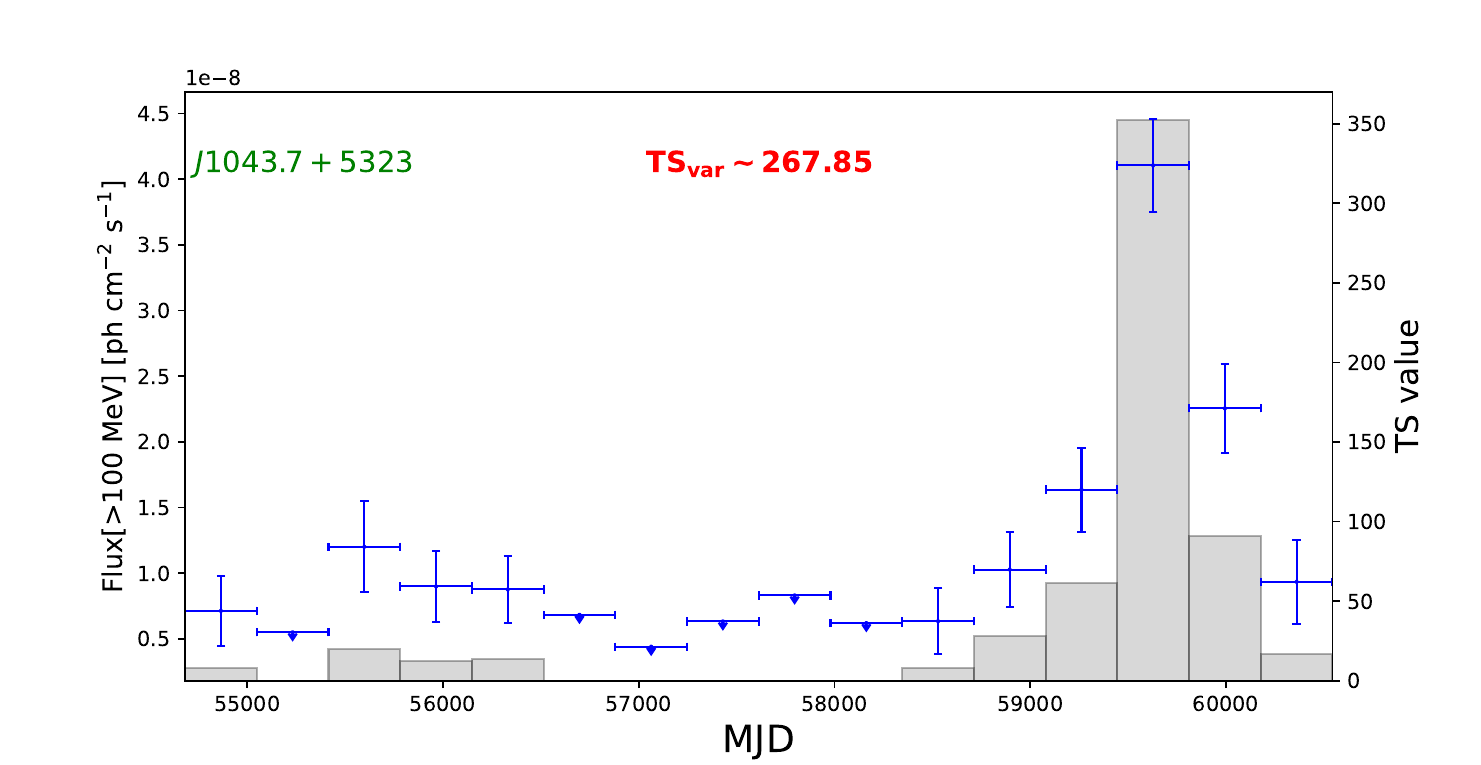}
		\caption{}
		\label{lc_fig:image1}
	\end{subfigure}
	\hfill
	\begin{subfigure}[b]{0.7\textwidth}
		\centering
		\includegraphics[width=\textwidth]{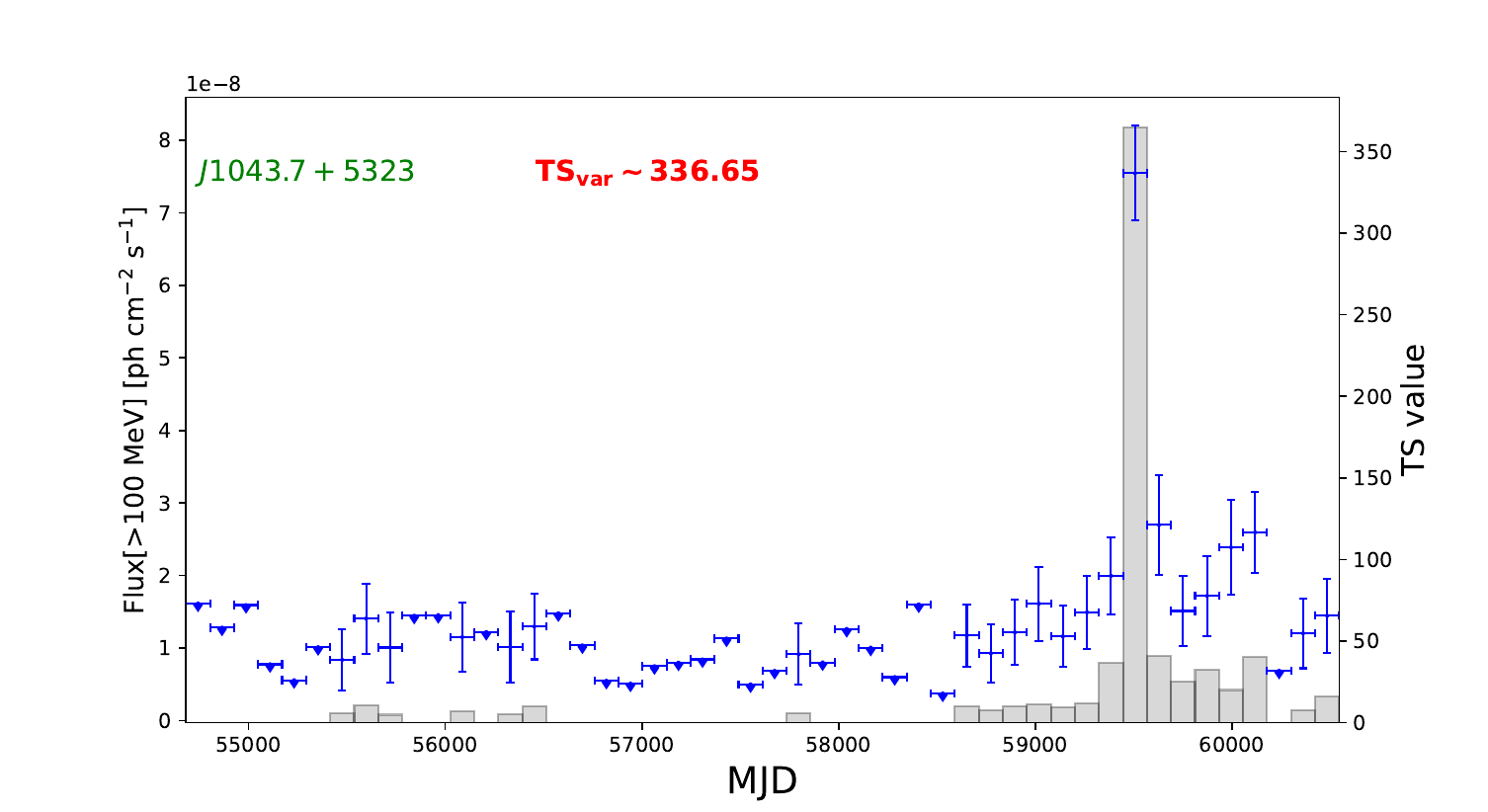}
		\caption{}
		\label{lc_fig:image2}
	\end{subfigure}
	\caption{LCs of J1043.7+5323 above 100 MeV using different time binnings. The x-axis represents the Modified Julian Date (MJD) in days. 
		(a) LC with yearly binning (16 time bins) yields a variability index of $\rm TS_{\mathrm{var}} \approx 267.85$;
		(b) LC with 4-month binning (48 time bins) shows a  variability index of $\rm TS_{\mathrm{var}} \approx 336.65$. 
		Blue points represent the photon fluxes with $1\sigma$ uncertainties, where data points with TS values $<$ 4 are replaced by 95\% confidence level upper limits.
		Gray bars indicate the TS values for each time bin.	}
	\label{fig:two_images}
\end{figure*}

\begin{table*}
	
	\centering
	\renewcommand{\arraystretch}{1.2}
	\setlength{\tabcolsep}{5pt}
	\small
	
	\begin{threeparttable}
	\caption{Fitting Results of Spatial Models for 17 Sources}
\centering
\label{tab3}
		\begin{tabular}{llccccccl}
			\hline\hline
			Name\_5BZCAT & Name\_DR2 & SpectrumType & $ \mathrm{Index}\ $ & $\rm TS_{cur,lp}$ & $\rm TS_{cur,plec}$ & $\rm TS_{ext,disk}$ & $\rm TS_{ext,gauss}$ & $\rm TS_{var}$ \\
			\hline
			5BZB J1411+3404 & J1411.6+3407 & PL & $2.02 \pm 0.19$ & 0 & 0.63 & 1.02 & 0.17 & 8.65 \\
			5BZG J1221+0821 & J1221.3+0825 & PL & $2.03 \pm 0.14$ & 5.06 & 5.07 & 3.17 & 2.62 & 12.46 \\
			5BZB J1004+3752 & J1004.4+3750 & PL & $1.84 \pm 0.18$ & 3.09 & 3.09 & 0 & -0.01 & 9.75 \\
			5BZB J1207-1746 & J1207.0-1746 & PL & $1.83 \pm 0.18$ & 0 & 7.58 & 0.09 & 0.09 & 10.83 \\
			5BZG J0643+4214 & J0643.4+4215 & PL & $1.81 \pm 0.20$ & 0 & 2.14 & 2.40 & 3.99 & 6.16 \\
			5BZB J0742+3018 & J0742.9+3021 & PL & $2.61 \pm 0.27$ & 0 & -0.01 & 2.89 & 2.29 & 9.56 \\
			5BZB J0250-2129 & J0250.4-2130 & PL & $2.07 \pm 0.25$ & 0 & 0.48 & 0 & 0 & 11.88 \\
			5BZQ J1215+3448 & J1216.2+3451 & PL & $1.95 \pm 0.22$ & 0 & 2.16 & 2.30 & 3.72 & 8.95 \\
			5BZQ J1220+0203 & J1219.9+0200 & PL & $1.99 \pm 0.19$ & 0 & 0.12 & 1.08 & 0.81 & 1.73 \\
			5BZQ J1342-2900 & J1342.3-2901 & PL & $2.17 \pm 0.15$ & 7.87 & 5.67 & 3.24 & 1.11 & 0.68 \\
			5BZQ J1012+2312 & J1012.0+2313 & PL & $2.34 \pm 0.19$ & 0 & 6.15 & 0.05 & 1.51 & 16.94 \\
			5BZB J1337+0035 & J1336.8+0040 & PL & $2.16 \pm 0.17$ & 0 & 4.33 & 0 & 0 & 5.34 \\
			5BZQ J1044+5322 & J1043.7+5323 & PL & $2.70 \pm 0.08$ & 0 & 0.84 & 0 & 0 & 267.85 \\
			5BZG J1531+0852 & J1531.7+0845 & PL & $2.13 \pm 0.20$ & 0 & 0 & 0 & 0 & 11.11 \\
			5BZQ J0756-1542 & J0756.6-1534 & PL & $2.47 \pm 0.13$ & 0 & 0 & 0 & 0 & 24.73 \\
			5BZQ J1243+4043 & J1243.9+4046 & PL & $1.98 \pm 0.23$ & 0 & 1.23 & -0.01 & -0.01 & 14.08 \\
			5BZQ J0205+1444 & J0205.4+1441 & PL & $2.19 \pm 0.23$ & 0 & 0.09 & 0.18 & 0.14 & 19.29 \\
			\hline
		\end{tabular}
		
		\begin{tablenotes}
			\small   
			\item[] \textbf{Note:}\ Name\_5BZCAT : ... : Source name from 5BZCAT  \\
			Name\_DR2 : ... : Source name in DR2  \\
			SpectrumType   : ... : The best-fit  Spectral model  \\
			Index  : ... :  The best-fit  Spectral Index \\    
			$\rm TS_{cur,lp}$ : ... : Curvature TS value for the LogParabola spectral model  \\
			$\rm TS_{cur,plec}$ : ... : Curvature TS value for PowerLaw with exponential cutoff model  \\
			$\rm TS_{ext,disk}$ : ... : Extension TS value for Disk model  \\
			$\rm TS_{ext,gauss}$ : ... : Extension TS value for 2D-Gaussian model  \\
			$\rm TS_{var}$ : ... : Variability index  \\
		\end{tablenotes}
	\end{threeparttable}
	
\end{table*}

\subsubsection{Variability Analysis} \label{sec2.4.3}

In this study, we performed variability analyses for 17 sources using the \textbf{lightcurve()} function. Each source's light curve (LC) was divided into 16 time bins, with each bin corresponding to one year. 	
Following the method proposed by  \citet{Nolan2012}, we calculated the variability index $\rm TS_{\text{var}}$ for each source (see Table~\ref{tab3}). 
According to statistical criteria, a source is considered significantly variable at the 99\% confidence level if $\rm TS_{\text{var}} > 30.58$, in which case the source can be identified as variable.
Our analysis revealed that among the 17 sources, J1043.7+5323 exhibited a high variability index of $\rm TS_{\text{var}} = 267.85$,  above the threshold, indicating significant flux variations in its LC, see Figure~8(\subref{lc_fig:image1}).
To further investigate this behavior, we generated a LC for J1043.7+5323 with a finer time binning of four months per bin. The result still showed significant variability, with $\rm TS_{\text{var}} = 336.65$, exceeding the corresponding 99\% confidence threshold of 72.44, see Figure~8(\subref{lc_fig:image2}). 
For the remaining 16 sources, no significant variability was observed in their LCs. All LCs are compiled in the file \texttt{LC17.tar.xz}, accessible via Section~\ref{sec4}.

\subsection{Statistical Analysis of Multi-Wavelength Flux Distributions of 17 New GeV Blazar Candidates} \label{sec2.5}

This section presents our in-depth statistical investigation of blazar flux distributions. First, we describe the statistical methods used. We then apply these methods to the BZB and BZQ samples for a quantitative comparison. Based on these findings, we model the multi-band flux distributions for both subclasses. Finally, we explore using this statistical model to construct a new classification criterion.

\subsubsection{Investigating the differences in multi-Wavelength flux distributions Between BZBs and BZQs} \label{sec2.6.1}

To ensure comprehensive spectral coverage and sufficient observational data, this study selected flux measurements spanning from radio to $\gamma$-ray bands as the targets of analysis. Specifically, fluxes at 74 MHz, 365 MHz, 843 MHz, 1.4 GHz, 5 GHz, 15 GHz, 20 GHz, 143 GHz, 0.1–2.4 keV, 0.3–10 keV, and 0.1–100 GeV were included.  
To systematically investigate the statistical properties of blazar fluxes  across multiple wavebands,  we selected the 5BZCAT\_err sample for analysis and designed a standardized data extraction procedure.  
Following the pipeline in Appendix \ref{appendixC.1}, we retrieved the available flux measurements for 3,442 sources across the aforementioned bands, with per-source coverage ranging from one to eleven bands.

To investigate the differences in the multi-wavelength flux distributions of BZBs and BZQs, we have designed a specialized analysis pipeline aimed at in-depth statistical characterization (see Appendix \ref{appendixC.2} for detailed procedures).  We first performed a logarithmic transformation on the flux data of the 5BZCAT\_err blazars and calculated eight common statistical parameters, such as the mean, median, skewness, and kurtosis, to comprehensively characterize their distribution properties. Subsequently, we introduced the Median Absolute Deviation (MAD) to quantify the degree of difference in various statistical indicators across different bands. Based on this, we analyzed the differences between BZBs and BZQs in terms of their distribution center, dispersion, and shape. The MAD analysis results show that BZBs and BZQs exhibit varying degrees of statistical differences in their multi-wavelength flux distributions, with the difference in kurtosis being the most prominent (MAD = 1.64).

In addition to using MAD analysis to assess differences in statistical parameters, this study also introduces the Jensen-Shannon distance (JSD) as a supplementary metric to evaluate the differences between BZBs and BZQs in their overall multi-wavelength flux distribution characteristics. The JSD is capable of assessing the overall difference between two distributions based on their probability density distributions (PDDs). The JSD analysis reveals that the flux distribution differences between BZBs and BZQs are highly significant in the 1.4 GHz, 843 MHz, 5 GHz, 0.1–2.4 keV, and 0.3–10 keV bands (JSD > 0.3). In other bands, although the differences are relatively smaller, the JSD values remain positive, indicating that the two source types exhibit systematic distributional differences of varying degrees across all bands (see Appendix \ref{appendixC.2} for detailed procedures).

In summary, both the MAD values of the eight classical statistical indicators and the JSD analysis results are  consistently greater than zero, providing strong evidence that BZBs and BZQs exhibit systematic differences in their flux distributions.

\subsubsection{Modeling multi-band flux distributions via Box–Cox transformation and truncated normal distributions}  \label{sec2.6.2}

According to the analysis in Section~\ref{sec2.6.1}, the flux distributions of BZQs and BZBs show varying degrees of divergence across the radio to $\gamma$-ray bands, with kurtosis emerging as the most discriminative statistical metric. 
The flux kurtosis of BZQs is consistently higher than that of BZBs across all wavebands, with the most significant difference occurring at 143 GHz. 
However, if we solely consider employing the kurtosis at 143 GHz as a classification criterion for blazars, it presents two key challenges:

\textbf{(1)} Although higher kurtosis values were observed in the 143 GHz band, it is important to note that the sample size in this band is extremely limited, with only 8 data points for BZBs and 50 for BZQs. 
Given such a small sample, the estimation of the distributional characteristics of BZBs is likely subject to considerable statistical uncertainty. 
Moreover, kurtosis captures only a single aspect of a distribution and cannot comprehensively characterize its overall characteristics. 
Therefore, relying on the kurtosis at 143\,GHz as a classification criterion for BZBs and BZQs is not statistically robust.

\textbf{(2)} Except for the 143\,GHz band, differences in flux distributions across other wavebands are not taken into account.
Relying solely on a single-band criterion is insufficient to fully characterize the multi-band radiative properties arising from the jets and accretion disks of BZBs and BZQs.

In summary, to evaluate the common statistical characteristics of the multi-band fluxes of the 17 new GeV blazars and the 5BZCAT\_err sample, and to provide robust statistical support for classification and identification, two key issues should be addressed: 
\begin{enumerate}[label=(\alph*)]
	\item The limitations of using single-band criteria for classification; 
	\item The lack of a unified statistical model and reliable indicators for quantifying the shared characteristics of flux distributions.
\end{enumerate}

To address the first problem, we adopted an approach that focuses on commonly available multi-band flux data, thereby avoiding the limitations of relying on single-waveband criteria for classification and identification. 

To address the second problem, we designed a statistical modeling approach that combines the Box–Cox transformation with the truncated normal distribution (TND) to systematically characterize the multi-band flux distributions of BZBs and BZQs. 
To achieve a unified modeling of multi-band flux distributions, we propose a statistical method that combines the Box-Cox transformation and the TND. 
The procedure first applies the Box-Cox transformation to normalize the data, and then fits the data with the TND model. 
The goodness-of-fit (GoF) is evaluated using the Kolmogorov--Smirnov (KS) and Cram\'{e}r--von Mises (CVM) tests, with a requirement of $p > 0.05$.  
Additionally, we use model fit plots and Q-Q plots as auxiliary diagnostic tools to visually assess the GoF  between the observed data and the theoretical distribution.

The initial analysis indicates that certain BZB/BZQ subsamples in 5BZCAT---across the radio (74\,MHz, 843\,MHz, 1.4\,GHz, 5\,GHz) and X-ray (0.1--2.4\,keV, 0.3--10\,keV) bands---initially failed the KS or CVM tests due to the presence of outliers. 
To address this, we applied the Local Outlier Factor (LOF) algorithm to remove a few outliers, after which the cleaned  data successfully passed the test. 
Notably, the data from most bands met the fitting criteria without requiring additional processing. 
This result confirms the general applicability and robustness of our proposed ``$\mathrm{Box\text{-}Cox+TND}$'' model. 
The complete details of this modeling procedure are provided in Appendix~\ref{appendixC.3}.

\subsubsection{Application of the model} \label{sec2.6.3}

In Section \ref{sec2.1}, data analysis was performed for the 100 MeV to 1 TeV energy band, yielding the corresponding photon flux. However, in Section \ref{sec2.6.1}, the high-energy band for blazars is defined as 100 MeV to 100 GeV. 
To maintain consistency in the energy range, \textbf{Fermipy} was employed to recalculate the photon flux for the 17 blazar candidates within the 100 MeV to 100 GeV band, in preparation for subsequent statistical analysis. 
Then, based on the Box–Cox transformation parameters \( \lambda \) listed in Table~\ref{tab11}, we uniformly transformed the flux values of the 17 blazar candidates across each observational waveband using Equation~\ref{equ5}.

\begin{figure*}
	\captionsetup[subfigure]{skip=2pt} 
	\captionsetup{aboveskip=2pt, belowskip=2pt} 
	\centering
	\begin{subfigure}[b]{0.30\textwidth}
		\includegraphics[width=\linewidth]{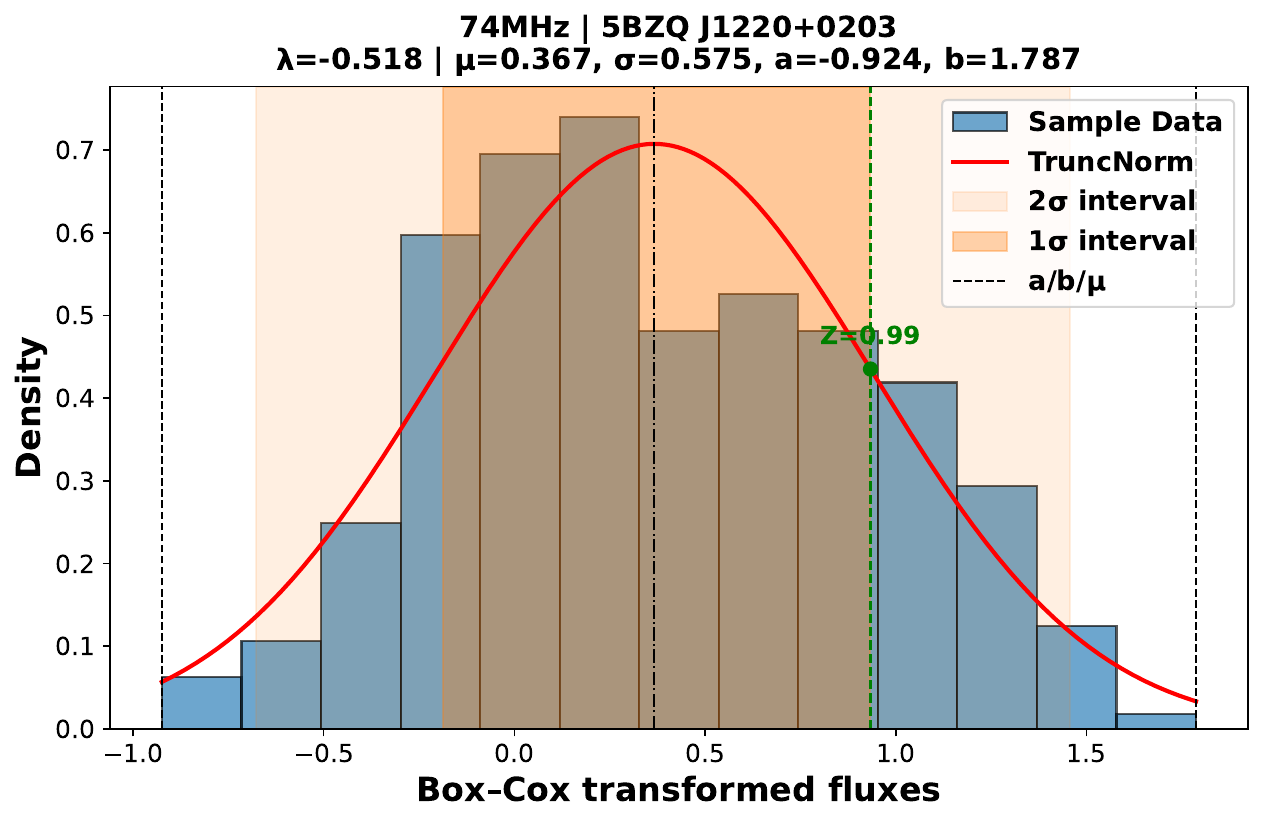}
		\caption{}
		\label{fig_box:image1}
	\end{subfigure}\hspace{0.02\textwidth}
	\begin{subfigure}[b]{0.30\textwidth}
		\includegraphics[width=\linewidth]{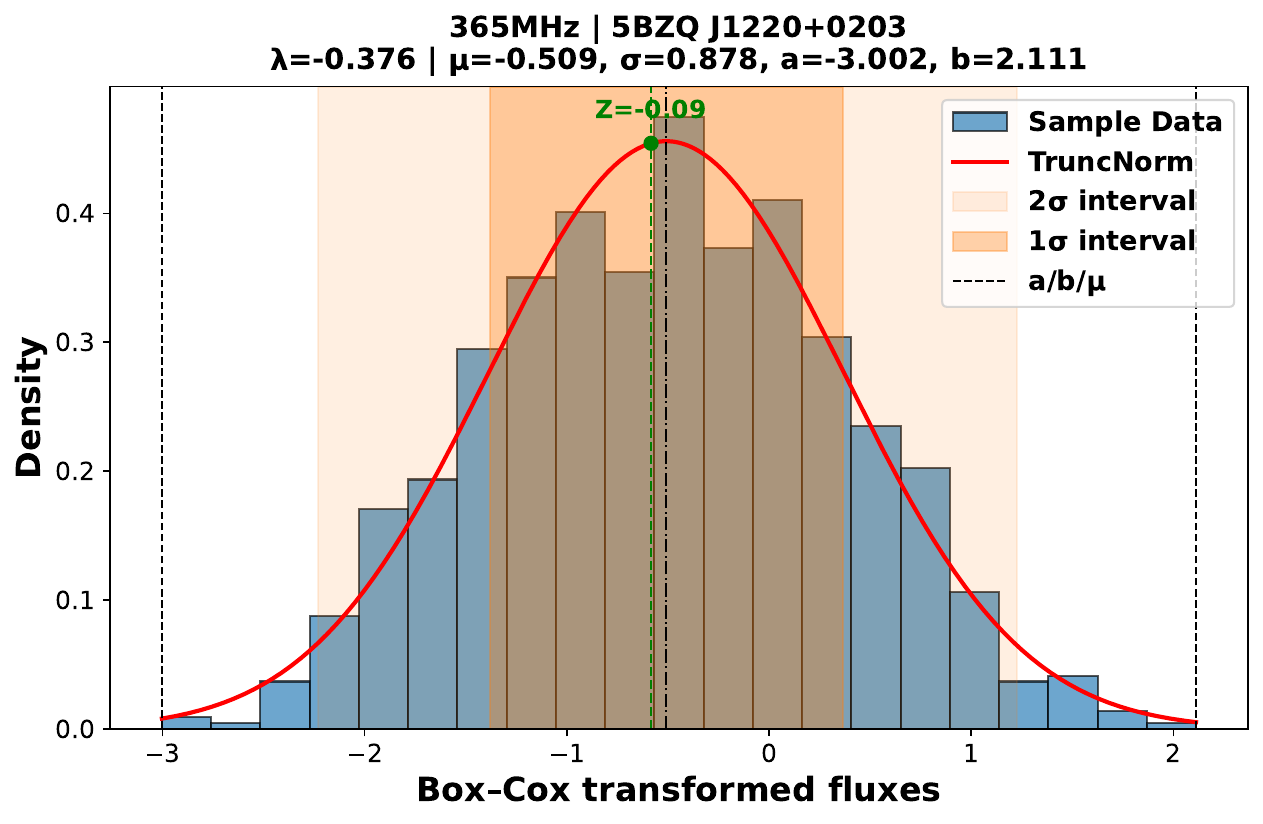}
		\caption{}
		\label{fig_box:image2}
	\end{subfigure}\hspace{0.02\textwidth}
	\begin{subfigure}[b]{0.30\textwidth}
		\includegraphics[width=\linewidth]{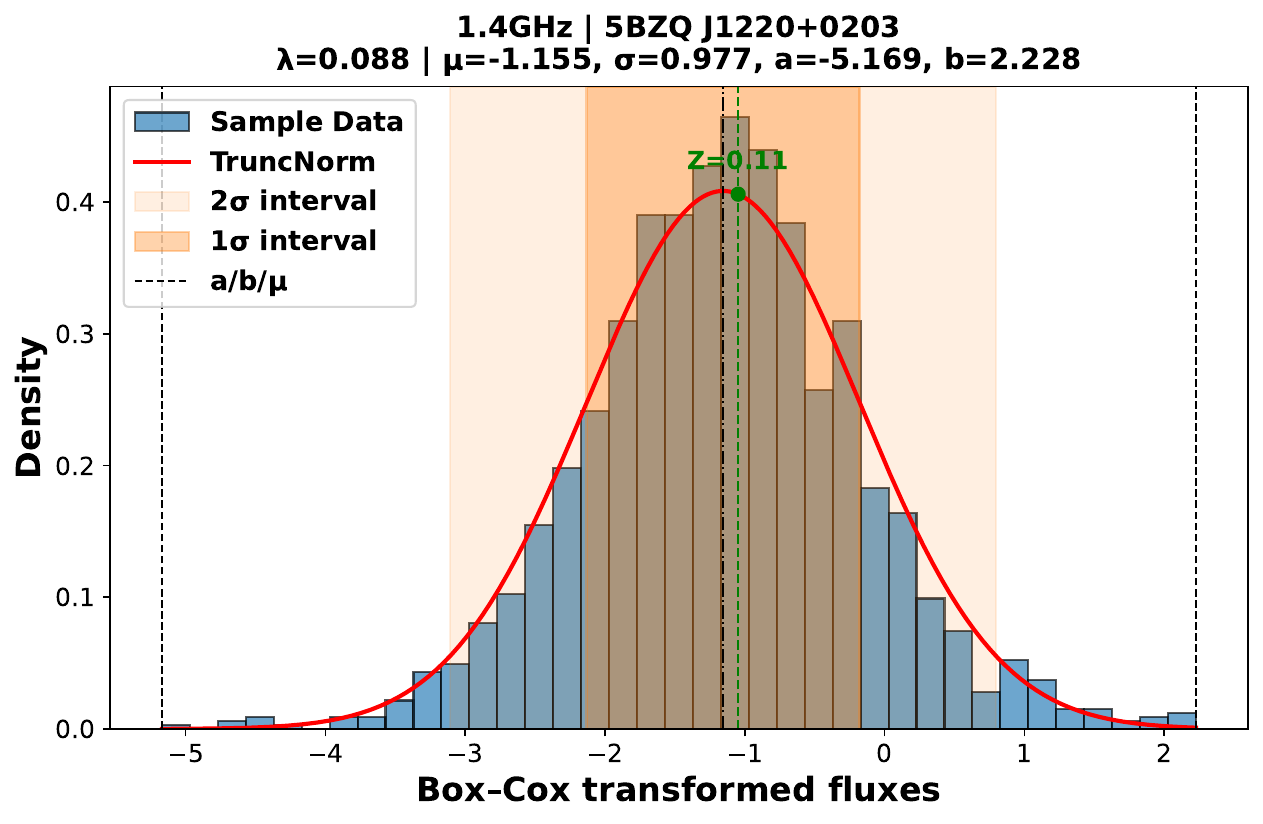}
		\caption{}
		\label{fig_box:image3}
	\end{subfigure}
	
	\vspace{0.25em}
	
	\begin{subfigure}[b]{0.30\textwidth}
		\includegraphics[width=\linewidth]{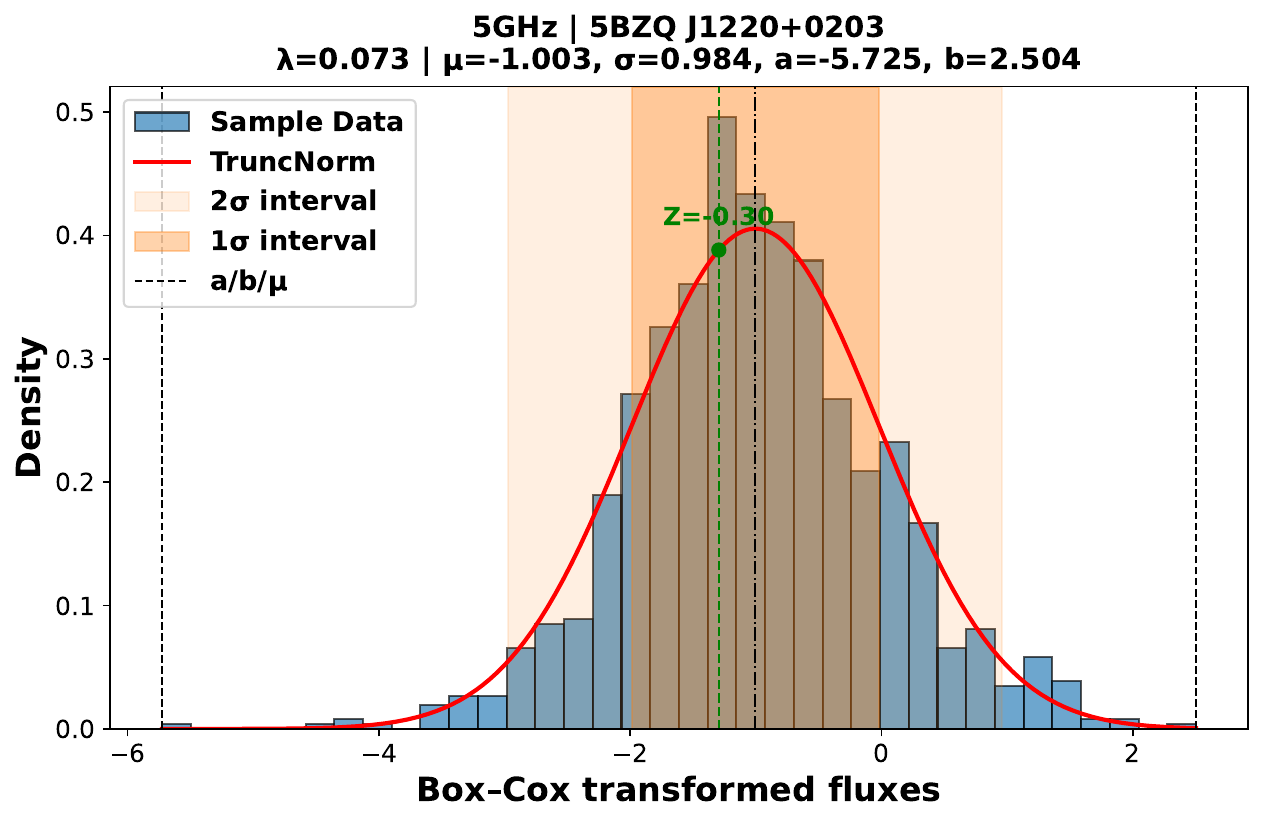}
		\caption{}
		\label{fig_box:image4}
	\end{subfigure}\hspace{0.02\textwidth}
	\begin{subfigure}[b]{0.30\textwidth}
		\includegraphics[width=\linewidth]{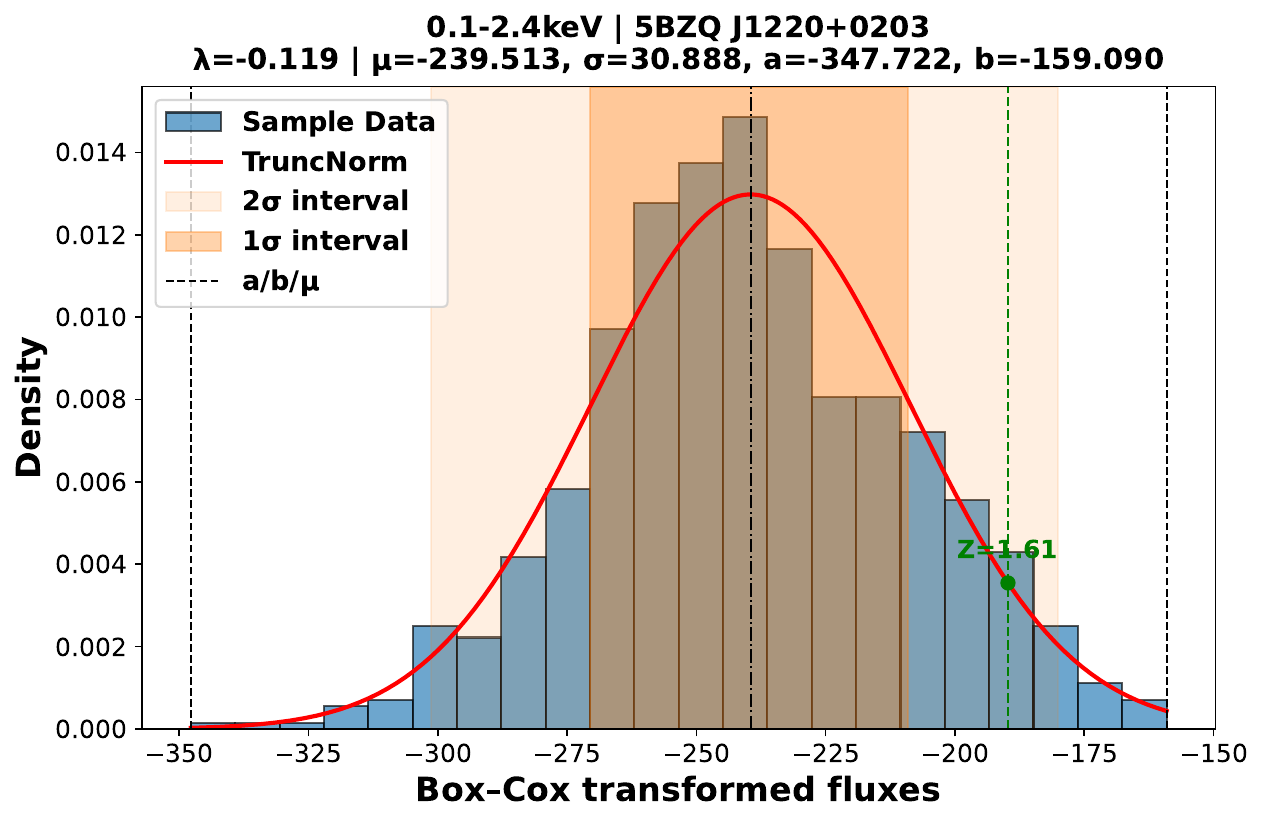}
		\caption{}
		\label{fig_box:image5}
	\end{subfigure}\hspace{0.02\textwidth}
	\begin{subfigure}[b]{0.30\textwidth}
		\includegraphics[width=\linewidth]{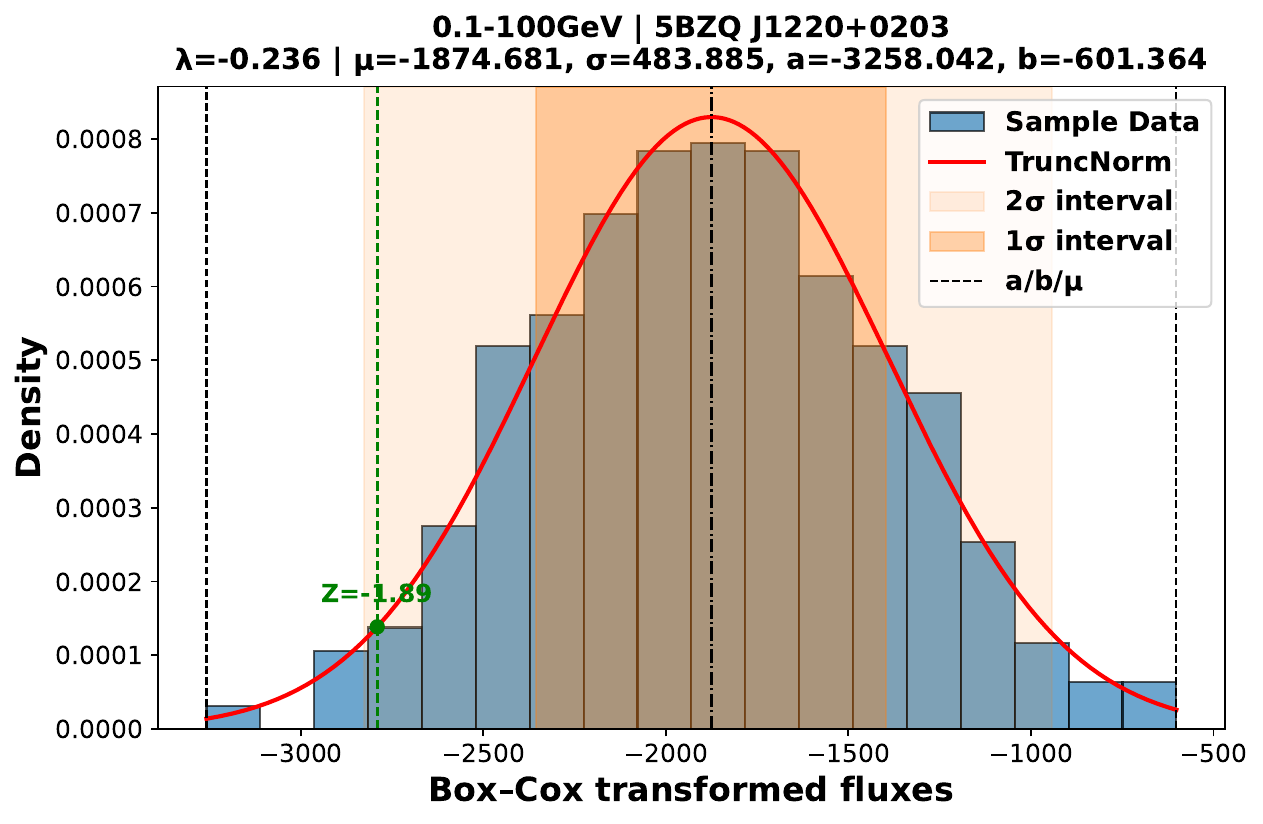}
		\caption{}
		\label{fig_box:image6}
	\end{subfigure}
	
	\caption{The figure displays the flux distribution analysis results for 5BZQ J1220+0203 across the 74 MHz, 365 MHz, 1.4 GHz, 5 GHz, 0.1–2.4 keV, and 0.1–100 GeV bands.   
		The header data added in each figure caption are taken from Table \ref{tab11}.
		The blue histogram depicts the density distribution of Box-Cox transformed flux values from a large sample, while the red curve represents the best-fit TND. The black dashed lines on both sides denote the truncation of the TND, the black dashed line in the middle represents the sample mean, the dark orange region represents the 1$\sigma$ CI, and the light orange region represents the 2$\sigma$ CI. The green point marks the position of 5BZQ J1220+0203, accompanied by its annotated z-score value.}
	\label{modle_fit:0203}
\end{figure*}

The resulting Box–Cox transformed values are plotted on the optimal TND fitting curves corresponding to each waveband and source type,  facilitating an intuitive comparison of the relative positions of each candidate within the overall distribution. 
Using 5BZQ J1220+0203 as an example, we present the statistical analysis results of its multi-band flux distribution in Figure~\ref{modle_fit:0203}, 
with the transformed data points annotated on the optimal fitted curve to highlight their positions relative to the 1$\sigma$ and 2$\sigma$  confidence intervals (CIs). 
Similar plots for the remaining 16 candidates are included in the archive \texttt{Model\_fitting\_results.tar.xz}, as referenced in Section~\ref{sec4}. The corresponding z-score statistics for each candidate are summarized in Table~\ref{tab4}.

As shown in Figure \ref{modle_fit:0203}, based on the z-score analysis of 5BZQ J1220+0203 across six wavebands, we found that the transformed fluxes in the four bands—74 MHz, 365 MHz, 1.4 GHz, and 5 GHz—all lie within the 1$\sigma$ CI of the modeled distribution. This indicates that, in these wavebands, the flux of 5BZQ J1220+0203 is relatively typical and shows no significant deviations.  
In the 0.1–2.4 keV and 0.1–100 GeV bands, the transformed fluxes lie within the 2$\sigma$ CI of the modeled distribution, suggesting moderate deviations that nonetheless remain within a statistically acceptable range. 
This indicates that the radiation characteristics of 5BZQ J1220+0203 in these wavebands exhibit strong similarity to those observed in larger samples.  
As discussed in Section \ref{sec3.1}, the differences in multi-band flux distributions between BZBs and BZQs are highly likely governed by redshift and three physical mechanisms. Therefore, we speculate that 5BZQ J1220+0203 likely shares similar intrinsic properties and emission behaviors with the larger BZQ population.



\begin{longtable}{lllllr}	
	\caption{Results of Statistical Analysis for 17 New GeV Blazars} 
	\label{tab4} \\
	\toprule
	Name & Waveband & Type &  Box–Cox Flux& CI  & Z-score \\
	\midrule
	\endfirsthead
	\toprule
	Name & Waveband & Type & Box–Cox Flux& CI  & Z-score \\
	\midrule
	\endhead
	\endfoot

	5BZB J0250-2129 & 0.1-2.4keV & BZB & -34.047 & $1\sigma$ & 0.575 \\
	5BZB J0250-2129 & 0.1-100GeV & BZB & -3341.606 & $2\sigma$ & -1.757 \\
	
	5BZB J0742+3018 & 1.4GHz & BZB & -4.114 & $1\sigma$ & -0.511 \\
	5BZB J0742+3018 & 0.1-100GeV & BZB & -2987.876 & $2\sigma$ & -1.174 \\
	
	5BZB J1004+3752 & 1.4GHz & BZB & -3.278 & $1\sigma$ & -0.049 \\
	5BZB J1004+3752 & 0.1-2.4keV & BZB & -37.332 & $1\sigma$ & -0.693 \\
	5BZB J1004+3752 & 0.1-100GeV & BZB & -3299.560 & $2\sigma$ & -1.687 \\
	
	5BZB J1207-1746 & 1.4GHz & BZB & -6.318 & $2\sigma$ & -1.727 \\
	5BZB J1207-1746 & 0.1-2.4keV & BZB & -33.848 & $1\sigma$ & 0.651 \\
	5BZB J1207-1746 & 0.1-100GeV & BZB & -2814.007 & $1\sigma$ & -0.887 \\
	
	5BZB J1337+0035 & 1.4GHz & BZB & -2.123 & $1\sigma$ & 0.588 \\
	5BZB J1337+0035 & 0.1-2.4keV & BZB & -38.838 & $2\sigma$ & -1.275 \\
	5BZB J1337+0035 & 0.1-100GeV & BZB & -3051.207 & $2\sigma$ & -1.278 \\
	
	5BZB J1411+3404 & 0.1-2.4keV & BZB & -37.816 & $1\sigma$ & -0.880 \\
	5BZB J1411+3404 & 0.1-100GeV & BZB & -3310.612 & $2\sigma$ & -1.706 \\
	
	5BZG J0643+4214 & 5GHz & BZG & -3.327 & $1\sigma$ & -0.865 \\
	5BZG J0643+4214 & 0.1-2.4keV & BZG & -32.470 & $2\sigma$ & 1.183 \\
	5BZG J0643+4214 & 0.1-100GeV & BZG & -3174.972 & $2\sigma$ & -1.482 \\
	
	5BZG J1221+0821 & 365MHz & BZG & -1.401 & $1\sigma$ & -0.777 \\
	5BZG J1221+0821 & 1.4GHz & BZG & -1.819 & $1\sigma$ & 0.756 \\
	5BZG J1221+0821 & 5GHz & BZG & -3.418 & $1\sigma$ & -0.925 \\
	5BZG J1221+0821 & 0.1-2.4keV & BZG & -36.161 & $1\sigma$ & -0.242 \\
	5BZG J1221+0821 & 0.1-100GeV & BZG & -2854.699 & $1\sigma$ & -0.954 \\
	
	5BZG J1531+0852 & 0.1-100GeV & BZG & -3048.081 & $2\sigma$ & -1.273 \\
	
	5BZQ J0205+1444 & 1.4GHz & BZQ & -1.510 & $1\sigma$ & -0.363 \\
	5BZQ J0205+1444 & 0.1-100GeV & BZQ & -2736.900 & $2\sigma$ & -1.782 \\
	
	5BZQ J0756-1542 & 74MHz & BZQ & 0.565 & $1\sigma$ & 0.344 \\
	5BZQ J0756-1542 & 365MHz & BZQ & -0.075 & $1\sigma$ & 0.495 \\
	5BZQ J0756-1542 & 1.4GHz & BZQ & -0.179 & $2\sigma$ & 0.999 \\
	5BZQ J0756-1542 & 5GHz & BZQ & 0.150 & $2\sigma$ & 1.172 \\
	5BZQ J0756-1542 & 20GHz & BZQ & -0.332 & $1\sigma$ & 0.852 \\
	5BZQ J0756-1542 & 0.1-100GeV & BZQ & -2230.560 & $1\sigma$ & -0.735 \\
	
	5BZQ J1012+2312 & 1.4GHz & BZQ & -1.766 & $1\sigma$ & -0.626 \\
	5BZQ J1012+2312 & 5GHz & BZQ & -1.237 & $1\sigma$ & -0.238 \\
	5BZQ J1012+2312 & 15GHz & BZQ & -1.264 & $1\sigma$ & -0.109 \\
	5BZQ J1012+2312 & 0.1-2.4keV & BZQ & -244.877 & $1\sigma$ & -0.174 \\
	5BZQ J1012+2312 & 0.3-10keV & BZQ & -16.937 & $1\sigma$ & -0.363 \\
	5BZQ J1012+2312 & 0.1-100GeV & BZQ & -2690.863 & $2\sigma$ & -1.687 \\
	
	5BZQ J1044+5322 & 365MHz & BZQ & -0.810 & $1\sigma$ & -0.343 \\
	5BZQ J1044+5322 & 1.4GHz & BZQ & -0.837 & $1\sigma$ & 0.326 \\
	5BZQ J1044+5322 & 5GHz & BZQ & -1.149 & $1\sigma$ & -0.149 \\
	5BZQ J1044+5322 & 15GHz & BZQ & -1.334 & $1\sigma$ & -0.167 \\
	5BZQ J1044+5322 & 0.1-2.4keV & BZQ & -247.188 & $1\sigma$ & -0.248 \\
	5BZQ J1044+5322 & 0.1-100GeV & BZQ & -2053.905 & $1\sigma$ & -0.370 \\
	
	5BZQ J1215+3448 & 74MHz & BZQ & 0.954 & $2\sigma$ & 1.022 \\
	5BZQ J1215+3448 & 365MHz & BZQ & 0.720 & $2\sigma$ & 1.400 \\
	5BZQ J1215+3448 & 1.4GHz & BZQ & 0.347 & $2\sigma$ & 1.538 \\
	5BZQ J1215+3448 & 5GHz & BZQ & -0.311 & $1\sigma$ & 0.703 \\
	5BZQ J1215+3448 & 0.1-2.4keV & BZQ & -259.501 & $1\sigma$ & -0.647 \\
	5BZQ J1215+3448 & 0.1-100GeV & BZQ & -3013.069 & $>2\sigma$ & -2.353 \\
	
	5BZQ J1220+0203 & 74MHz & BZQ & 0.933 & $2\sigma$ & 0.986 \\
	5BZQ J1220+0203 & 365MHz & BZQ & -0.584 & $1\sigma$ & -0.085 \\
	5BZQ J1220+0203 & 1.4GHz & BZQ & -1.048 & $1\sigma$ & 0.110 \\
	5BZQ J1220+0203 & 5GHz & BZQ & -1.293 & $1\sigma$ & -0.295 \\
	5BZQ J1220+0203 & 0.1-2.4keV & BZQ & -189.758 & $2\sigma$ & 1.611 \\
	5BZQ J1220+0203 & 0.1-100GeV & BZQ & -2790.174 & $2\sigma$ & -1.892 \\
	
	5BZQ J1243+4043 & 1.4GHz & BZQ & -1.539 & $1\sigma$ & -0.393 \\
	5BZQ J1243+4043 & 5GHz & BZQ & -1.688 & $1\sigma$ & -0.697 \\
	5BZQ J1243+4043 & 0.1-100GeV & BZQ & -3293.402 & $>2\sigma$ & -2.932 \\
	
	5BZQ J1342-2900 & 365MHz & BZQ & -1.560 & $2\sigma$ & -1.197 \\
	5BZQ J1342-2900 & 1.4GHz & BZQ & -1.123 & $1\sigma$ & 0.033 \\
	5BZQ J1342-2900 & 5GHz & BZQ & -0.836 & $1\sigma$ & 0.169 \\
	5BZQ J1342-2900 & 20GHz & BZQ & -0.855 & $1\sigma$ & 0.425 \\
	5BZQ J1342-2900 & 0.1-100GeV & BZQ & -2774.527 & $2\sigma$ & -1.860 \\
	
	\bottomrule
	\multicolumn{6}{@{}l}{
		\parbox[t]{0.65\textwidth}{
			\textbf{Note:} Name : ... : Name of the source \\
			Waveband : ... : Observational frequency band \\ 
			Type : ... : Source class   \\
			Box–Cox Flux : ... : Box–Cox transformed flux value   \\
			CI : ... : The range where the target object lies within the confidence interval  \\
			Z-score : ... : The Z-score of the target source is based on TND  \\.
		}
	}   
\end{longtable}

\subsection{Exploring the Likely Physical Origins of Differences in Multi-Wavelength Flux Distribution Between BZBs and BZQs}
\label{sec3.1}

The statistical differences observed in the multi-wavelength flux distributions between BZBs and BZQs likely stem from fundamental distinctions in their underlying physical properties, such as radiation mechanisms, jet formation processes, and relativistic beaming effects. 
In the following sections, we provide a comprehensive analysis of the intrinsic relationships between these physical mechanisms and the multi-wavelength flux characteristics of BZBs and BZQs.

(1) Different Radiation Mechanisms

\begin{figure*}
	\centering   
	\includegraphics[width=0.8\textwidth]{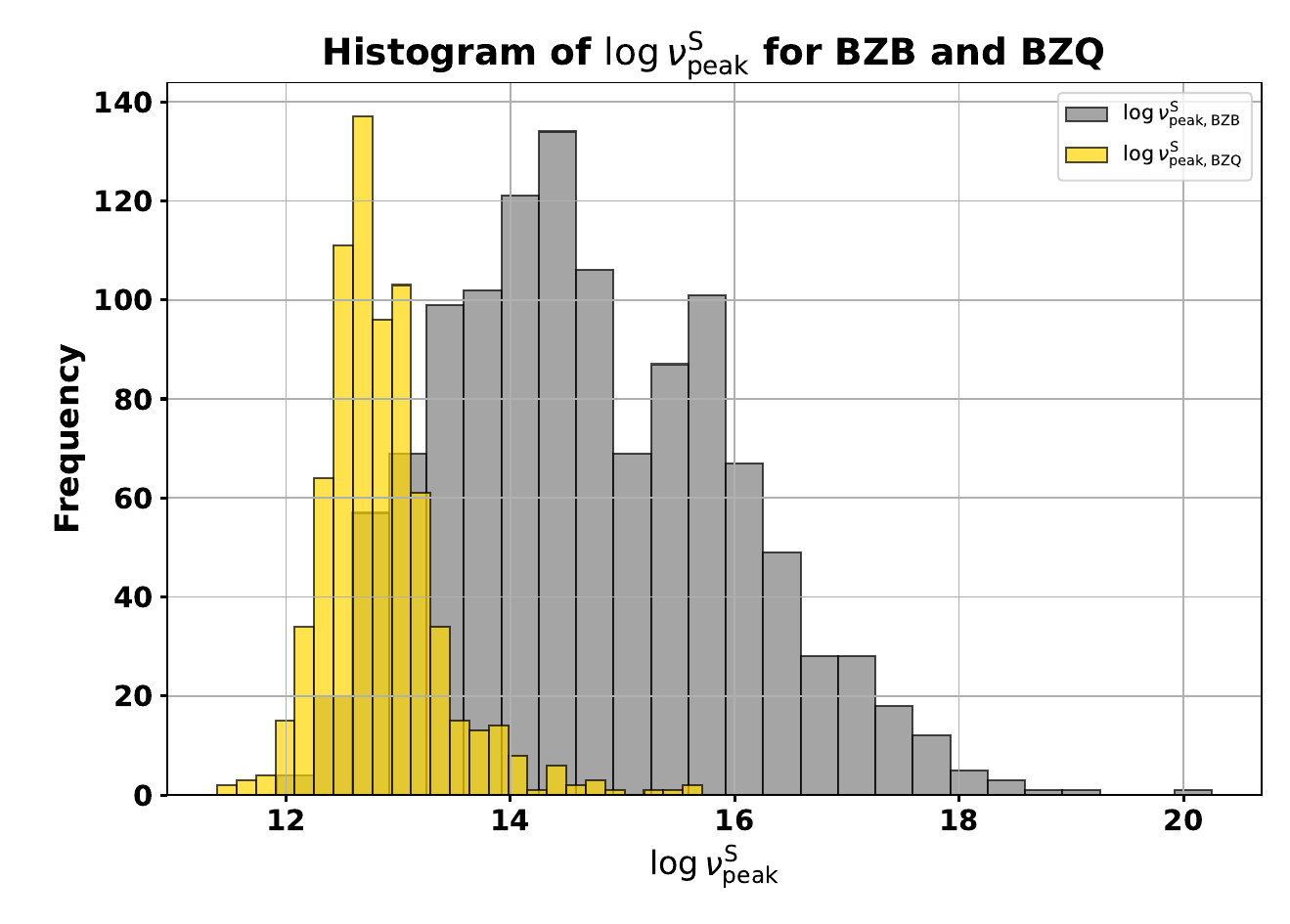}
	\caption{This figure shows the distributions of \(\log \nu_{\mathrm{peak,BZB}}^{\mathrm{S}}\) and \(\log \nu_{\mathrm{peak,BZQ}}^{\mathrm{S}}\), representing the synchrotron peak frequencies of BZBs and BZQs, shown in grey and yellow, respectively. The relative data are taken from \citet{ajello2022fourth}.
	}    
	\label{histogram_plot_Vpeak}
\end{figure*}

Studies by \citet{Bottcher2013} and \citet{Abdo2010c} indicate that the leptonic model effectively explains the quasi-simultaneous SEDs of most blazars in the high-energy band, thereby offering strong support for the hypothesis that their high-energy emission stems from leptonic processes. 
Within the framework of leptonic models, different radiation mechanisms dominate at various frequency bands. 
For BZBs, the emission from radio to X-ray bands is primarily produced by synchrotron radiation from relativistic electrons in the jet’s magnetic field, while their SEDs in the GeV to TeV band are typically dominated by the synchrotron self-Compton (SSC)  process.  
In contrast, BZQs often exhibit prominent EC components, resulting from the scattering of external photon fields—such as those originating from the BLR or the dusty torus—by relativistic electrons. 
The dominance of EC processes not only contributes significantly to the low-frequency emission but also enhances the high-energy component and shifts the peak frequency of their SEDs to lower values \citep{Fossati1998, Ghisellini1998, Bottcher2013, Fan2016}. 
Systematic differences in radiation mechanisms are likely to result in variations in the synchrotron peak frequencies (\( \nu_{\mathrm{peak}}^{\mathrm{S}} \)) . The study by \citet{Abdo2010c} provides robust evidence in support of this inference. 
Through a comprehensive analysis of the multi-wavelength radiation properties of 48 bright $\gamma$-ray blazars observed by Fermi-LAT, 
they demonstrated that the distributions of \( \nu_{\mathrm{peak}}^{\mathrm{S}} \) differ significantly between BZBs and BZQs. 
To provide stronger statistical support for this difference, we first extracted the \(\log(\nu_{\mathrm{peak}}^{\mathrm{S}})\) data from 4LAC-DR3 \citep{ajello2022fourth} and constructed histograms of the  \(\log(\nu_{\mathrm{peak}}^{\mathrm{S}})\) for both BZBs and BZQs. 
As shown in Figure~\ref{histogram_plot_Vpeak}, a clear distinction is observed in their distribution characteristics: the peak frequencies of BZQs are primarily concentrated in the lower-frequency range with a relatively narrow spread, whereas those of BZBs exhibit a broader range  and are generally shifted toward higher frequencies.

\begin{table*}
	
	\centering
	\scriptsize
	\renewcommand{\arraystretch}{1.3}
	
	\begin{threeparttable}
		\caption{Statistical Comparison of Physical Properties between BZBs and BZQs}  \label{tab5}
		\begin{tabular}{cccccccccc} 
			\toprule
			Sample & Min & Max & Mean & Median & Mode & Std & Skewness & Kurtosis & JSD \\
			\midrule
			\multicolumn{10}{c}{\(\log \nu_{\mathrm{peak,BZB}}^{\mathrm{S}}\) vs \(\log \nu_{\mathrm{peak,BZQ}}^{\mathrm{S}}\)} \\
			\midrule
			\(\log \nu_{\mathrm{peak,BZB}}^{\mathrm{S}}\) & 11.92 & 20.26 & 14.72 & 14.55 & 16.03 & 1.29 & 0.44 & -0.16 & \multirow{2}{*}{0.58} \\
			\(\log \nu_{\mathrm{peak,BZQ}}^{\mathrm{S}}\) & 11.39 & 15.71 & 12.85 & 12.77 & 13.04 & 0.53 & 1.38 & 4.30 &  \\ 
			\midrule
			\multicolumn{10}{c}{$\log P_{\mathrm{jet,BZ,BZB}}$ vs $\log P_{\mathrm{jet,BP,BZQ}}$} \\
			\midrule
			$\log P_{\mathrm{jet,BZ,BZB}}$ & 43.17 & 46.86 & 44.52 & 44.48 & 44.34 & 0.58 & 0.67 & 1.39 & \multirow{2}{*}{0.67} \\
			$\log P_{\mathrm{jet,BP,BZQ}}$ & 41.08 & 49.17 & 46.67 & 46.85 & 46.85 & 1.15 & -1.21 & 2.56 & \\
			\midrule
			\multicolumn{10}{c}{$\log R_{\mathrm{v,BZB}}$ vs $\log R_{\mathrm{v,BZQ}}$} \\
			\midrule
			$\log R_{\mathrm{v,BZB}}$ & 0.15 & 3.99 & 2.23 & 2.15 & 0.15 & 0.81 & 0.01 & -0.16 & \multirow{2}{*}{0.51} \\
			$\log R_{\mathrm{v,BZQ}}$ & 1.90 & 4.83 & 3.51 & 3.57 & 1.90 & 0.60 & -0.47 & -0.19 & \\
			\midrule
			\multicolumn{10}{c}{$z_{\mathrm{BZB}}$ vs $z_{\mathrm{BZQ}}$} \\
			\midrule
			$z_{\text{BZB}}$ & 0.00 & 3.53 & 0.42 & 0.31 & 0.70 & 0.39 & 2.68 & 10.66 & \multirow{2}{*}{0.48} \\
			$z_{\text{BZQ}}$ & 0.03 & 4.31 & 1.19 & 1.11 & 0.72 & 0.66 & 0.81 & 0.86 & \\
			\bottomrule
		\end{tabular}
		
		\renewcommand{\arraystretch}{1}
		\vspace{2mm}
		
		\begin{tablenotes}
			
			\footnotesize
			\item[] \textbf{Note:}\ 
			Sample : ... : Research sample \\
			Min : ... : Minimum of the  distribution  \\
			Max : ... : Maximum of the  distribution  \\
			Mean : ... : Mean of the distribution  \\
			Median : ... : Median of the distribution  \\
			Mode : ... : Mode of the distribution  \\
			Std : ... : Standard deviation  \\
			Skewness : ... :Skewness of the distribution  \\
			Kurtosis : ... : Kurtosis of the distribution  \\
			JSD : ... : Jensen–Shannon distance  \\
		\end{tablenotes}
	\end{threeparttable}
\end{table*}

Subsequently, we conducted a comparative analysis following the technical procedure outlined in Content~4 of Figure~\ref{technical-route3}. 
As shown in Table~\ref{tab5}, BZBs exhibit significantly higher values of \( \nu_{\mathrm{peak}}^{\mathrm{S}} \) than BZQs, characterized by a broader distribution range (min = 11.92, max = 20.26) and a higher mean value (14.72 for BZBs vs.  12.85 for BZQs). 
The std of \(\log \nu_{\mathrm{peak,BZB}}^{\mathrm{S}}\) distribution (std = 1.29) is more than approximately twice that of \(\log \nu_{\mathrm{peak,BZQ}}^{\mathrm{S}}\) (std = 0.53), 
indicating  substantially greater dispersion. Moreover, the skewness and kurtosis values suggest that \(\log \nu_{\mathrm{peak,BZB}}^{\mathrm{S}}\) distribution is more symmetric and flatter, while that of BZQs displays strong positive skewness (Skewness = 1.38) and a sharply peaked distribution (Kurtosis = 4.30). 
Meanwhile, the JSD value of 0.58 further confirms the significant difference between the \(\log(\nu_{\mathrm{peak}}^{\mathrm{S}})\) distributions between BZBs and BZQs  from the perspective of the overall probability density profiles. 

Based on the above analysis, we suggest that differences in the underlying radiation mechanisms play a key role in producing the differences observed in the multi-band SEDs, thereby influencing the distribution characteristics of the  \(\log(\nu_{\mathrm{peak}}^{\mathrm{S}})\). 
The significant discrepancy between \(\log \nu_{\mathrm{peak,BZB}}^{\mathrm{S}}\) and \(\log \nu_{\mathrm{peak,BZQ}}^{\mathrm{S}}\) likely contributes to the differences observed in the flux distributions at specific bands, such as 1.4\,GHz, 0.1--2.4\,keV, and other wavebands.

(2) Different Jet Formation Mechanisms

Currently, two mainstream models are widely accepted to explain the formation of blazar jets: the Blandford–Znajek (BZ) and Blandford–Payne (BP) models. 
In the BZ model, the jet is powered by the rotational energy of the black hole, with its energy output governed by both the spin and mass of the black hole \citep{BZ1977}. In contrast, the BP model drives the jet by extracting energy from the rotational motion of the accretion disk \citep{BP1982}.  Based on the analysis of the relationship between jet power and normalized disk luminosity, \citet{Xiao2022} suggested that the jets of BL Lacs are more likely powered by black hole spin, consistent with the BZ mechanism, where jet energy is extracted through the interaction between a rotating black hole and its magnetic field. 
In contrast, the jets of BZQs are more likely driven by the rotational energy of the accretion disk, consistent with the BP mechanism, in which magnetic field lines anchored in the disk extract and accelerate material to form jets \citep{Xiao2022}. 
Furthermore, \citet{Xie2024} found that the jet powers of BZQs, as estimated from radio flux density, closely match the theoretical predictions of the BP mechanism, whereas those of BZBs  are  more consistent with the expectations of the BZ mechanism. 
This finding further highlights that BZBs and BZQs  differ markedly in their jet formation mechanisms.

\begin{figure*}
	\centering
	\includegraphics[width=0.8\textwidth]{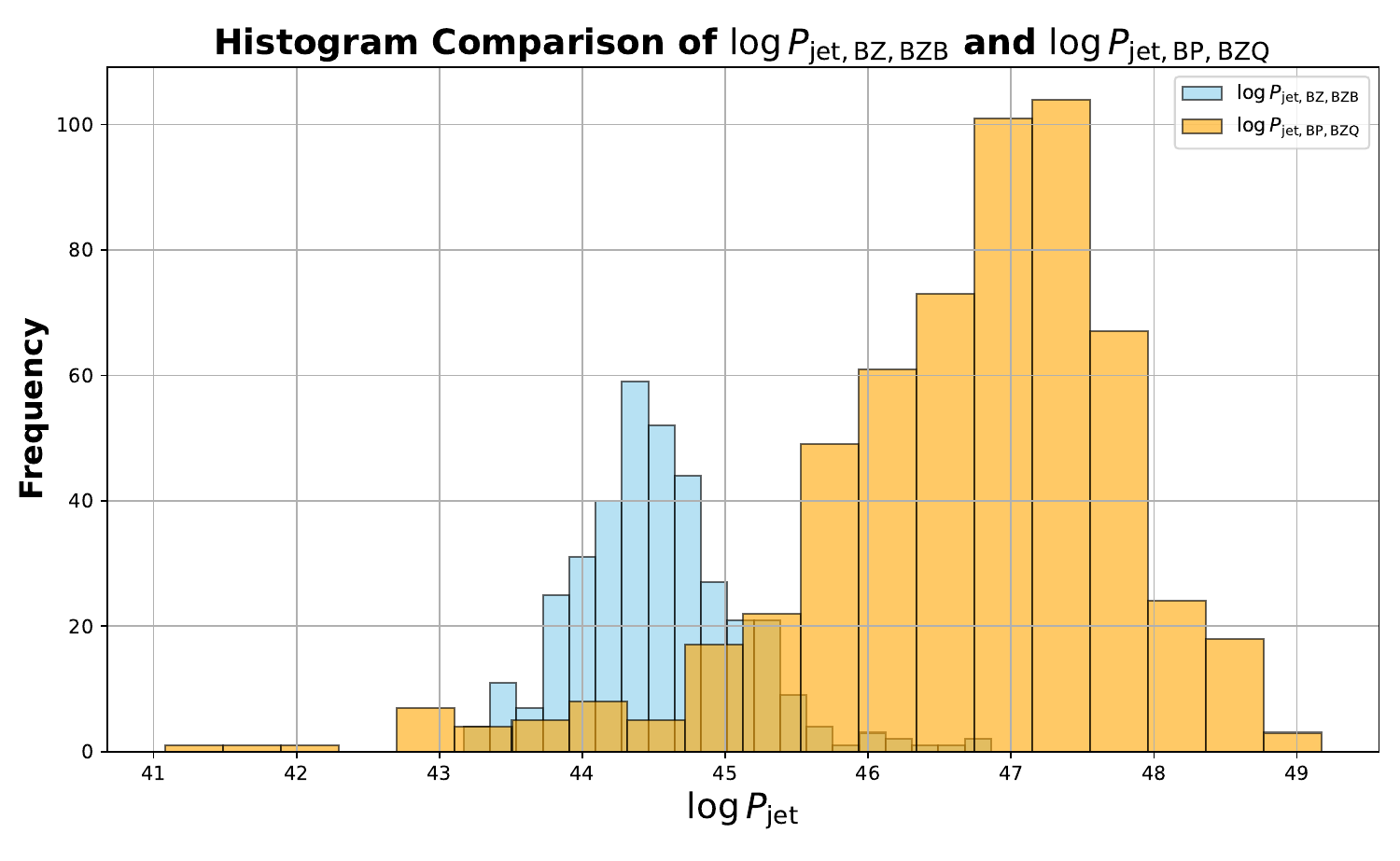}
	\caption{This figure presents the distributions of \(\log P_{\mathrm{jet,BZ,BZB}}\) and \(\log P_{\mathrm{jet,BP,BZQ}}\), where \(P_{\mathrm{jet,BZ,BZB}}\) represents the jet power derived from the BZ model for BZBs, depicted in blue, and \(P_{\mathrm{jet,BP,BZQ}}\) represents the jet power derived from the BP model for BZQs, depicted in orange. The relevant data are taken from Table 3 of \citet{Xie2024}. }
	\label{jet_power_distributions}
\end{figure*}

To further quantify the impact of the BZ and BP mechanisms on the jet power of the two blazar subclasses, we extracted the relevant data from Table 3 of \citet{Xie2024} and constructed the frequency histograms shown in Figure~\ref{jet_power_distributions} to visualize the distribution characteristics of jet power for each class. 
Additionally, we performed a quantitative analysis of the two distributions following the methodology outlined in Content 4 of Figure \ref{technical-route3}. The results, summarized in Table~\ref{tab5}, reveal significant differences in the jet power  (\(\log P_{\mathrm{jet}}\)) distributions between BZBs and BZQs.  
These differences are evident in several key statistical characteristics, as detailed below:

First, in terms of central tendency, the mean and median of \(\log P_{\mathrm{jet,BZ,BZB}}\) distribution are 44.52 and 44.48, respectively, while those of \(\log P_{\mathrm{jet,BP,BZQ}}\) distribution are 46.67 and 46.85. This reflects a difference of approximately two orders of magnitude,  suggesting that BZQs generally exhibit significantly higher jet power than BZBs.

Second, with respect to std, BZBs show a value of 0.59, while BZQs reach 1.15, indicating that the jet power distribution for BZQs is more dispersed and covers a broader energy range.  
Furthermore, regarding skewness, BZBs display a slight positive skew with a skewness value of 0.66, whereas BZQs show a significant negative skew of -1.21. This suggests that BZBs are mainly concentrated in the lower-power regime, while BZQs include a larger proportion of high-power sources. 
In addition, the kurtosis of BZQs is 2.55, higher than the value of 1.39 observed for BZBs, indicating that the jet power distribution of BZQs is more sharply peaked and exhibits a higher likelihood of extreme values.  
The JSD value of 0.67 quantifies the difference between the PDDs of \(\log P_{\mathrm{jet,BZ,BZB}}\) and \(\log P_{\mathrm{jet,BP,BZQ}}\), revealing a significant overall discrepancy.  
As shown in Figure~\ref{jet_power_distributions}, the jet power distributions of BZBs and BZQs exhibit marked differences: 
BZQs display a significantly higher distribution peak, a broader range, and a sharper profile, while BZBs display a more concentrated distribution skewed toward the lower-power regime. 

According to the empirical formula (\ref{equ7}) proposed by \citet{Foschini2024} and \citet{Xie2024}, derived from the \citet{Blandford1979} model, 

\begin{equation}
	P_{\mathrm{jet}} = \left(4.5 \times 10^{44}\right) \left( \frac{S_{\mathrm{obs}}  d_{L,9}^2}{1 + z} \right)^{\frac{12}{17}} 
	\label{equ7}
\end{equation}
the relationship between the jet power \(P_{\mathrm{jet}}\) and observed flux \(S_{\mathrm{obs}}\) indicates that systematic differences in jet power between BZBs and BZQs are likely a key factor contributing to the observed discrepancies in their flux distributions.

In summary, differences in jet formation mechanisms likely result in  variations in the jet power distributions between BZBs and BZQs. Via the coupling relationship between jet power and flux described in Equation~\ref{equ7}, these variations are likely to further contribute to the statistical differences observed in their multi-wavelength flux distributions.

(3) Difference in Doppler boosting effect

The Doppler beaming effect is a key factor explaining why blazars display rapid variability, high luminosity, and strong, variable polarization \citep{Urry1995,Fan2017,Yang2022}.  
Its macroscopic manifestation is Doppler boosting, which refers to the apparent  enhancement of radiation toward the observer due to the relativistic motion of a jet  approaching the speed of light \citep{Urry1995,Mei1999,Zhou2006,Capdessus2018}.  
To quantify the strength of this effect, \cite{Wills1995} introduced a robust and widely used parameter—the beaming factor $R_{\rm v}$, defined as the ratio of the radio core luminosity to the optical continuum luminosity in the V band. The formula is given by: 

\begin{equation}
	\log R_{\rm v} = \log\left( \frac{L_{\mathrm{core}}}{L_{\mathrm{opt}}} \right) = \log L_{\mathrm{core}} + \frac{M_{\mathrm{abs}}}{2.5} - 13.7
	\label{equ8}
\end{equation} 
In Equation \ref{equ8}, \( L_{\mathrm{core}} \) denotes the radio core luminosity, and \( L_{\mathrm{opt}} \) represents the optical continuum luminosity in the V band. The absolute optical magnitude with k-correction, \( M_{\mathrm{abs}} \), is defined as
\( M_{\mathrm{abs}} = M_{\rm v} - k \),
where \( M_{\rm \mathrm{v}} \) is the apparent magnitude in the V band, and the k-correction term $\rm  k$ is given by:

\begin{equation}
	k = -2.5 \log(1 + z)^{1 - \alpha_{\mathrm{opt}}}
	\label{equ9}
\end{equation}
A typical value of the optical spectral index is assumed to be \( \alpha_{\mathrm{opt}} = 0.5 \).
A higher \( R_{\rm v} \) value generally indicates that the jet is more  aligned with the observer’s line of sight, implying a stronger Doppler boosting effect.

In a study utilizing the 3FGL sample, \citet{Chen2016} examined the relationship between $R_{\rm v}$ and the $\gamma$-ray luminosity $L_\gamma$ for BZBs and BZQs, revealing a statistically significant positive correlation between $\log R_{\rm v}$ and $\log L_\gamma$. 
For the full blazar sample, a Pearson correlation coefficient of 0.72 indicates a strong linear relationship between the two physical quantities. 
Based on this, we extracted the pertinent \(\log R_{\rm \mathrm{v}}\) and \(\log L_{\gamma}\) data from \citet{Chen2016} to further examine the linear correlation between these two variables.   
Assuming that $\log L_\gamma$ and $\log R_{\rm v}$ adhere to the linear relationship outlined in Equation~\ref{equ10}, we conducted a linear regression analysis employing the \texttt{LinearRegression} class from the \texttt{scikit-learn} machine learning library \citep{Pedregosa2011}. 
The optimal parameters $a$ and $b$ were determined to be 0.90 and 43.78, respectively.

\begin{equation}
	\log L_{\gamma} = a \log R_{\rm \mathrm{v}} + b 
	\label{equ10}
\end{equation}
\begin{figure*}
	\centering
	\includegraphics[width=0.8\textwidth]{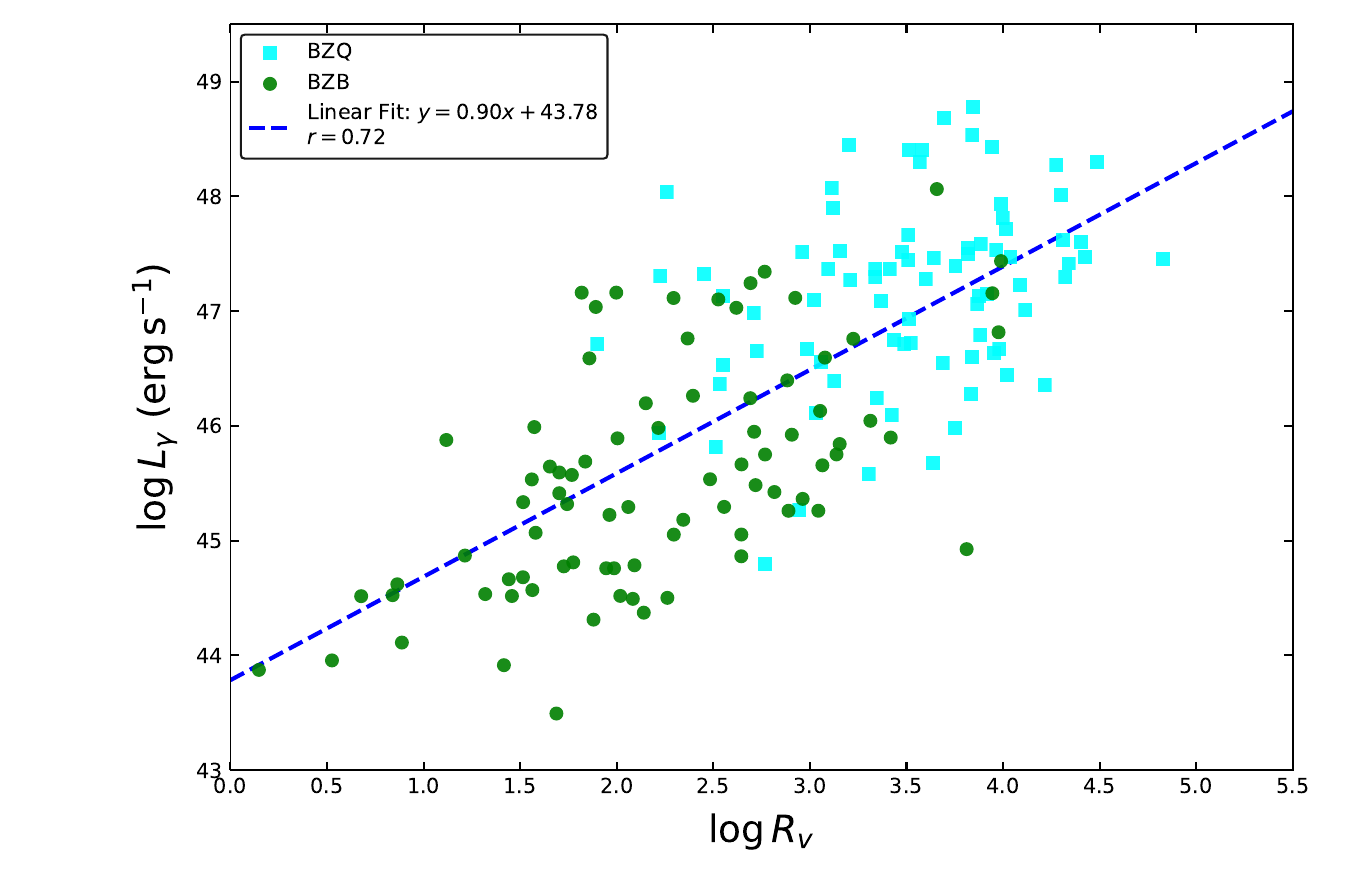}
	\caption{The relationship between the logarithmic beaming factor $\log R_{\rm v}$ and the logarithmic $\gamma$-ray luminosity $\log L_\gamma$ is depicted, with cyan points denoting BZQs, green points denoting BZBs, and the blue dashed line representing the best-fit curve derived from Equation~\ref{equ10}. The Pearson correlation coefficient is indicated by $r$. The relevant data are taken from \citet{Chen2016}.} 
	\label{Rv-Lgamma}
\end{figure*}

Subsequently, we utilized the \texttt{pearsonr} function from the \textbf{SciPy} statistics module \citep{Virtanen2020} to compute the Pearson correlation coefficient between the two variables, obtaining a value of 0.72, consistent with the findings reported by \citet{Chen2016}. 
The correlation fitting results, as shown in Figure~\ref{Rv-Lgamma}, indicate that Equation~\ref{equ10} adequately accounts for the linear relationship between the two variables. 
Additionally, the data reveal distinct trends for BZBs and BZQs: $\log R_{\rm \mathrm{v,BZB}}$ values are mainly clustered in the lower-left region, while $\log R_{\rm \mathrm{v,BZQ}}$ values are predominantly situated in the upper-right region.  

Following the methodology detailed in Content 3 of Figure~\ref{technical-route3}, we conducted a systematic evaluation of the distributional differences in $\log R_{\rm v}$ between BZBs and BZQs.
\begin{figure*}
	\centering\rm
	\includegraphics[width=0.8\textwidth]{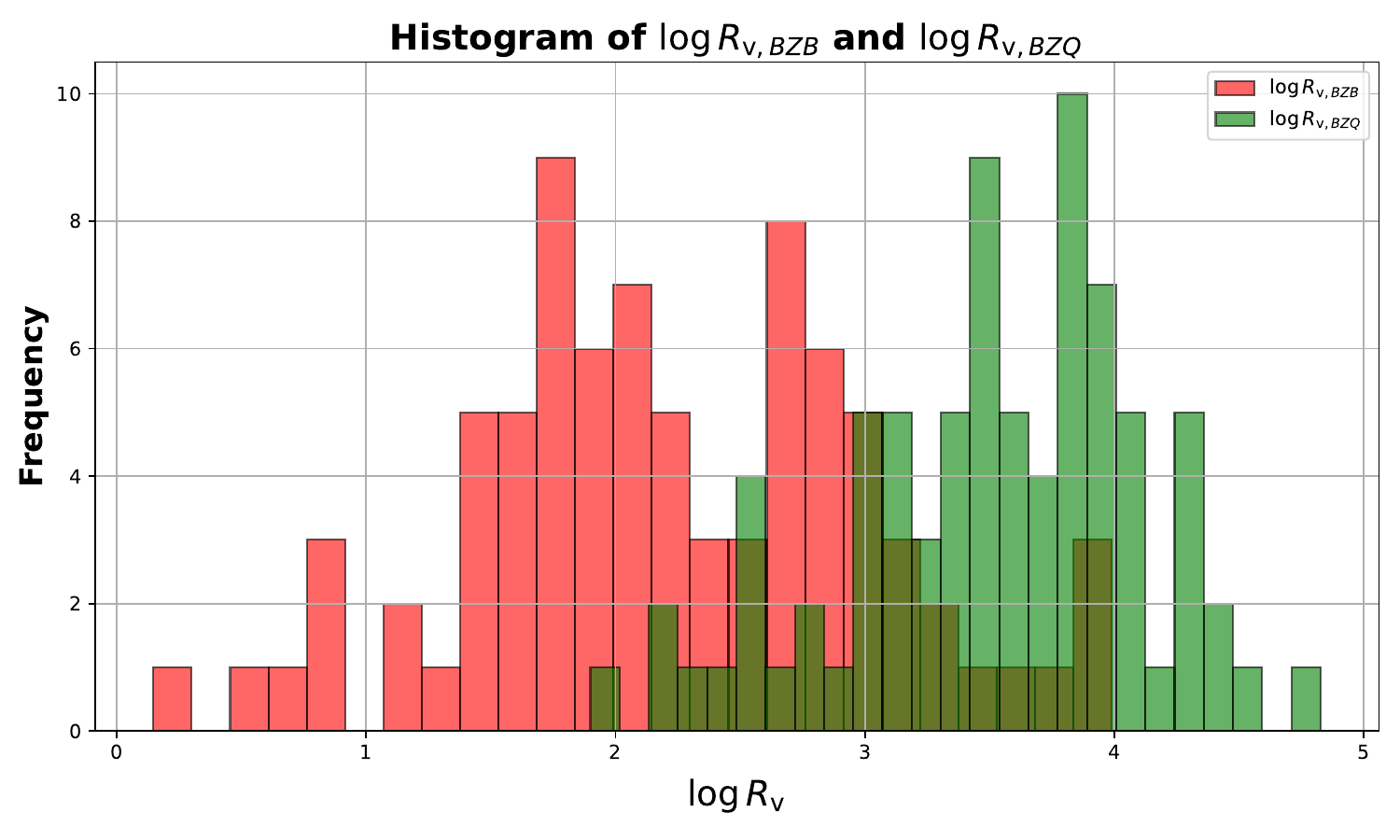}
	\caption{This figure displays the distributions of \(\log R_{\rm v,\mathrm{BZB}}\) and \(\log R_{\rm v,\mathrm{BZQ}}\), representing the logarithmic beaming factors of BZBs and BZQs, respectively. The red histogram represents the distribution of \(\log R_{\rm v,\mathrm{BZB}}\), while the green histogram depicts that of \(\log R_{\rm v,\mathrm{BZQ}}\). The relevant data are taken from \citet{Chen2016}.} 
	\label{Rv_distributions}
\end{figure*}  
Specifically, we generated frequency histograms for each subclass based on their respective $\log R_{\rm \mathrm{v}}$ values. 
As shown in Figure~\ref{Rv_distributions}, the \(\log R_{\rm \mathrm{v}}\) distributions for BZBs and BZQs display marked differences.
The $\log R_{\rm v,\mathrm{BZB}}$ values are mainly concentrated in the lower range, whereas the $\log R_{\rm v,\mathrm{BZQ}}$ distribution is notably shifted toward higher values. 
Moreover, as evidenced by the analytical results in Table \ref{tab5}, BZBs and BZQs exhibit systematic differences in the distributional characteristics of $\log R_{\rm \mathrm{v}}$. 
Both the mean and median values of \(\log R_{\rm \mathrm{v,BZQ}}\) are markedly higher than those of \(\log R_{\rm \mathrm{v,BZB}}\), indicating that BZQs generally exhibit stronger Doppler beaming effects than BZBs. 
Furthermore, we note that the range of $\log R_{\rm v,\mathrm{BZQ}}$ is 1.90–4.83, whereas that of $\log R_{\rm v,\mathrm{BZB}}$ is 0.15–3.99. 
	The two distributions overlap only between 1.90 and 3.99, underscoring a clear separation between BZBs and BZQs. 
Moreover, the std of $\log R_{\rm v,\mathrm{BZB}}$ is 0.81, exceeding that of $\log R_{\rm v,\mathrm{BZQ}}$ at 0.60.  
This indicates that the $\log R_{\rm \mathrm{v,BZB}}$ distribution is more dispersed, reflecting greater variability among individual sources, whereas the $\log R_{\rm \mathrm{v,BZQ}}$ values are more tightly clustered.  
Skewness analysis reveals that the distribution of \(\log R_{\rm \mathrm{v,BZB}}\) is approximately symmetric, with a skewness of 0.01, whereas \(\log R_{\rm \mathrm{v,BZQ}}\) exhibits a slight negative skew with a skewness of -0.47, indicating a concentration of \(\log R_{\rm \mathrm{v,BZQ}}\) values toward higher ranges. 
Both source classes exhibit negative kurtosis values, with $\log R_{\rm \mathrm{v,BZB}}$ at -0.16 and $\log R_{\rm \mathrm{v,BZQ}}$ at -0.19, indicating that their distributions are flatter than the SND. 
The JSD value of 0.51 quantifies the distinction of the PDDs between $\log R_{\rm \mathrm{v,BZB}}$ and $\log R_{\rm \mathrm{v,BZQ}}$, revealing a significant difference in their overall distributional characteristics.

For blazars, the luminosity \( L \) in a specific waveband and the flux \( S \) satisfy the following relationship:

\begin{equation}
	L = 4\pi d_L^2 S
	\label{equ11}
\end{equation}
where \( d_L \) is the luminosity distance, \( S \) needs to be corrected for redshift using the k-correction, given by: 

\begin{equation}
	S = S_{\mathrm{obs}} (1 + z)^{\alpha - 1}
	\label{equ12}
\end{equation}
where \( z \) is the redshift and \( \alpha \) is the spectral index.
Drawing from Equations~\ref{equ8}, \ref{equ10}, and \ref{equ11}, we infer the existence of a coupling relationship between $R_{\rm \mathrm{v}}$ and the flux spanning the radio to $\gamma$-ray bands. 
This suggests that the difference in jet beaming effects between BZBs and BZQs is likely an important factor contributing to the observed  differences in their flux distributions.  

(4) Different Redshift distributions

Redshift (\( z \)), as an observable quantity reflecting a source's relative motion or cosmological distance from us \citep{freedman2010hubble}, is inextricably linked to the magnitude of the observed flux. 
Light from a source radiates outwards in all directions. As the distance of the source increases, this light spreads over a larger spherical area, causing the flux (the number of photons per unit area) to decrease. This relationship is known as the inverse square law\footnote{http://hyperphysics.phy-astr.gsu.edu/hbase/vision/isql.html}.
Secondly, due to the expansion of the universe, light waves from a distant celestial object are stretched, causing their wavelength to become longer. This means the light we receive in a specific waveband was actually emitted by the object at a shorter, bluer wavelength in its rest frame.
To accurately calculate the intrinsic flux of a celestial object, it is essential to 
apply a K-correction based on its \( z \) . This demonstrates the tight correlation between \( z \) and observed flux \citep{oke1968energy}. 

\begin{figure*}
	
	\centering
	
	\includegraphics[width=0.8\textwidth]{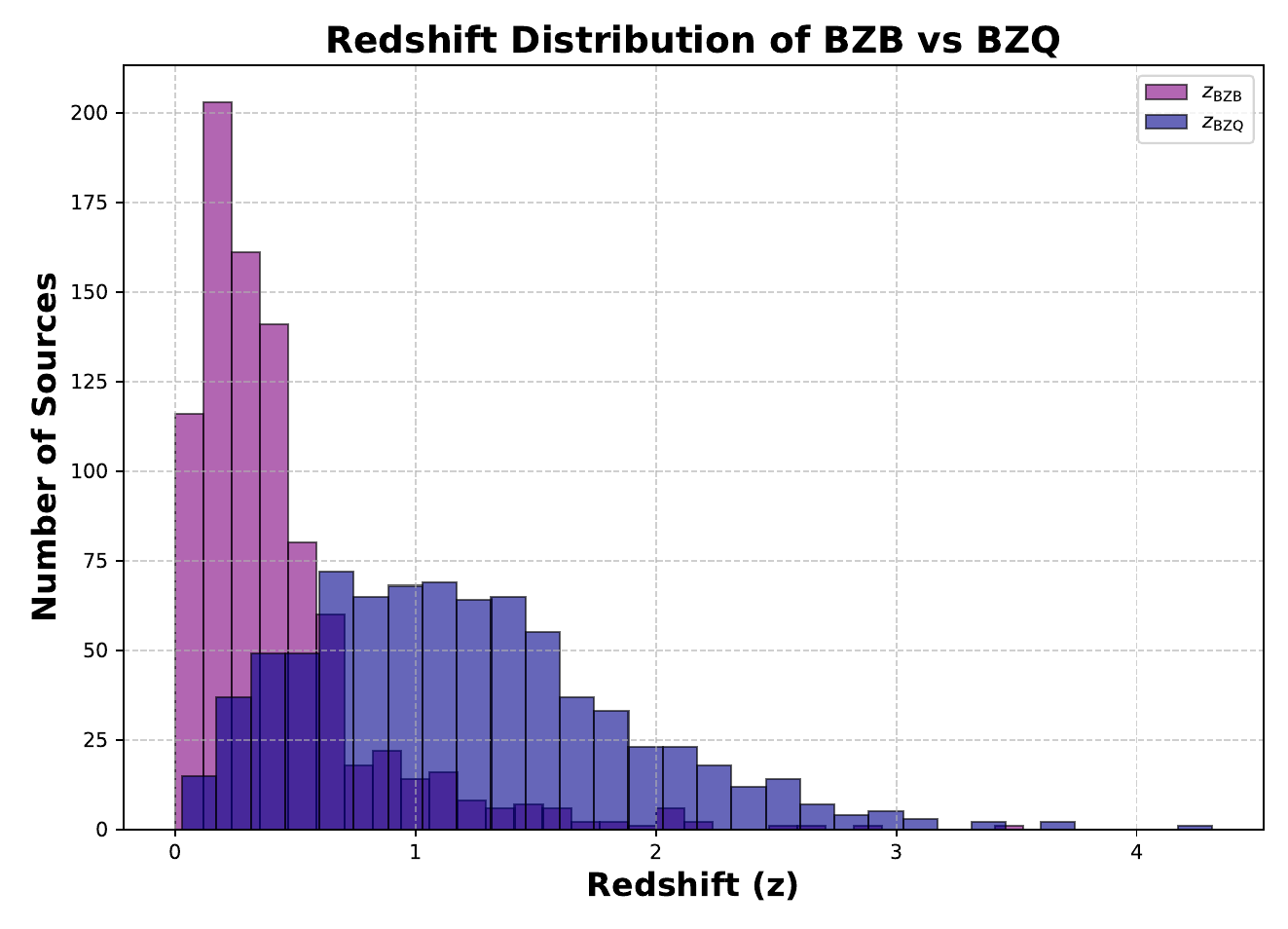}
	
	\caption{This figure presents the distributions of \(\log P_{\mathrm{jet,BZ,BZB}}\) and \(\log P_{\mathrm{jet,BP,BZQ}}\), where \(P_{\mathrm{jet,BZ,BZB}}\) represents the jet power derived from the BZ model for BZBs, depicted in blue, and \(P_{\mathrm{jet,BP,BZQ}}\) represents the jet power derived from the BP model for BZQs, depicted in orange. The relevant data are taken from \citet{ajello2022fourth}.}
	
	\label{redshift_hist_bzb_bzq}
	
\end{figure*}

Building upon the understanding of the relationship between \( z \) and observed flux, and to investigate the intrinsic physical differences between BZBs and BZQs, we conducted an in-depth analysis of the differences in their \( z \) distributions from the 4LAC-DR3 catalog \citep{ajello2022fourth}, utilizing the technical procedure outlined in Content 4 of Figure ~\ref{technical-route3}.  
Our statistical analysis of the \( z \) distributions for BZBs and BZQs reveals a significant difference between these two populations.
The histogram in Figure \ref{redshift_hist_bzb_bzq} clearly illustrates the differences in the central tendencies and overall spreads of the two \( z \) distributions. 

The quantitative metrics in Table \ref{tab5} show that the mean (0.42) and median (0.31) \( z \) of the BZB sample are substantially lower than those of the BZQ sample (mean = 1.19, median = 1.11). This indicates that BZQ objects are, on average, located at greater cosmological distances. As indicated in Table  \ref{tab5}, the BZB sample displays a lower mean (0.42) and median (0.31) \( z \) compared to the BZQ sample (mean = 1.19, median = 1.11), suggesting that BZQ objects are, on average, located at greater cosmological distances. The BZQ sample covers a broader \( z \) range (0.03 – 4.31) than the BZB sample (0 – 3.53), indicating that BZQs include more distant, high-z objects. The BZB sample's \( z \) distribution is sharply peaked at low \( z \) and has a long tail, as evidenced by its high skewness (2.68) and kurtosis (10.66). In contrast, the distribution of BZQs is flatter, with a skewness of 0.81 and a kurtosis of 0.86, making it closer to a symmetrical, low-kurtosis distribution. Moreover, a JSD of 0.479 further confirms a fundamental difference between the \( z \) distributions of the two samples.

Finally, it should be noted that the optical spectra of BZBs often lack prominent emission lines or display only extremely weak lines, which makes spectroscopic redshift measurements inherently more difficult \citep{landoni2014}. 
Taking 4LAC-DR3 as an example, there are 1,458 sources classified as BZB, of which 875 have measured redshifts, indicating that roughly 40\% of the BZBs still lack redshift information \citep{Ajello2022}. Given this sample incompleteness, any firm assertion that the redshift distributions of BZBs and BZQs are significantly different cannot be considered fully robust. We therefore look forward to obtaining more redshift measurements in the future to further support this statistical result.

In conclusion, the histogram, statistical metrics, and JSD value all clearly indicate that the redshift distributions of the BZB and BZQ samples are significantly different. Given the tight correlation between redshift and observed flux shown in Equation (\ref{equ12}), this disparity in redshift is likely a key factor driving the observed differences in the flux distributions of these two populations.

\section{Discussion}  \label{sec3}

\subsection{On the Low Cross-match Rate between DR2 and 5BZCAT\_err} \label{sec3.1x}

It is well established that blazars constitute a large fraction of the sources in 4FGL-DR4 \citep{Ballet2023}.  However, in our cross-match we find that blazars account for only about 3\% of DR2 sources with TS $\geq$ 25 ($\approx$5$\sigma$). 
To investigate the origin of this low fraction, we cross-matched 5BZCAT with 4FGL-DR4 using TOPCAT and obtained 1,647 matched sources. Since 5BZCAT contains 3,562 sources in total, this means that 1,915 5BZCAT sources remain unmatched, suggesting that a substantial fraction of blazars in 5BZCAT are not detected as gamma-ray sources. 
On the other hand, 4FGL-DR4 contains 3,933 sources labeled as blazars; subtracting the 1,647 originating from 5BZCAT leaves 2,286 that do not come from 5BZCAT. Subsequently, we conducted a rapid survey of the provenance of blazar candidates in 4FGL-DR4. The results show that the blazar candidates in 4FGL-DR4 are not only from 5BZCAT, but also from other catalogs such as \citet{veron2010}  and \citet{d2014wise}. Upon further inspection, we find that the blazar fraction in 4FGL-DR4 is indeed high, accounting for about 54.66\% of all 4FGL-DR4 sources.  However, the remaining 45.34\% of GeV sources originate from other classes, accounting for nearly half of the total. Therefore, based on the current evidence, we cannot claim that the majority of newly detected sources in DR2 are likely to be blazars. Likewise, on the basis of the cross-match results of DR2 and 5BZCAT\_err, we also  cannot conclude that the unmatched sources are definitively non-blazars.

\subsection{Exploring the Relationship of the Blazar Sequence with Distributional Differences in Section \ref{sec3.1} \label{sec3.1xx}} 

Based on  \( \nu_{\mathrm{peak}}^{\mathrm{S}} \),  blazars can be further classified into three subtypes: low-synchrotron-peaked BL Lac (LBL, $\log \nu_p < 14$ Hz), intermediate-synchrotron-peaked BL Lac (IBL, $14 \leq \log \nu_p \leq 15$ Hz), and high-synchrotron-peaked BL Lac (HBL, $\log \nu_p > 15$ Hz) \citep{Ackermann2011}. 
 According to the “blazar sequence” proposed by \citet{Fossati1998}, their Figure~8 shows that as $ \nu_{\mathrm{peak}}^{\mathrm{S}} $ increases, both spectral indices $\alpha_{\mathrm{RO}}$ (radio–optical) and $\alpha_{\mathrm{RX}}$ (radio–X) gradually decline. This trend indicates a continuous spectral evolution from FSRQs/LBLs to HBLs.
The quantities $\alpha_{\mathrm{RO}}$, $\alpha_{\mathrm{RX}}$, and \( \nu_{\mathrm{peak}}^{\mathrm{S}} \) are key parameters that characterize the multiwavelength SED morphology of blazars. 
When $\alpha_{\mathrm{RO}}$ or $\alpha_{\mathrm{RX}}$ varies, while holding other parameters in the radiation model fixed, the multiwavelength SED morphology correspondingly changes. For example, the synchrotron and IC peak frequencies shift, the luminosities in the corresponding energy/frequency bands change, and the relative dominance of synchrotron or IC emission in the X-ray band also varies. 
Along the FSRQ/LBL → IBL → HBL sequence, the source luminosity in the radio or optical bands gradually decreases, and the external radiation field becomes weaker. As a result, the radiative cooling of high-energy electrons becomes less efficient, allowing them to accumulate and be accelerated to higher energies. 
Consequently, \( \nu_{\mathrm{peak}}^{\mathrm{S}} \) shifts from lower values (FSRQs/LBLs: $<10^{14}$ Hz) to higher ones (HBLs: $>10^{15}$ Hz), and the dominant high-energy emission mechanism transitions from external Compton (EC) to SSC \citep{Fossati1998,Ghisellini1998,Abdo2010c,Bottcher2013}. 
As a result of these differing radiation mechanisms, BZQs and BZBs exhibit a pronounced bimodality in the distribution of \( \nu_{\mathrm{peak}}^{\mathrm{S}} \), which is fully consistent with the results we observe in Figure \ref{histogram_plot_Vpeak}.

In Section \ref{sec3.1}, in addition to discussing the differences in the \( \nu_{\mathrm{peak}}^{\mathrm{S}} \) distributions, we also examine the differences in the distributions of redshift, jet power, and Doppler boosting factor. First, we can directly rule out the possibility that the blazar sequence is responsible for the difference in the redshift distributions. 
According to \citet{Fossati1998}, particularly the Kendall’s $\tau$ test results in their Table 4, the P-values between \( \nu_{\mathrm{peak}}^{\mathrm{S}} \) and $\alpha_{\mathrm{RO}}$  (or $\alpha_{\mathrm{RX}}$)   are very small (far below 0.05) after removing the influence of redshift. This shows that the correlations between \( \nu_{\mathrm{peak}}^{\mathrm{S}} \) and $\alpha_{\mathrm{RO}}$, $\alpha_{\mathrm{RX}}$ remain significant, indicating that the blazar sequence is not intrinsically related to redshift. 
Moreover, from equations (\ref{equ7}), (\ref{equ8}), and (\ref{equ11}), we see that the jet power and Doppler boosting factor are not only related to the flux but also closely linked to redshift, distance, and optical magnitude. Therefore, from a mathematical standpoint, it remains challenging to explain the distributional differences in jet power and Doppler boosting factor using only the inverse correlation between the spectral index and \( \nu_{\mathrm{peak}}^{\mathrm{S}} \) in the blazar sequence.

\subsection{Evaluating the Rationality  and Merits of Statistical Modeling} \label{sec3.2}


The SND, due to its symmetry, desirable mathematical properties, and strong theoretical grounding in the central limit theorem, offers fundamental advantages in statistical modeling. 
It is widely used to describe natural phenomena and to facilitate statistical inference \citep{Cheng2024,DasGupta2008}. 
However, in practical applications, especially when dealing with flux data from astronomical observations, the ideal assumption of an SND is often difficult to fulfill.  
Specifically, observed fluxes in astrophysics are typically constrained by both physical mechanisms and instrumental limitations, such as the non-negativity of flux, and the sensitivity threshold of detectors,  resulting in intrinsic truncation at both lower and upper limits. 
The domain of the independent variable in an SND spans the entire range of real numbers, and its PDF remains nonzero across both positive and negative infinitesimals \citep{Rice2007}. 
Consequently, directly applying the SND to model data with inherent lower and upper bounds, such as non-negative flux, inevitably assigns nonzero probabilities to physically invalid negative values. 
This not only violates the physical constraints of the variable but also undermines the normalization integrity of the probability model. 
In contrast, the TND sets zero probability density outside the domain of the variable and renormalizes the density function within the valid interval. This approach preserves the favorable mathematical properties of the SND while strictly adhering to the physical boundaries of the variable.  
Therefore, the TND model is well-suited for modeling flux data in astronomical observations, where non-negativity and bounded-domain constraints are inherently present.

In this study, we applied the LOF algorithm \citep{Breunig2000} to identify and remove outliers from raw samples that had initially failed the KS or CVM tests. 
By assessing variations in local density, the LOF method sensitively detects locally anomalous data points. 
To optimize the performance of the \texttt{scikit-learn} LOF implementation \citep{Pedregosa2011} used in this study, we systematically investigate various parameter combinations, including the \textbf{n\_neighbors} ($\boldsymbol{k}$)  and \textbf{contamination} ($\boldsymbol{\alpha}$), to determine the optimal configuration that minimizes the contamination proportion while satisfying the KS or CVM tests.  
This approach ensures effective outlier removal while preserving the overall distribution characteristics of the original dataset to the greatest extent possible.


To quantify the statistical confidence of each source within the fitted flux distribution, we computed the 68.27\% (1$\sigma$) and 95.45\% (2$\sigma$) CIs based on the TND. 
Instead of conventional symmetric interval definitions, we adopted a CDF-based approach, which is appropriate for distributions exhibiting asymmetry due to truncation  \citep{Hyndman1996,Ialongo2019}.  
Specifically, the 1$\sigma$ interval spans the 15.87\% to 84.14\% quantiles, and  the 2$\sigma$ interval spans from the 2.28\% to 97.73\% quantiles. 
This method does not rely on the assumption of symmetry around the mean and more accurately reflects the true  probability distribution within the truncated domain. 
Samples that fall within the 2$\sigma$ CI are generally regarded as statistically consistent with the overall population, whereas those outside this range are likely to exhibit distinct physical properties.

The z-score offers a quantitative measure of how far a given sample deviates from the population mean.
A larger absolute z-score indicates more extreme radiative behavior relative to the overall population, reflecting a greater departure from the distribution's central tendency. 
Conversely, a z-score near zero suggests that the sample lies close to the population mean and is more representative of the overall distribution. 
By integrating z-score analysis with CIs, this study offers a precise characterization of each sample’s relative position and degree of deviation within the distribution. 
The magnitude of this deviation is particularly important for identifying potential outliers or candidate sources \citep{Rousseeuw2011, Shaikh2014, Roy2024}.

	Due to the limited spatial resolution of Fermi-LAT, multiwavelength spatial cross-identification faces significant challenges. Currently, only 362 objects can be firmly identified as counterparts associated with other catalogs (denoted by uppercase letters in 4FGL-DR4), while candidate sources account for approximately 
	(7195-362)/7195$\approx$ 95\% 
	of all $\gamma$-ray sources (denoted by lowercase letters in 4FGL-DR4)\footnote{\url{https://heasarc.gsfc.nasa.gov/W3Browse/fermi/fermilpsc.html}}. 
	Thus, as early as the construction of the 1FGL catalog, the Fermi team proposed that, in addition to relying on positional consistency with reference candidates, additional information such as spectral properties and variability should be used together to improve the reliability of source associations \citep{Abdo2010a}. 
	In past Fermi-LAT data analyses, Bayesian probability methods could yield high-probability association results, but this does not imply that all paired sources can be confirmed with 100\% certainty as true counterparts \citep{Abdo2010a, Nolan2012, Acero2015, Abdollahi2022}. 
	The fundamental reason is the limited spatial resolution of Fermi-LAT, which results in relatively large positional uncertainties. In regions of high source density, it is common for a single GeV source to correspond to multiple, mutually independent candidates, creating a “one-to-many” situation \citep{fioc2014,mayer2024}. 
	In this case, when the association probabilities of these candidates differ only slightly, it is often difficult to uniquely identify the optimal counterpart. 
	Strictly speaking, the 17 blazar associations we report should be regarded only as high-probability candidates derived under the common-source hypothesis, rather than as counterparts confirmed with 100\% certainty.

	Building on the previous discussion and the close relationship between the multi-wavelength flux of blazars and their intrinsic physical properties, we choose to use the statistical consistency of multi-wavelength flux as additional evidence for the common-source hypothesis in cross-identification analyses. 
	By utilizing the pipeline of Content 4 of Figure \ref{technical-route3}, we assess whether the multi-wavelength flux characteristics of a given blazar candidate are statistically consistent with the flux distribution models derived from a large sample. This study determines the candidate’s relative position within the flux distribution of each band and computes its corresponding z-score. 
For a specific band, the combined use of the z-score and CIs intuitively reflects the position of the target source within the overall distribution, thereby enabling an assessment of whether its flux characteristics are consistent with those of the overall sample. 
If the z-score of the target source falls within the 2$\sigma$ CI, it is considered statistically consistent with the flux distribution of the large sample.  
	Additionally, the pipeline offers effective visualization, presenting the analytical results in a clear and  intuitive manner that enhances statistical interpretability.

In summary, the pipeline framework from Appendix~\ref{appendixC} exhibits two main advantages in terms of model performance: 
First, it enables rapid screening and batch processing of multi-source flux data under a unified standard, substantially improving the efficiency of large-sample statistical analyses. 
Second, it possesses excellent robustness and generalizability. This is specifically validated: for the few bands in Table~\ref{tab10} that initially failed the tests, we removed only a very small number of outliers ($\alpha$ between 0.01 and 0.03) by utilizing the pipeline, after which all samples passed the KS and CVM tests, as shown in Table~\ref{tab11}. 
This blend of efficient, unified processing and adaptability (via minor outlier removal) confirms the framework's robustness and generalizability.

\subsection{Statistical Consistency Analysis of Multiband Fluxes for 17 New GeV Blazar Candidates} \label{sec3.4}

In this section, we provide a detailed discussion of the results presented in Section \ref{sec2.6.3}.
Our goals are twofold. First, we place the newly identified 17 candidates into a multi-band flux distribution model and determine their statistical positions individually. 
Second, we provide explanatory analyses for those cases that show significant deviations in certain bands while remaining consistent with the overall distribution in others.

In Sections \ref{sec2.1}–\ref{sec2.4}, 
we constructed two catalogs (DR2 and 5BZCAT\_err) incorporating high-precision positional errors and successfully cross-matched 17 new candidates. 
In Section \ref{sec2.5}, we then  extracted and compared the flux distributions of BZB and BZQ across multiple wavebands, uncovering significant differences in their distributions across the various bands. Subsequently, we applied a Box-Cox transformation combined with the TND to statistically model the log-transformed flux distributions for each waveband. 
Based on the established distribution models and the corresponding transformation parameters detailed in Table~\ref{tab11}, we applied the Box–Cox transformation to the flux data of the 17 selected blazars. 
The transformed values were then mapped onto the TND models for their respective source classes, and their relative positions within the population distributions were assessed using z-scores and CIs. 
The mapping results are summarized in Table~\ref{tab4}. Specifically, the multi-band flux z-scores for 9 BZBs and 6 BZQs are near zero and fall entirely within the $2\sigma$ CIs, indicating that their fluxes in the corresponding bands are close to the population mean and lie within the statistically ``normal'' range. 
As discussed in Section~\ref{sec3.2}, this statistical consistency suggests that these sources likely share similar physical mechanisms for their multi-wavelength radiative behavior, thereby providing further support for the common-source hypothesis.

\begin{figure*}
	\centering
	\begin{subfigure}[b]{0.48\textwidth}
		\centering
		\includegraphics[width=\textwidth]{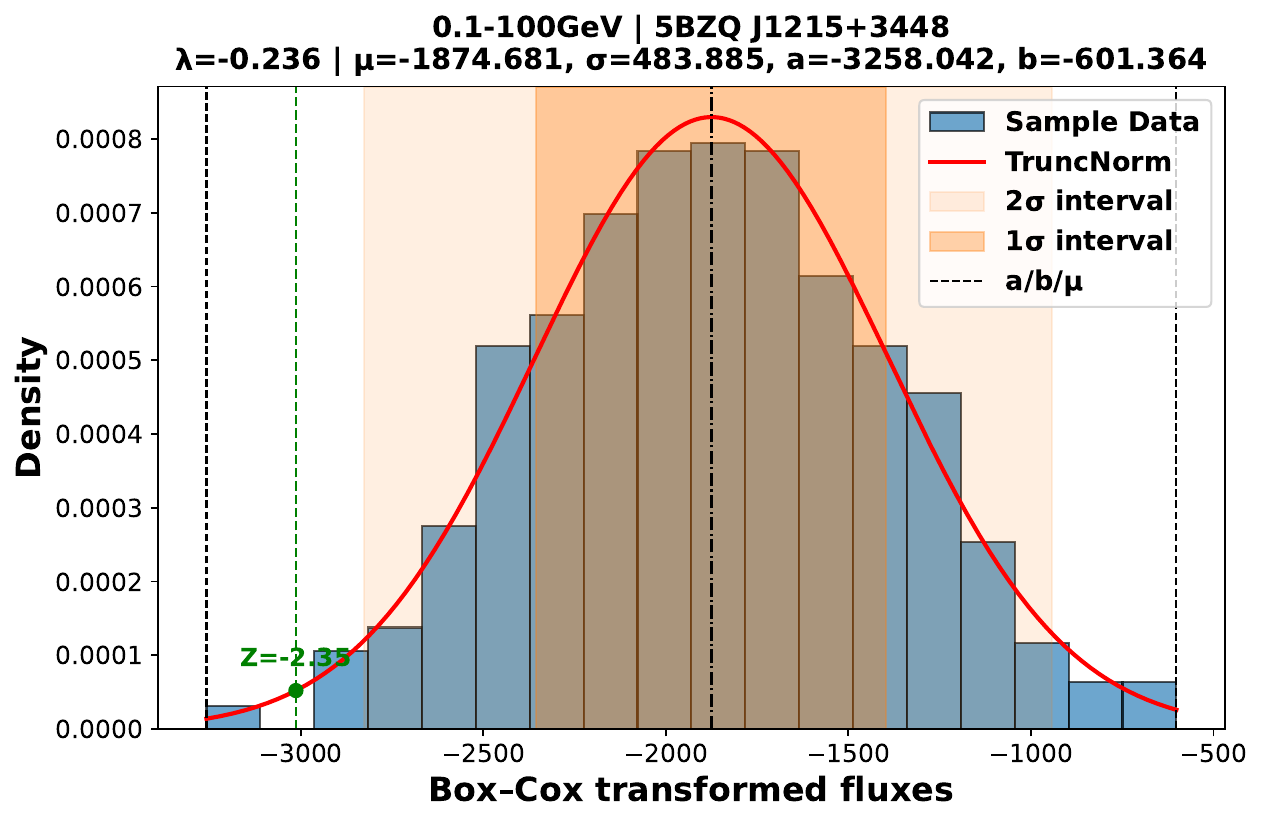}  
		
		\label{fig:subfig1}
	\end{subfigure}
	\hfill
	\begin{subfigure}[b]{0.48\textwidth}
		\centering
		\includegraphics[width=\textwidth]{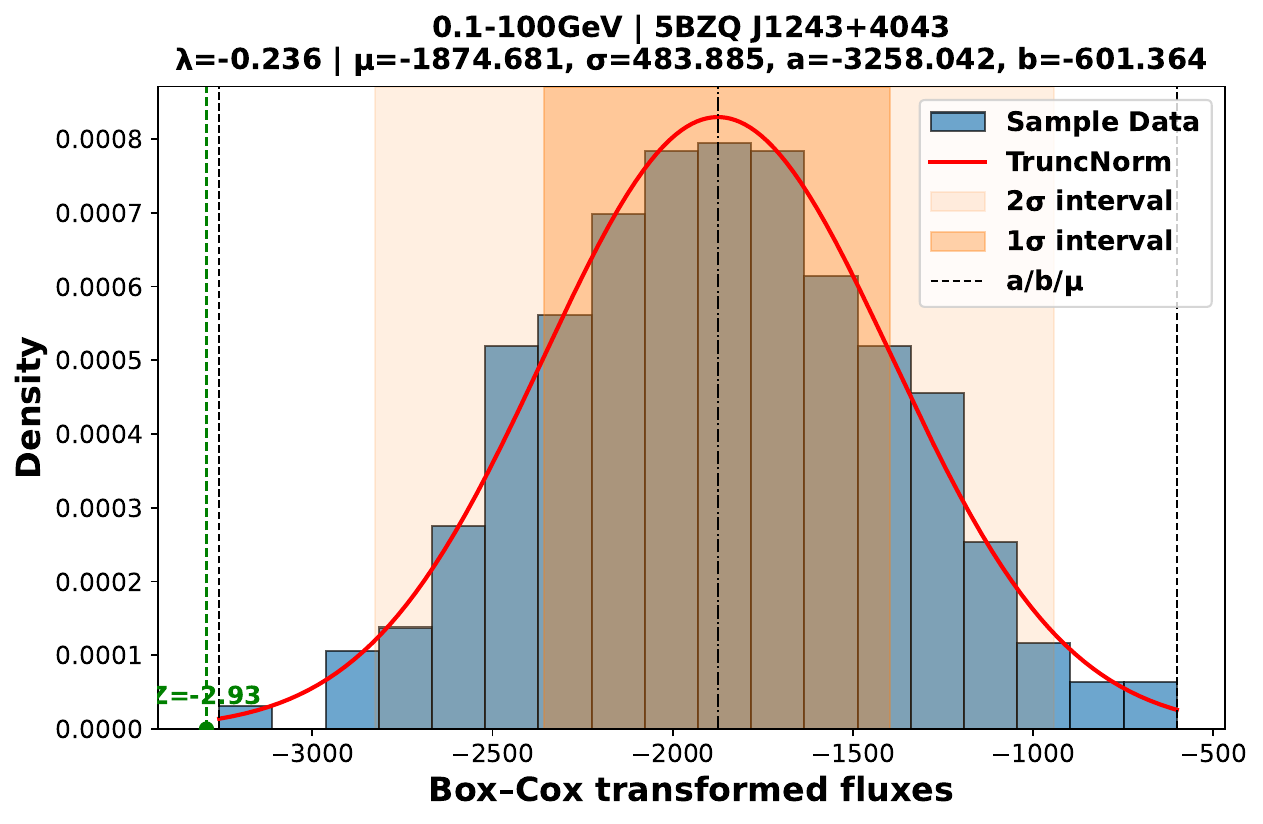}
		
		\label{fig:subfig2}
	\end{subfigure}
	\caption{Analysis results of the TND model for 5BZQ~J1215+3448 and 5BZQ~J1243+4043 in the 0.1--100\,GeV band. Similar details are illustrated in Figure~\ref{modle_fit:0203}.	
	}
	\label{model_fit_3448_4043}
\end{figure*}

Notably, in the results obtained in Section \ref{sec2.6.3}, we find that the 0.1–100 GeV converted fluxes of 5BZQ J1215+3448 (z-score = -2.417) and 5BZQ J1243+4043 (z-score = -3.102) show significant deviations from the distribution model, as illustrated in Figure \ref{model_fit_3448_4043}. 
These fluxes are significantly below the mean of the overall population, with z-scores falling outside the 2$\sigma$ CI, indicating a relatively weak activity state. 
The Box–Cox transformed flux of 5BZQ J1243+4043 falls slightly below the lower bound of the TND model. 
Although the flux of 5BZQ J1215+3448 remains within the truncated range, it still exhibits a significant deviation from the mean of the overall population. 

To explain the pronounced outlier behavior of 5BZQ J1215+3448 and 5BZQ J1243+4043 in the 0.1–100 GeV band, we further investigated how the observing time affects the GeV flux. Specifically, we compared the 100 MeV–1 TeV energy flux (Energy\_Flux100) between the earlier 4FGL-DR3 and the latest 4FGL-DR4 datasets. 
As shown in Table \ref{tab1}, we observed that with the accumulation of observational data, the Energy\_Flux100 of most sources exhibits an increase. 
Specifically, a total of 3,578 sources have a Ratio\_Eflux greater than 1, where Ratio\_Eflux is defined as the ratio of Energy\_Flux100\_v35 to Energy\_Flux100\_v31. 
This observational result provides important clues for understanding the states of the two new candidates, 5BZQ J1243+4043 and 5BZQ J1215+3448. 
In our analysis, their TS values above 100 MeV in DR2 are 17.45 and 20.33, respectively. 
Accordingly, the weak $\gamma$-ray signals currently observed may merely reflect their states during the initial stages of data accumulation. 
Given that the TS values of 5BZQ J1243+4043 and 5BZQ J1215+3448 are near the Fermi-LAT detection threshold and display distinct source features in their TS maps, we anticipate that these two sources will likely gradually approach the high-confidence region of the corresponding distribution model as observational data continue to accumulate.

Additionally, the z-score values of 5BZQ J1243+4043 at 1.4 GHz and 5 GHz are close to the means of the corresponding distributions and lie within the $1\sigma$ CIs. For 5BZQ J1215+3448, the z-scores across the 74 MHz to 2.4 keV bands fall within the $2\sigma$ CIs. These results indicate that, in the radio or X-ray bands, the characteristics of flux distributions of these two sources exhibit no systematic deviation from the statistical model.  Therefore, we retain these sources in the sample and designate them as GeV BZQ candidates positioned near the boundary of the distribution model. 
We recommend the continued monitoring and validation of the evolution of their $\gamma$-ray flux within the context of our distribution model.

Through this analysis, we found that 9 BZBs and 6 out of 8 BZQs exhibit distributional characteristics consistent with those of the larger sample, showing only minor deviations from the population mean. 
Considering the strong spatial association between these 15 $\gamma$-ray sources and the blazar objects in the 5BZCAT\_err, we suggest that they likely constitute a high-probability associated population.  
In particular, 5BZQ J1215+3448 and 5BZQ J1243+4043 exhibit flux distribution characteristics in the radio and/or X-ray bands that closely align with those of the larger population, along with strong spatial associations and significant GeV signals. 
Consequently, we propose them as promising candidates for GeV BZQs.  
Due to limited photon counts, the low flux levels in the 0.1–100 GeV band for these two sources necessitate ongoing observations to confirm the similarity of their GeV flux distribution characteristics.

\section{CONCLUSIONS} \label{sec4.0}

The classification and physical origin of blazars are among the core questions in high-energy astrophysics. To probe the intrinsic differences between BZBs and BZQs, this study develops an end-to-end analysis pipeline that spans data collection, new-source identification, statistical modeling, and investigation of physical origins. This pipeline not only systematically quantifies the significant differences in the observational and physical properties of BZBs and BZQs—providing robust, multidimensional evidence for the classic blazar dichotomy—but also offers strong statistical support for classifying newly identified or ambiguously classified blazar candidates. The main findings and conclusions are as follows:

\begin{enumerate} [label=\textup{(\arabic*)}, ref=\textup{(\arabic*)}]
	\item Based on 16 years of Fermi-LAT data, we reanalyzed the DR1 catalog using \textbf{Fermipy} to supplement the missing elliptical positional errors, and ultimately released the updated DR2 catalog, which provides more precise and comprehensive parameters.
	
	\item To remedy the lack of positional errors in \texttt{5BZCAT}, we integrated it with NED, populated uncertainties for all 3,561 sources, and constructed a new catalog, \texttt{5BZCAT\_err}, suitable for probabilistic cross-matching.
	
	\item By cross-matching DR2 with \texttt{5BZCAT\_err}, we identified 17 high-confidence GeV blazars. Their associations show high probabilities and were independently verified using \textbf{TOPCAT} and \textbf{DS9}.
	
	\item Using Fermipy, we analyzed the 17 new GeV blazars and found no significant spatial extension or spectral curvature. Most sources show no significant variability; the exception is J1043.7+5323, which exhibits strong variability across multiple timescales.
	
	\item To compare the flux distributions of BZBs and BZQs, we collected multi-wavelength flux data from \texttt{5BZCAT\_err} and quantified their differences using two methods: MAD for eight statistical indicators and JSD for overall probability distributions. Both methods consistently reveal systematic differences between BZBs and BZQs across all observed bands. Specifically, the MAD analysis identifies kurtosis as the most discriminative metric, while the JSD analysis highlights the 1.4 GHz, 843 MHz, 5 GHz, 0.1–2.4 keV, and 0.3–10 keV bands as showing larger distributional differences compared to other bands.
	
	\item To address the limitations of single-band metrics in blazar classification (e.g., small sample sizes and incomplete data), we propose a multi-messenger joint-analysis pipeline for classification and identification. Building on our ``$\mathrm{Box\text{-}Cox+TND}$'' model, the method quantifies the statistical consistency between a candidate and its parent population, yielding robust decisions. Modeling results indicate that the approach is broadly applicable and accurately fits the flux distributions of BZBs and BZQs.
	
	\item We apply a $\mathrm{Box\text{-}Cox+TND}$ model to test whether the multi-band fluxes of 17 new blazar candidates are consistent with the \texttt{5BZCAT} population; most candidates fall within $1$–$2\sigma$, supporting correct classifications and the common-source origin.
	
	\item For the first time, we use the JSD to compare BZBs and BZQs across synchrotron peak frequency, jet power, Doppler beaming factor, and redshift; all four show JSD $>$ 0.3, indicating strong separations and providing multi-dimensional statistical support for the optically based blazar dichotomy.
	
\end{enumerate}

\section{Acknowledgements}
We sincerely thank the referee for his/her invaluable comments, and appreciate the data and analysis software 
provided by the Fermi Science Support Center and also thank 
the support for this work from  the
Doctoral Initiation Fund of West China Normal University
(22kE040), the Open Fund of Key Laboratory of Astroparticle 
Physics of Yunnan Province (2022Zibian3), and the National Natural Science Foundation
of China (NSFC, Grant No. 12303048). XP is supported by Doctoral Initiation
Grant 493150 from China West Normal University. 
Key Scientific and Educational Joint Project of Sichuan Province (25LHJJ0097) \\

Facility: Fermi (LAT)\\

Software: \textbf{Fermipy} \citep{wood2017fermipy}, \textbf{DS9} \citep{Joye2003}, \textbf{SciPy} \citep{Virtanen2020}, \textbf{Matplotlib} \citep{Hunter2007}, \textbf{nway}  \citep{Salvato2018}, \textbf{astroquery} \citep{Ginsburg2019},
\textbf{sklearn} \citep{Pedregosa2011}, \textbf{TOPCAT} \citep{Taylor2005}.

\section{Data Availability} \label{sec4}

All data products generated in this study have been publicly released through the \href{https://nadc.china-vo.org/res/r101633/}{China-VO PaperData repository}. The released datasets include:

\begin{enumerate}[label=\textup{(\arabic*)}, ref=\textup{(\arabic*)}]
	\item Refined $\gamma$-ray source catalog, \texttt{4FGL-Xiang-DR2.fit}; described in Section~\ref{sec2.1}.
	\item Elliptical positional uncertainties and optical magnitudes for 5BZCAT sources, \texttt{5BZCAT\_err.fit}; described in Section~\ref{sec2.2}.
	\item Manually verified cross-matching results and spatial association plots for 62 sources lacking positional uncertainties, \texttt{62Figures.tar.xz}; described in Section~\ref{sec2.2}.
	\item GeV emission characteristic analysis for 17 new candidate blazars, including:
	\begin{itemize}
		\item  \texttt{TSmap17.tar.xz}; described in Appendix~\ref{appendixB}.
		\item  \texttt{SED17.tar.xz}; described in Section~\ref{sec2.4.2}.
		\item  \texttt{LC17.tar.xz}; described in Section~\ref{sec2.4.3}.
	\end{itemize}
	\item Standardized multi-wavelength flux data for 3,442 blazars, \texttt{5BZCAT\_multi-band\_flux.xlsx}; described in Section~\ref{appendixC.1}.
	\item The fitting plots of the TND model after Box–Cox transformation for 17 blazar candidates, \texttt{Model\_fitting\_results.tar.xz}; described in Section~\ref{sec2.6.3}
	\item  The outliers in the 74 MHz, 843 MHz, 1.4 GHz, 5 GHz, 0.1–2.4 keV, and 0.3–10 keV bands are packaged in \texttt{outliers.tar.xz}, as described in Appendix \ref{appendixC.3}.
	\item Comparison of Signif\_Avg and Energy\_Flux100 between 4FGL-DR3 and 4FGL-DR4, \texttt{Comparison\_v31\_v35.xlsx}; described in Section~\ref{sec3.1}.
	
\end{enumerate}

Detailed descriptions of each dataset and usage instructions are provided in the accompanying \texttt{Readme.txt}.



\bibliographystyle{mnras}
\bibliography{example} 




\appendix

\section{The 5BZCAT\_err Construction Pipeline} \label{appendixA}

Given that the 5BZCAT data had been uploaded to NED  \citep{Massaro2015}, we initially attempted to manually extract the  positional uncertainties for each source from the NED website  \citep{NED2019}. 
However, due to the large sample size, manual extraction proved to be inefficient and time-consuming.  
To efficiently compile a version of the 5BZCAT with complete positional uncertainty information, we developed a targeted analytical pipeline, which is illustrated in Figure~\ref{technical-route1}. The corresponding construction procedure is detailed below:

\begin{figure*}
	\centering
	\includegraphics[width=0.6\textwidth]{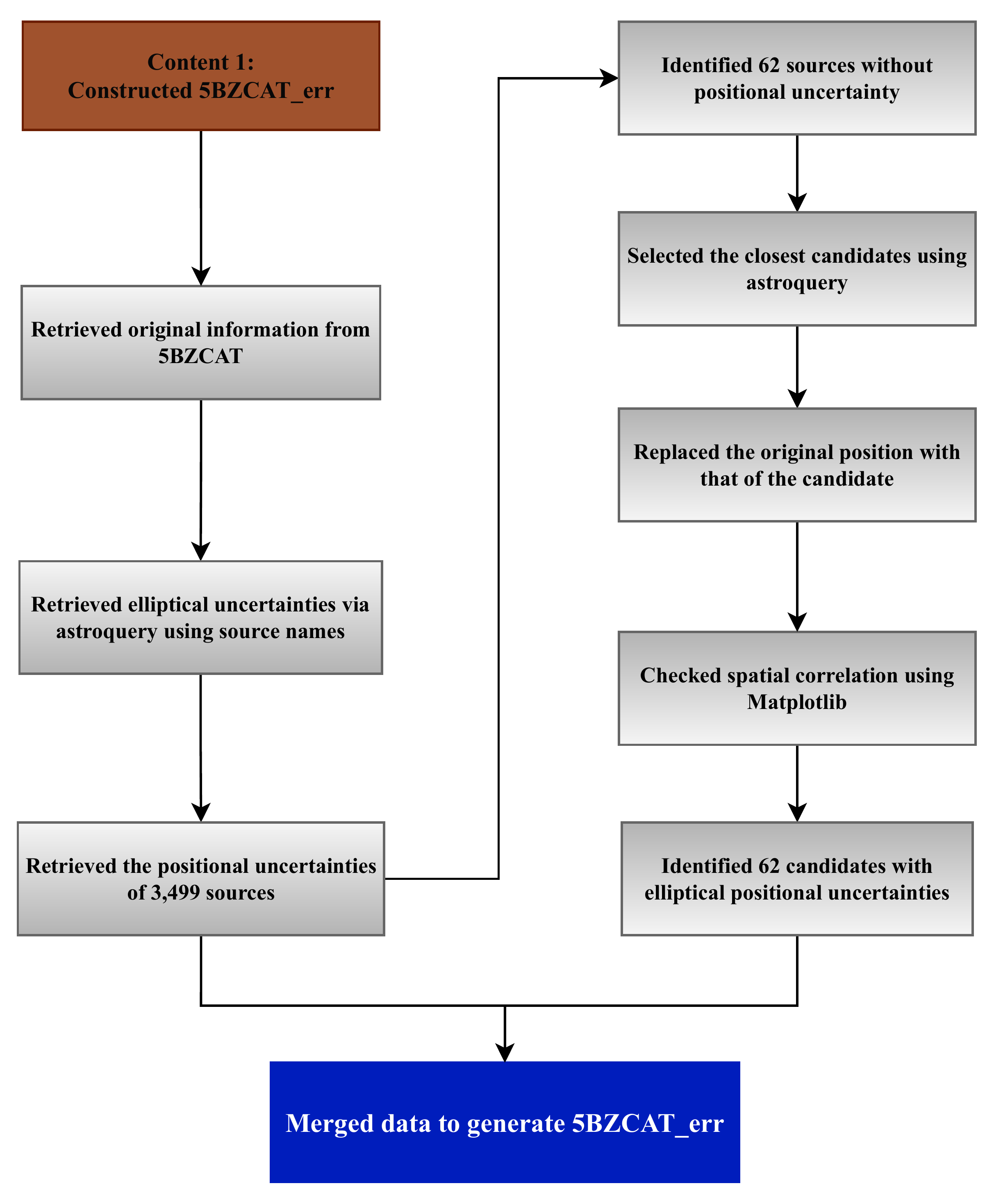}
	\caption{Workflow for constructing  5BZCAT\_err.}
	\label{technical-route1}
\end{figure*}

\textbf{Step 1}: We retrieved the complete parameter dataset for 3,561 sources from the official 5BZCAT website. Using the source names and their alternative designations as input, we extracted the positional  uncertainties from the NED via the \textbf{ned.get\_table()} function\footnote{\href{https://astroquery.readthedocs.io/en/latest/ipac/ned/ned.html}{\url{https://astroquery.readthedocs.io/en/latest/ipac/ned/ned.html}}} of the Python package \textbf{astroquery} \citep{Ginsburg2019}. In this step, we successfully obtained positional uncertainties for 3,499 sources.

\begin{figure*}
	\centering
	\includegraphics[width=0.5\textwidth]{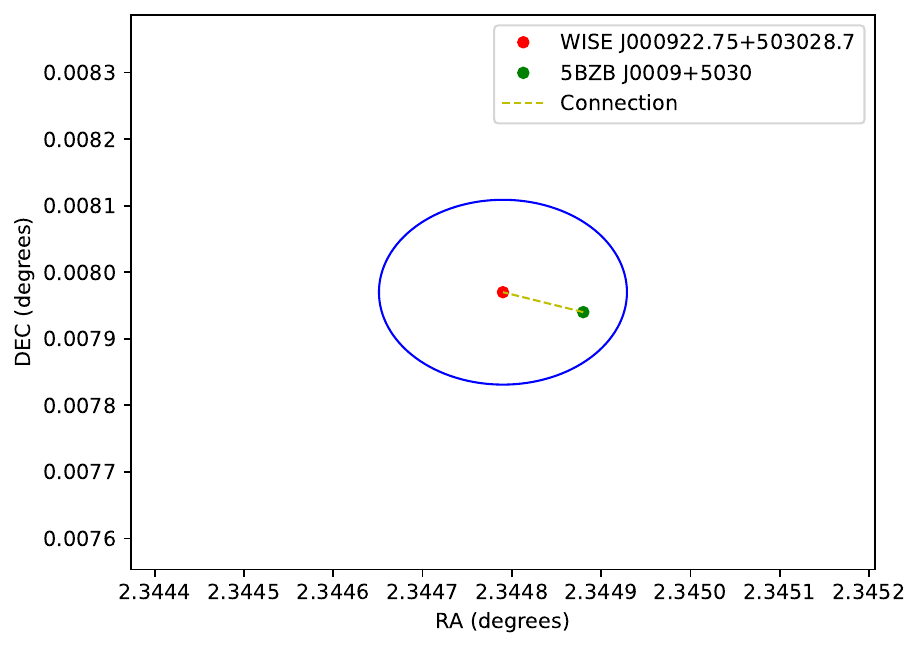}
	\caption{Spatial association between the 5BZCAT source (green) and the NED candidate (red),  exemplified by WISE J000922.75+503028.7 and 5BZB J0009+5030. The blue ellipse represents the positional uncertainty of the NED candidate, while the dashed yellow line indicates the angular separation between the two sources.}
	\label{fig2}
\end{figure*}

\textbf{Step 2}: In Step 1, we identified that positional uncertainty information was missing for 62 sources. To address this, we utilized the target positions provided by 5BZCAT as input and employed the \textbf{ned.query\_region()} function to retrieve relevant positional data. 
To identify potential matching candidates, we acquire NED sources within a circular region of 1° radius centered on each target source. Within this 1° region, we apply the  filtering criterion: the candidate has the smallest angular separation from the target source, and its positional uncertainty ellipse overlaps with that of the target source. 
We compiled the association results between NED and 5BZCAT into Table~\ref{tab7}. Subsequently, we utilized the \textbf{matplotlib.patches.Ellipse()} method \citep{Hunter2007} to generate spatial relationship plots between the target sources and their candidates.
Upon examining the associations for the 62 objects, we confirmed that the candidates  are spatially correlated with the target sources. For example, the spatial relationship between 5BZB~J0009+5030 and WISE~J000922.75+503028.7, as shown in Figure~\ref{fig2}, indicates a high degree of positional coincidence. 
The association result figures for the 62 objects are packaged in \texttt{62Figures.tar.xz}, as referenced in Section~\ref{sec4}.
The NED-derived candidate positions were subsequently adopted as the positional data for these 62 objects in the following analyses.

\textbf{Step 3:} We compiled and recorded the positional uncertainty  information of 3,561 objects into a new catalog, named 5BZCAT\_err. This catalog also incorporates the optical band magnitude data provided by 5BZCAT. Its basic format is shown in Table~\ref{tab8}, and the complete FITS version is available in Section \ref{sec4}.

\begin{table*}
	\centering
	\caption{Spatial Association Between 5BZCAT Source and NED  Candidate}
	\label{tab7}
	
	\resizebox{\textwidth}{!}{%
		\begin{tabular}{llllllllllll}
			\toprule
			5BZ\_Name &                       NED\_Name &        RA &       DEC &      RA\_m &     DEC\_m &   Sep &    Conf95\_Maj &     Conf95\_Min &  Unc\_PA &             Refcode &  In\_Ellipse \\ \\
			\midrule

			5BZG J0001-1031 & WISEA J000157.26-103117.4      & 0.489   & -10.521 & 0.489     & -10.522    & 0.003      & 0.22875                                 & 0.19675                                 & 0                        & 2013wise.rept....1C & TRUE            \\
			5BZQ J0006-4245 & {[}HB89{]} 0003-430            & 1.582   & -42.755 & 1.582     & -42.755    & 0.001      & 0.21325                                 & 0.213                                   & 0                        & 2013wise.rept....1C & TRUE            \\
			5BZB J0047+5657 & WISEA J004700.43+565742.4      & 11.752  & 56.962  & 11.752    & 56.962     & 0          & 0.00025                                 & 0.0002                                  & 0                        & 2019ApJS..242....5X & TRUE            \\
			5BZB J0116-6153 & WISEA J011619.62-615343.4      & 19.082  & -61.895 & 19.082    & -61.895    & 0.001      & 0.10925                                 & 0.10825                                 & 90                       & 2013wise.rept....1C & TRUE            \\
			5BZB J0133-4414 & WISEA J013306.37-441421.4      & 23.277  & -44.239 & 23.277    & -44.239    & 0.001      & 0.10325                                 & 0.10225                                 & 90                       & 2013wise.rept....1C & TRUE            \\
			5BZG J0320-7500 & WISEA J032052.20-750038.1      & 50.218  & -75.011 & 50.218    & -75.011    & 0          & 0.0975                                  & 0.09175                                 & 0                        & 2013wise.rept....1C & TRUE            \\
			5BZU J0416-4355 & SUMSS J041605-435516           & 64.024  & -43.921 & 64.024    & -43.921    & 0.002      & 0.5                                     & 0.5                                     & 0                        & 2013ApJS..206...13M & TRUE            \\
			5BZB J0425-5331 & WISEA J042504.27-533158.2      & 66.268  & -53.533 & 66.268    & -53.533    & 0          & 0.083                                   & 0.08125                                 & 90                       & 2013wise.rept....1C & TRUE            \\
			5BZB J0433-4502 & APMUKS(BJ) B043142.34-450909.3 & 68.308  & -45.049 & 68.308    & -45.049    & 0          & 0.5                                     & 0.5                                     & 0                        & 2011MNRAS.417.2651M & TRUE            \\
			5BZB J0434-2342 & WISEA J043428.98-234205.3      & 68.621  & -23.701 & 68.621    & -23.702    & 0.002      & 0.15                                    & 0.15                                    & 0                        & 2007ApJS..171...61H & TRUE            \\
			5BZU J0532-0307 & WISEA J053207.51-030707.5      & 83.031  & -3.119  & 83.031    & -3.119     & 0          & 0.15                                    & 0.15                                    & 0                        & 2007ApJS..171...61H & TRUE            \\
			5BZB J0556-4351 & WISEA J055618.74-435146.0      & 89.078  & -43.863 & 89.078    & -43.863    & 0.001      & 0.10375                                 & 0.0995                                  & 90                       & 2013wise.rept....1C & TRUE            \\
			5BZB J0737+3216 & WISEA J073705.58+321632.1      & 114.273 & 32.276  & 114.273   & 32.276     & 0.003      & 0.4065                                  & 0.37625                                 & 0                        & 2013wise.rept....1C & TRUE            \\
			5BZG J0748+2115 & SPIDERS 2\_11411               & 117.181 & 21.263  & 117.181   & 21.263     & 0.001      & 5                                       & 5                                       & 0                        & 2015ApJ...807..178W & TRUE            \\
			5BZB J0751+3928 & WISEA J075144.93+392817.5      & 117.937 & 39.472  & 117.937   & 39.472     & 0.001      & 0.306                                   & 0.2865                                  & 0                        & 2013wise.rept....1C & TRUE            \\
			5BZG J0835+1517 & WISEA J083548.14+151717.0      & 128.951 & 15.288  & 128.951   & 15.288     & 0.001      & 0.1145                                  & 0.11375                                 & 90                       & 2013wise.rept....1C & TRUE            \\
			5BZG J0906+4636 & B3 0902+468                    & 136.565 & 46.605  & 136.565   & 46.605     & 0          & 0.00025                                 & 0.00025                                 & 0                        & 2021AJ....162..121H & TRUE            \\
			5BZG J0950+1804 & WISEA J095000.30+180418.7      & 147.501 & 18.072  & 147.501   & 18.072     & 0.002      & 0.139                                   & 0.134                                   & 0                        & 2013wise.rept....1C & TRUE            \\
			5BZU J1010-0730 & PKS 1007-07                    & 152.521 & -7.506  & 152.521   & -7.506     & 0.001      & 0.18225                                 & 0.17225                                 & 0                        & 2013wise.rept....1C & TRUE            \\
			5BZU J1059+4051 & FIRST J105951.9+405113         & 164.966 & 40.854  & 164.966   & 40.854     & 0.004      & 0.342                                   & 0.319                                   & 172                      & 1995ApJ...450..559B & TRUE            \\
			5BZB J1110-1835 & WISEA J111027.77-183552.7      & 167.616 & -18.598 & 167.616   & -18.598    & 0.001      & 0.1365                                  & 0.13175                                 & 0                        & 2013wise.rept....1C & TRUE            \\
			5BZB J1242-1653 & GALEXASC J124220.51-165303.8   & 190.585 & -16.884 & 190.585   & -16.884    & 0          & 0.5                                     & 0.5                                     & 0                        & 2011MNRAS.417.2651M & TRUE            \\
			5BZG J1243+5212 & WISEA J124308.98+521245.0      & 190.787 & 52.213  & 190.787   & 52.213     & 0.001      & 0.13275                                 & 0.12925                                 & 0                        & 2013wise.rept....1C & TRUE            \\
			5BZG J1301+4634 & WHL J130132.6+463403           & 195.386 & 46.567  & 195.386   & 46.567     & 0          & 5                                       & 5                                       & 0                        & 2015ApJ...807..178W & TRUE            \\
			5BZU J1431-0052 & WISE J143112.91-005244.4       & 217.805 & -0.877  & 217.805   & -0.877     & 0.002      & 0.28                                    & 0.28                                    & 0                        & 2023arXiv230606308D & TRUE            \\
			5BZG J1451+5800 & SDSS J145111.70+580003.0       & 222.799 & 58.001  & 222.799   & 58.001     & 0.002      & 0.5                                     & 0.5                                     & 0                        & 2016SDSSD.C...0000: & TRUE            \\
			5BZG J1539+4143 & SDSS J153951.38+414325.4       & 234.964 & 41.724  & 234.964   & 41.724     & 0.002      & 0.5                                     & 0.5                                     & 0                        & 2007SDSS6.C...0000: & TRUE            \\
			5BZG J1548+6310 & WISEA J154828.42+631051.1      & 237.118 & 63.181  & 237.118   & 63.181     & 0          & 0.1175                                  & 0.114                                   & 0                        & 2013wise.rept....1C & TRUE            \\
			5BZG J1624+3726 & SDSS J162443.36+372642.4       & 246.181 & 37.445  & 246.181   & 37.445     & 0.001      & 0.5                                     & 0.5                                     & 0                        & 2016SDSSD.C...0000: & TRUE            \\
			5BZQ J1717-5155 & WISEA J171734.65-515532.0      & 259.394 & -51.926 & 259.394   & -51.926    & 0.001      & 0.21075                                 & 0.20675                                 & 0                        & 2013wise.rept....1C & TRUE            \\
			5BZB J1830+1324 & WISEA J183000.76+132414.3      & 277.503 & 13.404  & 277.503   & 13.404     & 0.001      & 0.10925                                 & 0.1055                                  & 90                       & 2013wise.rept....1C & TRUE            \\
			5BZB J1844+5709 & RBPL J1844+5709                & 281.212 & 57.161  & 281.212   & 57.161     & 0          & 0.5                                     & 0.5                                     & 0                        & 2014ApJS..215...14D & TRUE            \\
			5BZB J1849+2748 & WISEA J184931.74+274800.8      & 282.382 & 27.800  & 282.382   & 27.800     & 0.001      & 0.09425                                 & 0.09325                                 & 90                       & 2013wise.rept....1C & TRUE            \\
			5BZB J1925-2219 & WISEA J192539.77-221935.1      & 291.416 & -22.326 & 291.416   & -22.326    & 0          & 0.000275                                & 0.00025                                 & 90                       & 2019ApJS..242....5X & TRUE            \\
			5BZB J2036+6553 & WISE J203620.13+655314.5       & 309.084 & 65.887  & 309.084   & 65.887     & 0          & 0.5                                     & 0.5                                     & 0                        & 2014ApJS..215...14D & TRUE            \\
			5BZU J2208+6519 & WISEA J220803.09+651938.9      & 332.013 & 65.327  & 332.013   & 65.327     & 0          & 0.00025                                 & 0.00025                                 & 0                        & 2021AJ....162..121H & TRUE            \\
			5BZU J2250-4206 & WISEA J225022.22-420613.4      & 342.593 & -42.104 & 342.593   & -42.104    & 0.001      & 0.0955                                  & 0.0925                                  & 90                       & 2013wise.rept....1C & TRUE            \\
			5BZB J0105+3928 & 4FGL J0105.1+3929              & 16.289  & 39.471  & 16.291    & 39.496     & 1.527      & 274.14                                  & 228.06                                  & 137                      & 2020ApJS..247...33A & TRUE            \\
			5BZB J0110-0415 & GALEXASC J011030.92-041531.5   & 17.629  & -4.259  & 17.629    & -4.259     & 0.012      & 1.203323                                & 1.203323                                & 0                        & 2012GASC..C...0000S & TRUE            \\
			5BZB J0144+2705 & CRATES J0144+2705 NED02        & 26.140  & 27.084  & 26.140    & 27.084     & 0.002      & 0.15                                    & 0.15                                    & 0                        & 2007ApJS..171...61H & TRUE            \\
			5BZB J0309-3604 & GALEXASC J030902.41-360404.3   & 47.260  & -36.068 & 47.260    & -36.068    & 0.003      & 0.5                                     & 0.5                                     & 0                        & 2011MNRAS.417.2651M & TRUE            \\
			5BZB J0409-0400 & 4FGL J0409.8-0359              & 62.444  & -4.001  & 62.463    & -3.987     & 1.421      & 223.65                                  & 216.99                                  & 139                      & 2020ApJS..247...33A & TRUE            \\
			5BZB J0702-1951 & 3FHL J0702.6-1950              & 105.679 & -19.856 & 105.678   & -19.852    & 0.268      & 110.97                                  & 98.19                                   & 130                      & 2020ApJS..247...33A & TRUE            \\
			5BZB J1129+3756 & 3FGL J1129.0+3758              & 172.264 & 37.949  & 172.253   & 37.972     & 1.441      & 162                                     & 162                                     & 0                        & 2018ApJS..237...32A & TRUE            \\
			5BZB J1603-4904 & 2FGL J1603.8-4904              & 240.961 & -49.068 & 240.961   & -49.068    & 0.002      & 15                                      & 15                                      & 0                        & 2013ApJ...764..135S & TRUE            \\
			5BZB J2149+0322 & 3FHL J2149.8+0322              & 327.424 & 3.381   & 327.441   & 3.384      & 1.019      & 123.48                                  & 123.48                                  & 0                        & 2018ApJS..237...32A & TRUE            \\
			5BZB J2258-3644 & 3FHL J2258.1-3643              & 344.561 & -36.743 & 344.562   & -36.750    & 0.43       & 165.6                                   & 165.6                                   & 0                        & 2018ApJS..237...32A & TRUE            \\
			5BZG J0040-2340 & 4FGL J0040.4-2340              & 10.104  & -23.667 & 10.101    & -23.670    & 0.256      & 306.36                                  & 297.99                                  & 23                       & 2020ApJS..247...33A & TRUE            \\
			5BZG J0106+2539 & WHL J010611.0+253930           & 16.546  & 25.658  & 16.546    & 25.658     & 0.006      & 5                                       & 5                                       & 0                        & 2015ApJ...807..178W & TRUE            \\
			5BZG J0946+5819 & 6C B094238.5+583306            & 146.560 & 58.327  & 146.554   & 58.321     & 0.443      & 63.7                                    & 54.3                                    & 0                        & 1990MNRAS.246..256H & TRUE            \\
			5BZG J1147+3501 & WISEA J114722.13+350107.5      & 176.842 & 35.019  & 176.842   & 35.019     & 0.002      & 0.087                                   & 0.084                                   & 90                       & 2013wise.rept....1C & TRUE            \\
			5BZG J1531+0852 & RX J1531.5+0851                & 232.904 & 8.879   & 232.899   & 8.864      & 0.938      & 150                                     & 150                                     & 0                        & 2000ApJS..129..435B & TRUE            \\
			5BZG J1755+6236 & 6C B175524.9+623700            & 268.952 & 62.612  & 268.957   & 62.611     & 0.161      & 24.3                                    & 21.6                                    & 179                      & 1990MNRAS.246..256H & TRUE            \\
			5BZQ J0845-5555 & 4FGL J0845.7-5556              & 131.457 & -55.924 & 131.447   & -55.949    & 1.53       & 198.09                                  & 190.35                                  & 57                       & 2020ApJS..247...33A & TRUE            \\
			5BZU J0023+4456 & 6C B002051.6+444042            & 5.898   & 44.943  & 5.882     & 44.955     & 0.972      & 91.6                                    & 64.4                                    & 0                        & 1993MNRAS.263...25H & TRUE            \\
			5BZU J0729-1320 & CXO J072917.9-132002           & 112.324 & -13.334 & 112.325   & -13.334    & 0.034      & 8.78                                    & 8.78                                    & 0                        & 2010ApJS..189...37E & TRUE            \\
			5BZU J0804-1712 & CGMW 2-2082                    & 121.140 & -17.201 & 121.150   & -17.193    & 0.756      & 50                                      & 50                                      & 0                        & 1991PASJ...43..449S & TRUE            \\
			5BZU J1238-1959 & 3FGL J1238.2-1958              & 189.602 & -19.987 & 189.565   & -19.983    & 2.088      & 138.96                                  & 138.96                                  & 0                        & 2018ApJS..237...32A & TRUE            \\
			5BZU J1259-2310 & 3FHL J1259.0-2311              & 194.785 & -23.177 & 194.772   & -23.183    & 0.781      & 107.64                                  & 107.64                                  & 0                        & 2018ApJS..237...32A & TRUE            \\
			5BZU J1448+0402 & {[}WB92{]} 1446+0410           & 222.210 & 4.039   & 222.216   & 4.014      & 1.554      & 182                                     & 182                                     & 0                        & 1992ApJS...79..331W & TRUE            \\
			5BZU J1911+1611 & 4FGL J1912.0+1612              & 287.993 & 16.196  & 288.021   & 16.202     & 1.641      & 451.8                                   & 435.15                                  & 42                       & 2020ApJS..247...33A & TRUE            \\
			5BZU J2327+1524 & 1WGA J2327.3+1525              & 351.842 & 15.410  & 351.847   & 15.420     & 0.665      & 52.6                                    & 52.6                                    & 0                        & 2000WGA...C...0000W & TRUE       \\

			\hline
		\end{tabular}%
		
	}
	\begin{tablenotes}
		\small
		\item \textbf{{Note:}} 5BZ\_Name : ... : Source name from 5BZCAT \\
		NED\_Name : ... : Matched object name from NED  \\ 
		RA : deg : Right ascension  \\
		DEC : deg : Declination  \\
		RA\_m : deg : Right ascension of the matched NED  candidate  \\
		DEC\_m : deg : Declination of the matched NED candidate  \\
		Sep : arcmin : Angular separation between the source and the candidate  \\
		Conf95\_Maj : arcsec : Semi-major axis of the 95\% confidence positional  error uncertainty ellipse \\
		Conf95\_Min : arcsec : Semi-minor axis of the 95\% confidence positional error ellipse  \\
		Unc\_PA : deg : Position angle of the uncertainty ellipse  \\
		Refcode : ... : Reference code for the matched NED source  \\
		In\_Ellipse : ... : Within NED ellipse? (TRUE/FALSE)  \\
	\end{tablenotes}
	
\end{table*}

\begin{table*}[h]
	\centering
	\small
	\renewcommand{\arraystretch}{1.2}

	\begin{threeparttable}
			\caption{Positional Data and Uncertainty of 5BZCAT\_err}
		\label{tab8}
		\begin{tabular}{c @{\hspace{8pt}} c @{\hspace{8pt}} c @{\hspace{5pt}} c @{\hspace{5pt}} c @{\hspace{5pt}} c @{\hspace{5pt}} c @{\hspace{5pt}} c}
			\toprule
			Name & RA & DEC & Conf95\_Maj & Conf95\_Min & Unc\_PA & Refcode & Opt\_Mag \\
			\midrule
			5BZB J0001-0011 & 0.340 & -0.194 & 0.209 & 0.209 & 0   & 2013wise.rept....1C & 19.6 \\
			5BZB J0001-0746 & 0.325 & -7.774 & 0.105 & 0.101 & 90  & 2013wise.rept....1C & 17.9 \\
			5BZB J0002-0024 & 0.738 & -0.413 & 0.150 & 0.150 & 0   & 2007ApJS..171...61H & 19.7 \\
			\dots            & \dots & \dots  & \dots & \dots & \dots & \dots             & \dots \\
			\bottomrule
		\end{tabular}
		
		\vspace{2mm}
		\begin{tablenotes}[para]
		
			\small
			\item[] \textbf{Note:}\;
			Name : ... : Source name from 5BZCAT\_err \\
			RA : deg : Right Ascension  \\
			DEC : deg : Declination  \\
			Conf95\_Maj : arcsec : Semi-major axis of the 95\% confidence positional error ellipse  \\
			Conf95\_Min : arcsec : Semi-minor axis of the 95\% confidence positional error ellipse  \\
			Unc\_PA : deg : Position angle of the uncertainty ellipse   \\
			Refcode : ... : Reference code for the matched NED source  \\
			Opt\_Mag : ... : Apparent optical magnitude \\
		\end{tablenotes}
		
	\end{threeparttable}
\end{table*}

\section{The Cross-Identification Pipeline} \label{appendixB}

\begin{figure*}
	\centering
	\includegraphics[width=0.4\textwidth]{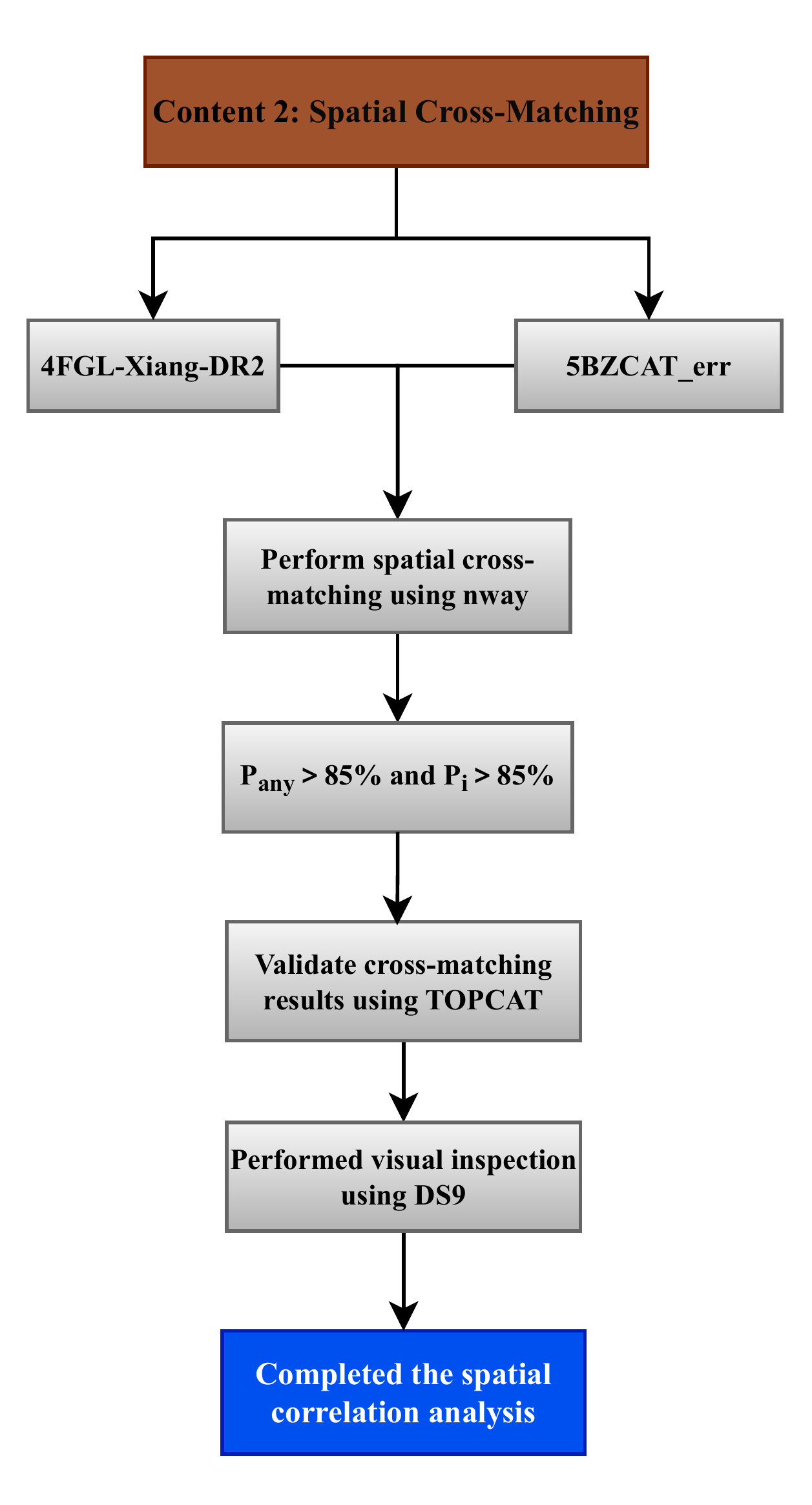}
	\caption{The analytical pipeline for spatial cross-matching between DR2 and 5BZCAT\_err.}
	\label{technical-route2}
\end{figure*}

We designed a cross-identification pipeline, as illustrated in Figure~\ref{technical-route2}. Following this pipeline, we conducted a spatial probability analysis based on the official \textbf{nway} tutorial. Specifically, DR2 was used as the primary catalog and \texttt{5BZCAT\_err} as the matching catalog.
To improve the matching accuracy, we adopted an elliptical error model that incorporates the semi-major axis, semi-minor axis, and position angle.
Furthermore, to improve the precision of posterior probability estimates in the cross-matching process, we incorporated the optical magnitude information (from the \texttt{Opt\_Mag} column of \texttt{5BZCAT\_err}) as a prior distribution.
To identify high-probability matches, all results were filtered using stringent probabilistic thresholds, requiring the any-match probability ($\rm p_{\text{any}}$) and the individual relative match probability ($\rm p_i$) to each exceed 85\%.
The definitions of these two probabilities are detailed below:

\begin{itemize}
	
	\item \( \rm P_{\text{any}} \): The probability that a given source in the primary catalog has at least one counterpart in the secondary catalog.
	
	\item \( \rm P_i \): The probability that a candidate is the true counterpart of the target source, based on its positional match and incorporating other prior information.
	
\end{itemize}

The final results satisfying the matching criteria are recorded in Table~\ref{tab2}. The results indicate that 17 matched  pairs exhibit significant spatial correlation. 
For the remaining matching candidates with lower probabilities, we performed a visual inspection using the plotting script \textbf{nway-explain.py} provided by \textbf{nway}, revealing that they exhibit a low degree of positional coincidence. 
Given the substantial amount of data requiring manual inspection in this step, and to prevent the oversight of cross-identified results due to  complicated procedures, we utilized the efficient cross-matching functionality provided by \textbf{TOPCAT} (see \citeauthor{Taylor2005} \citeyear{Taylor2005}) to perform an additional verification of the cross-matching results between DR2 and 5BZCAT\_err.
\textbf{TOPCAT} is a widely used data analysis tool for spatial cross-matching, providing an efficient method for matching sources based on elliptical positional uncertainties. 
In this study, its primary advantage lies in its ability to rapidly perform batch cross-matching between two catalogs. 
Using this approach, we identified 17 matched pairs, which are fully consistent with the results obtained using the probabilistic thresholds applied in \textbf{nway}.

\begin{figure*}
	\centering
	\begin{subfigure}[b]{0.33\textwidth}
		\centering
		\includegraphics[width=\textwidth]{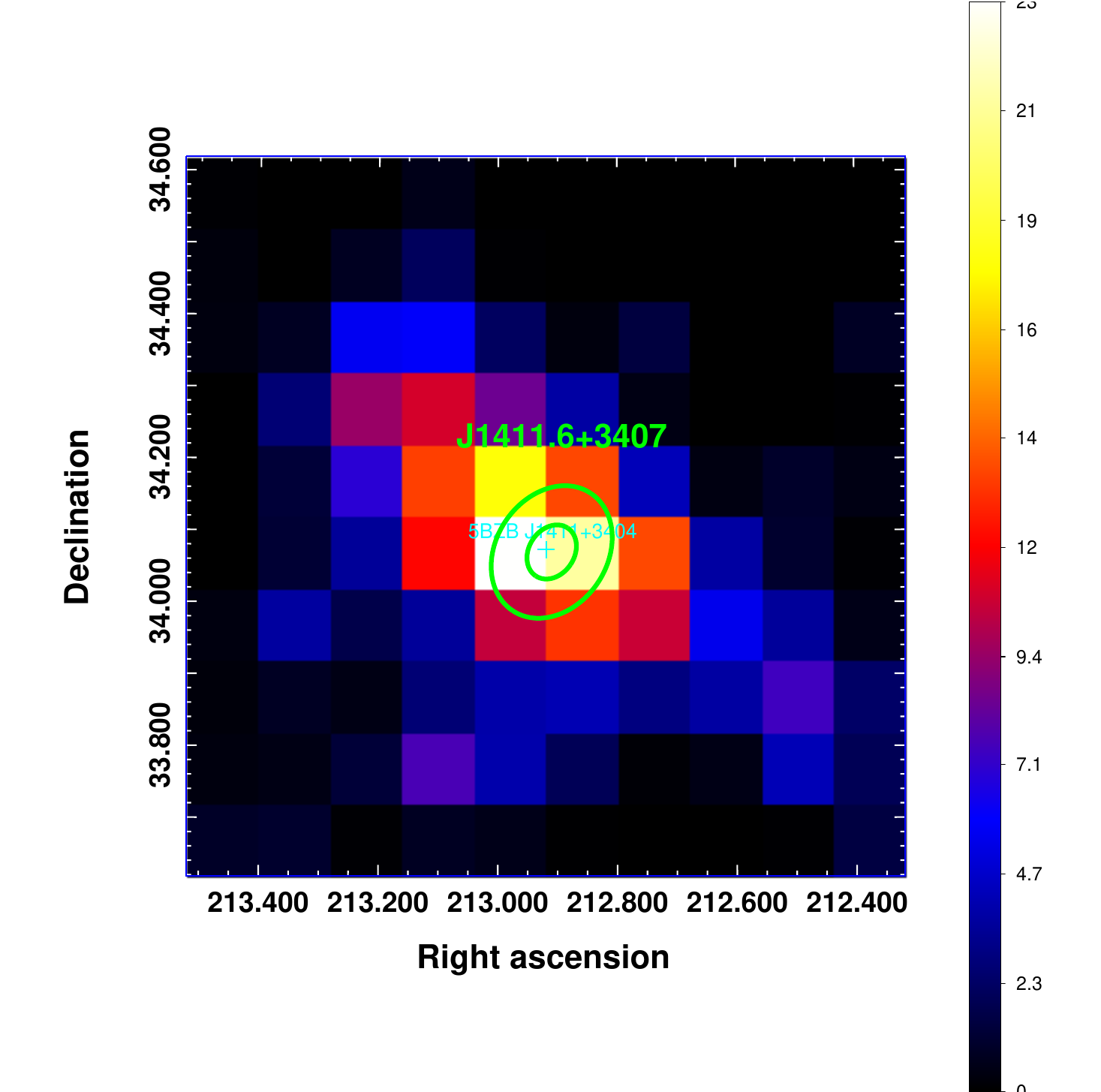}
		\caption{}
		\label{fig:image1}
	\end{subfigure}
	\hfill
	\begin{subfigure}[b]{0.33\textwidth}
		\centering
		\includegraphics[width=\textwidth]{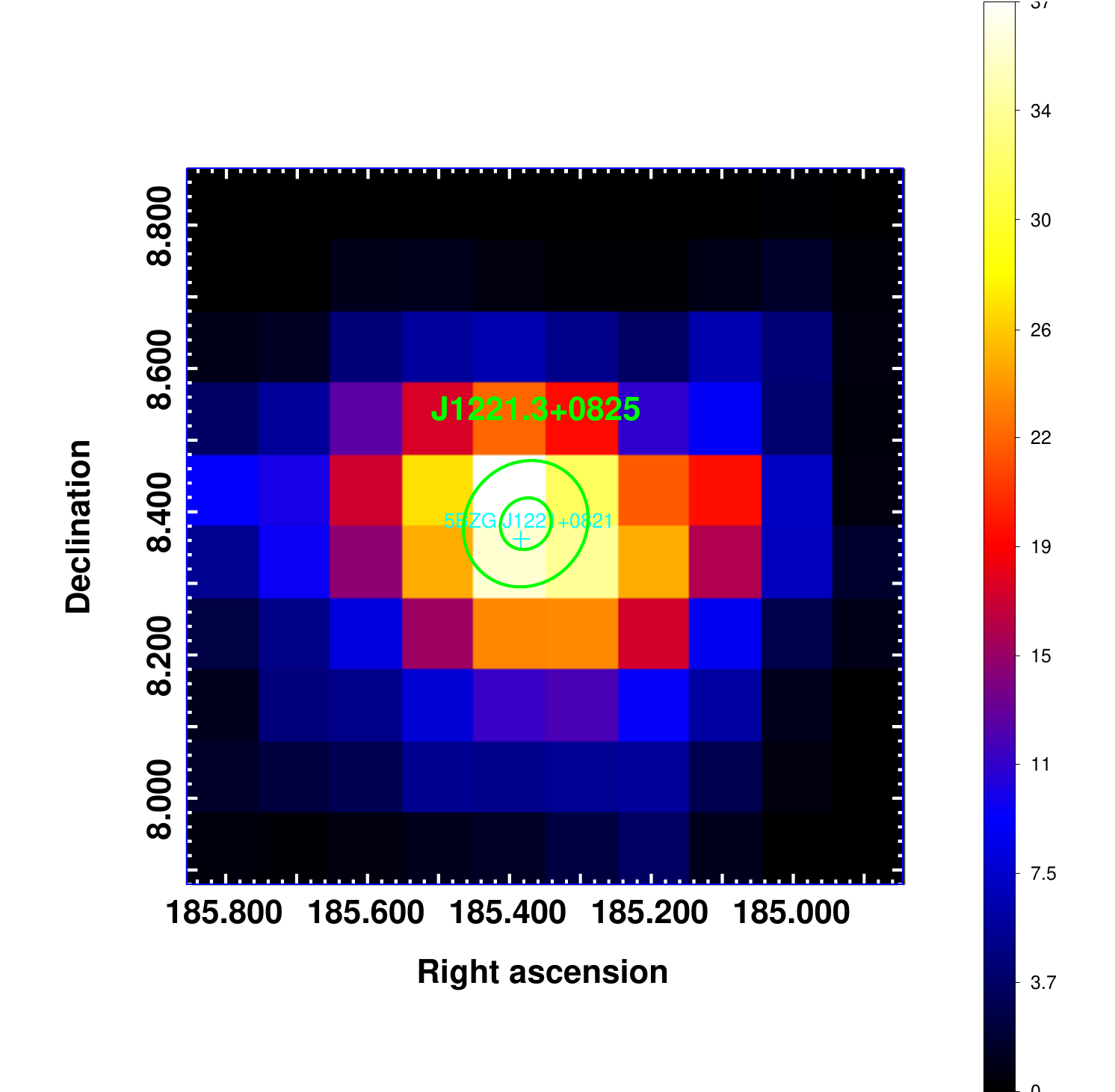}
		\caption{}
		\label{fig:image2}
	\end{subfigure}
	\hfill
	\begin{subfigure}[b]{0.33\textwidth}
		\centering
		\includegraphics[width=\textwidth]{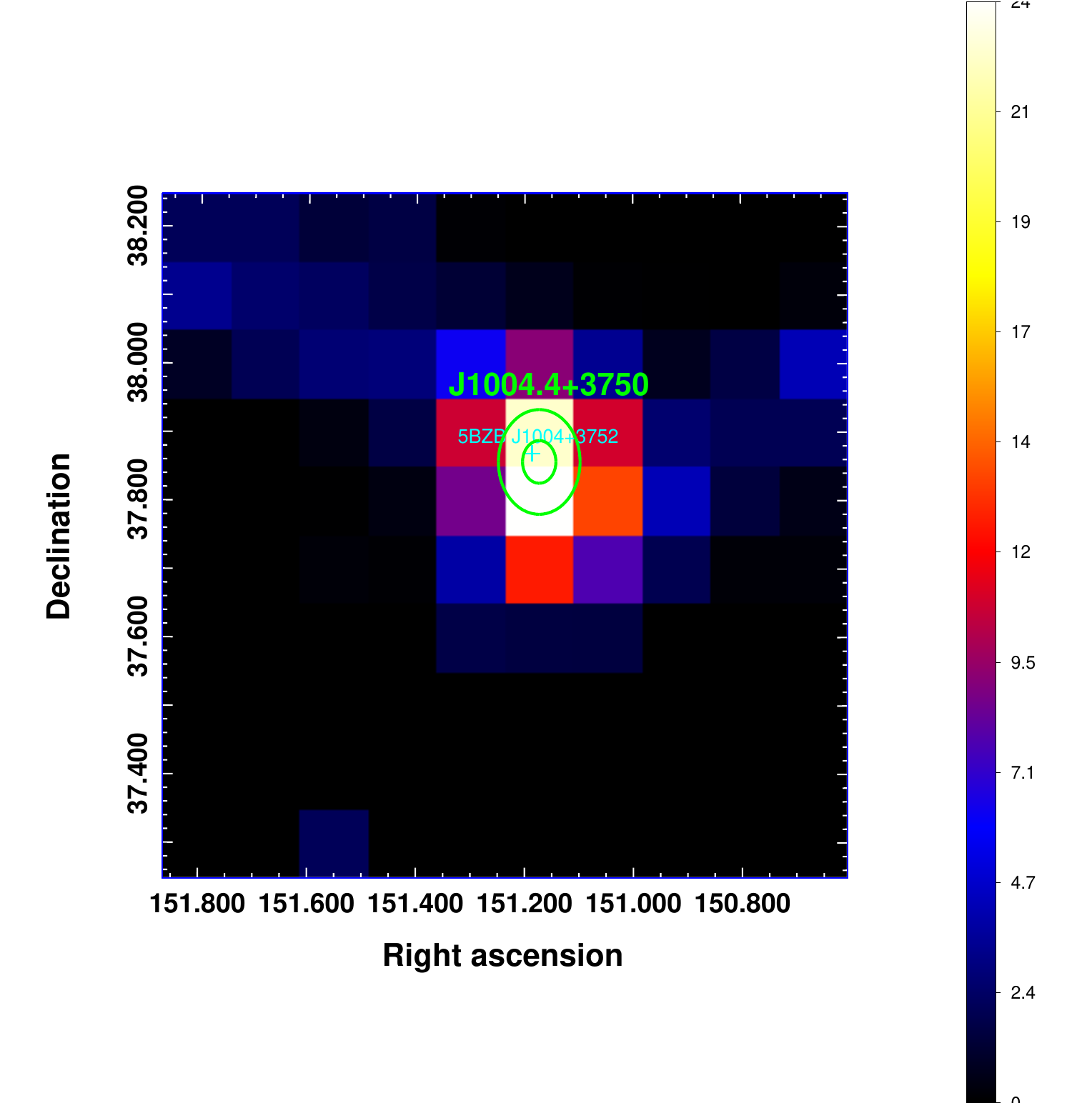}
		\caption{}
		\label{fig:image3}
	\end{subfigure}
	\caption{The TS map visualizes source cross-association, where the color scale reflects the TS value, serving as a measure of detection significance.}
	\label{fig:three_images0}
\end{figure*}

To more comprehensively verify the spatial association between the  $\gamma$-ray emission of target sources in DR2 and the 17 blazar objects in 5BZCAT\_err,  we employed the \textbf{tsmap()} function in \textbf{Fermipy} to generate TS  maps. Subsequently, we conducted a visual inspection of the 17 cross-matched results using the \textbf{DS9} software \citep{Joye2003}.
Here, we present the TS maps of three representative objects, as shown in  Figure~\ref{fig:three_images0}. 
The TS maps for the remaining 16 GeV sources are archived in the file \texttt{TSmap17.tar.xz}; see Section~\ref{sec4} for details.
The results indicate that these objects exhibit a high degree of spatial correlation, further supporting the reliability of our probabilistic cross-matching approach.

\section{Statistical Analysis Pipeline for Multi-wavelength Flux} \label{appendixC}

\subsection{Acquisition and Preparation of the Multi-wavelength Flux Dataset} \label{appendixC.1}

Following the pipeline from Content 3 in Figure~\ref{technical-route2}, we efficiently
extract multi-wavelength flux data for blazars in \texttt{5BZCAT\_err}.  First, we used the \textbf{astroquery.ipac.ned.get\_table()} function with the parameter \texttt{table=`photometry'} to retrieve flux measurements from NED. 
During data collection, we found that 119 sources lacked flux data in the required bands. 
To rule out potential issues such as network interruptions, we re-executed the \textbf{get\_table()} function for these sources and confirmed that the relevant flux data were indeed unavailable. 
Ultimately, we successfully retrieved multi-wavelength flux data for 3,442 sources. The collected data were subsequently standardized in terms of format and units to ensure consistency for subsequent analyses. 
The processed dataset was uniformly stored in the file  \texttt{5BZCAT\_multi-band\_flux.xlsx}, as referenced in Section~\ref{sec4}. 
Next, based on blazar classification, the data were grouped into two categories by energy band: BZB  (including both BZB and BZG) and BZQ. Sources classified as BZU were excluded from subsequent analysis due to their ambiguous classification.

\subsection{Comparative Statistical Analysis of Flux Distributions} \label{appendixC.2}

\begin{figure*}
	\centering
	\begin{subfigure}[b]{0.33\textwidth}
		\centering
		\includegraphics[width=\textwidth]{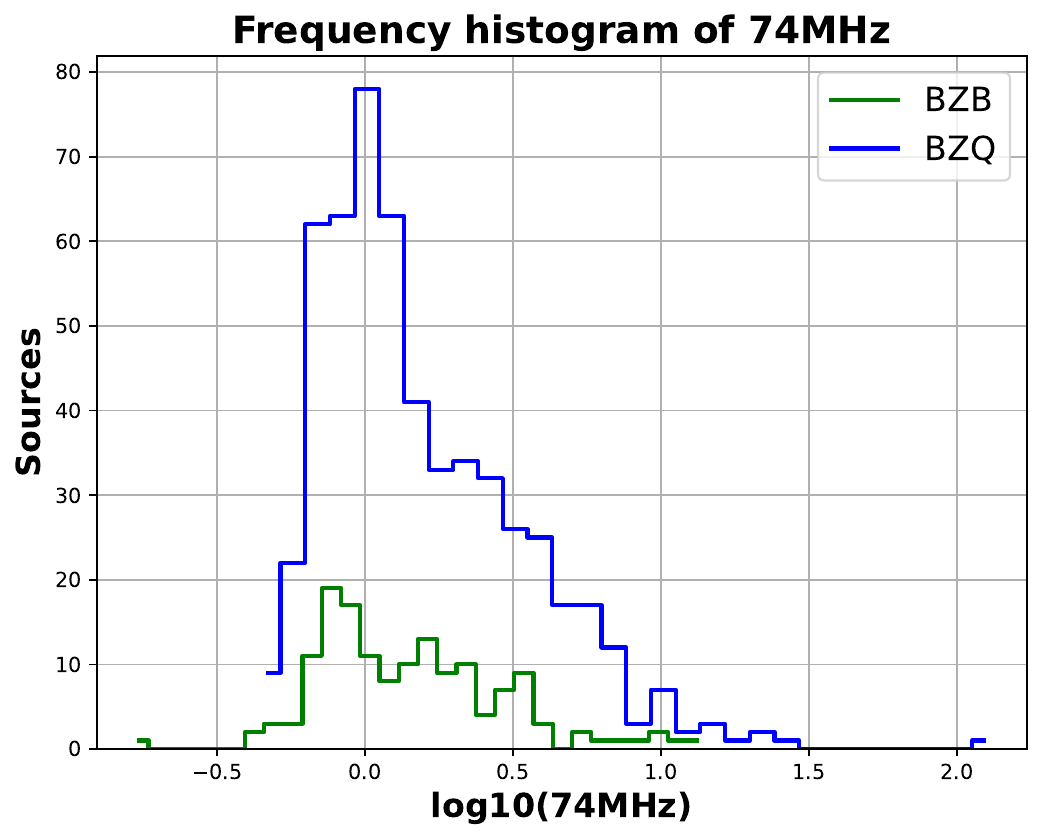}
		\caption{}
		\label{fig_hist:image1}
	\end{subfigure}
	\hfill
	\begin{subfigure}[b]{0.33\textwidth}
		\centering
		\includegraphics[width=\textwidth]{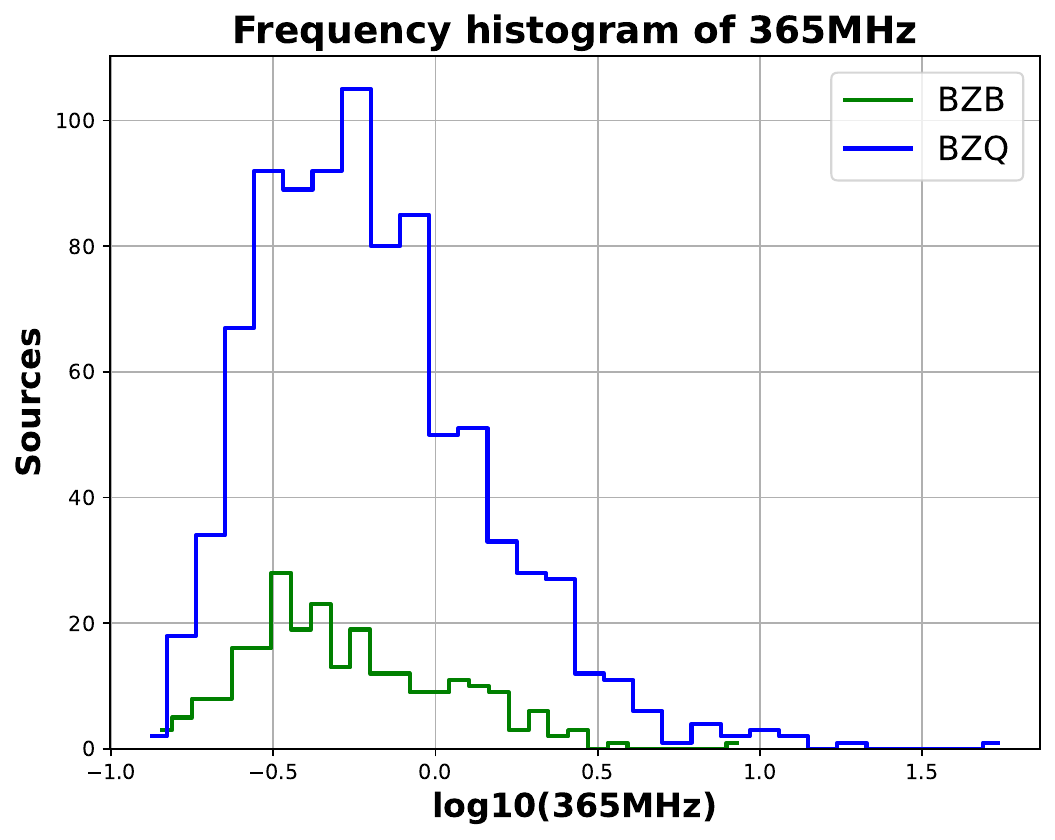}
		\caption{}
		\label{fig_hist:image2}
	\end{subfigure}
	\hfill
	\begin{subfigure}[b]{0.33\textwidth}
		\centering
		\includegraphics[width=\textwidth]{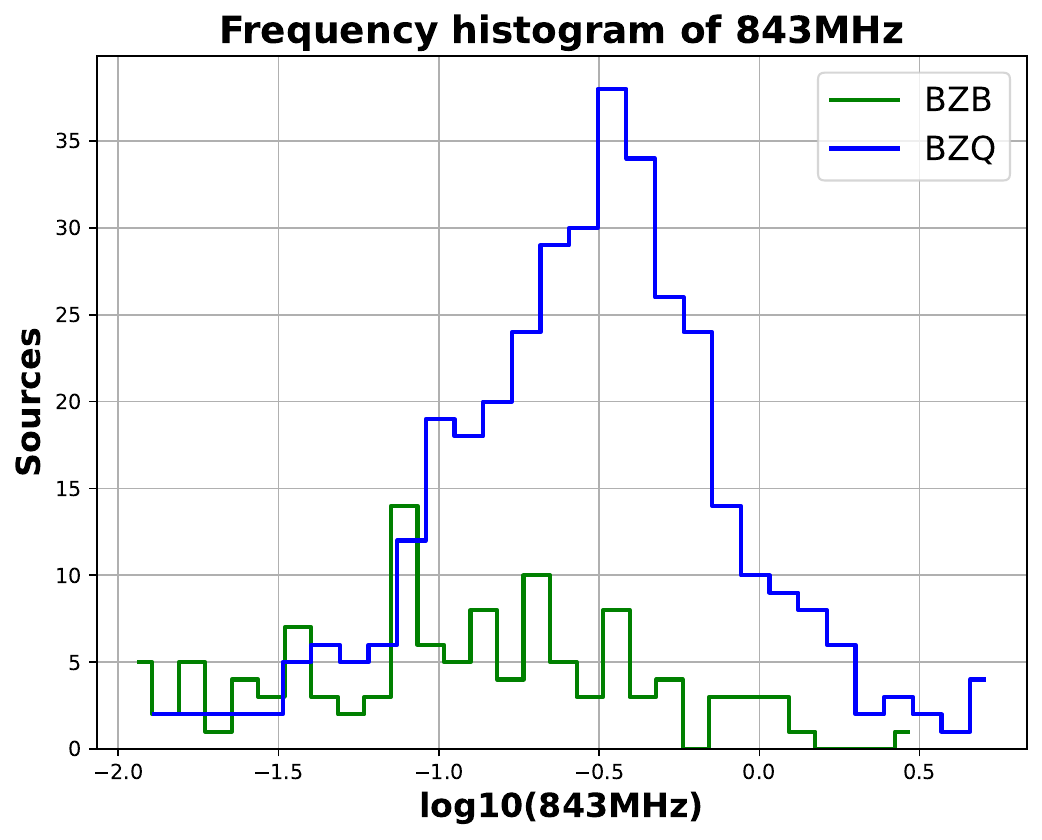}
		\caption{}
		\label{fig_hist:image3}
	\end{subfigure}
	
	\vspace{0.25em}
	
	\begin{subfigure}[b]{0.33\textwidth}
		\centering
		\includegraphics[width=\textwidth]{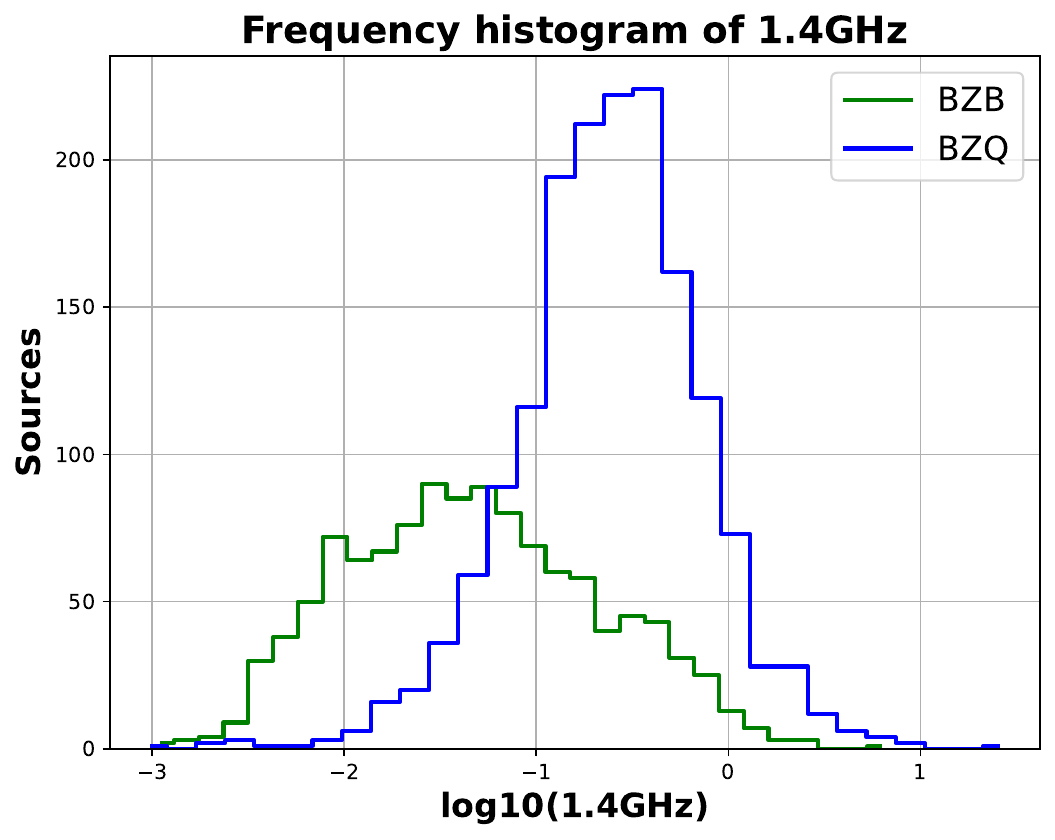}
		\caption{}
		\label{fig_hist:image4}
	\end{subfigure}
	\hfill
	\begin{subfigure}[b]{0.33\textwidth}
		\centering
		\includegraphics[width=\textwidth]{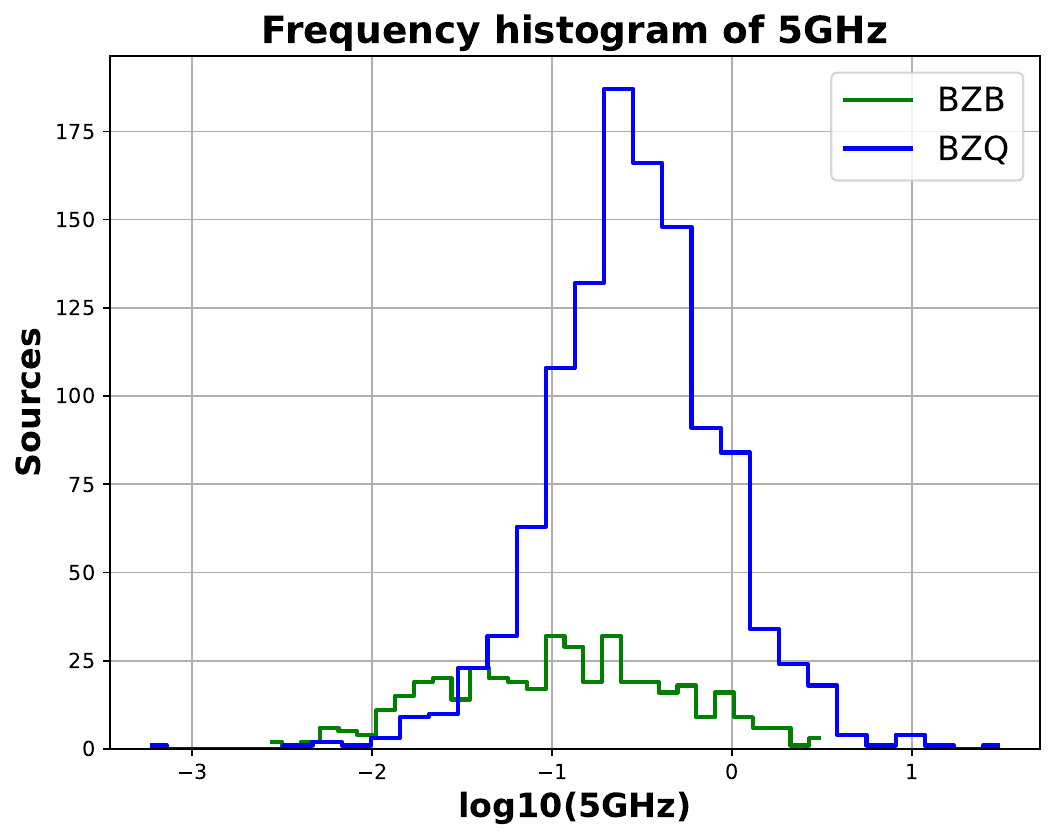}
		\caption{}
		\label{fig_hist:image5}
	\end{subfigure}
	\hfill
	\begin{subfigure}[b]{0.33\textwidth}
		\centering
		\includegraphics[width=\textwidth]{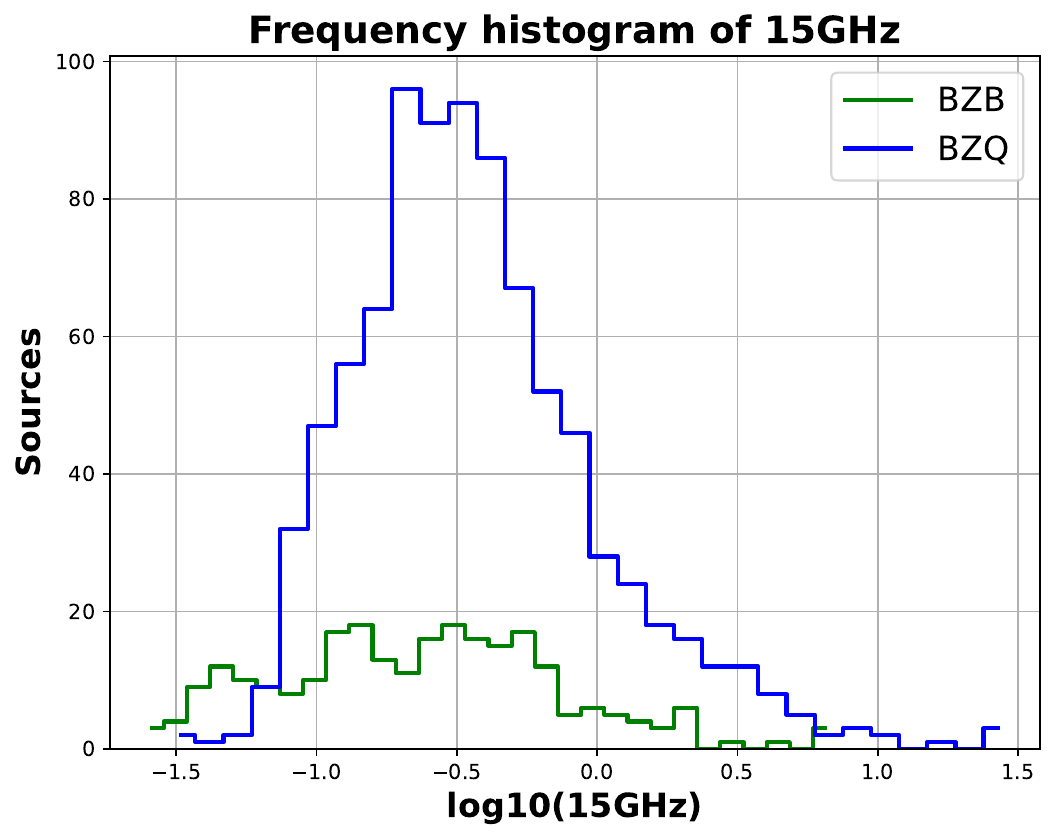}
		\caption{}
		\label{fig_hist:image6}
	\end{subfigure}
	
	\vspace{0.25em}
	
	\begin{subfigure}[b]{0.33\textwidth}
		\centering
		\includegraphics[width=\textwidth]{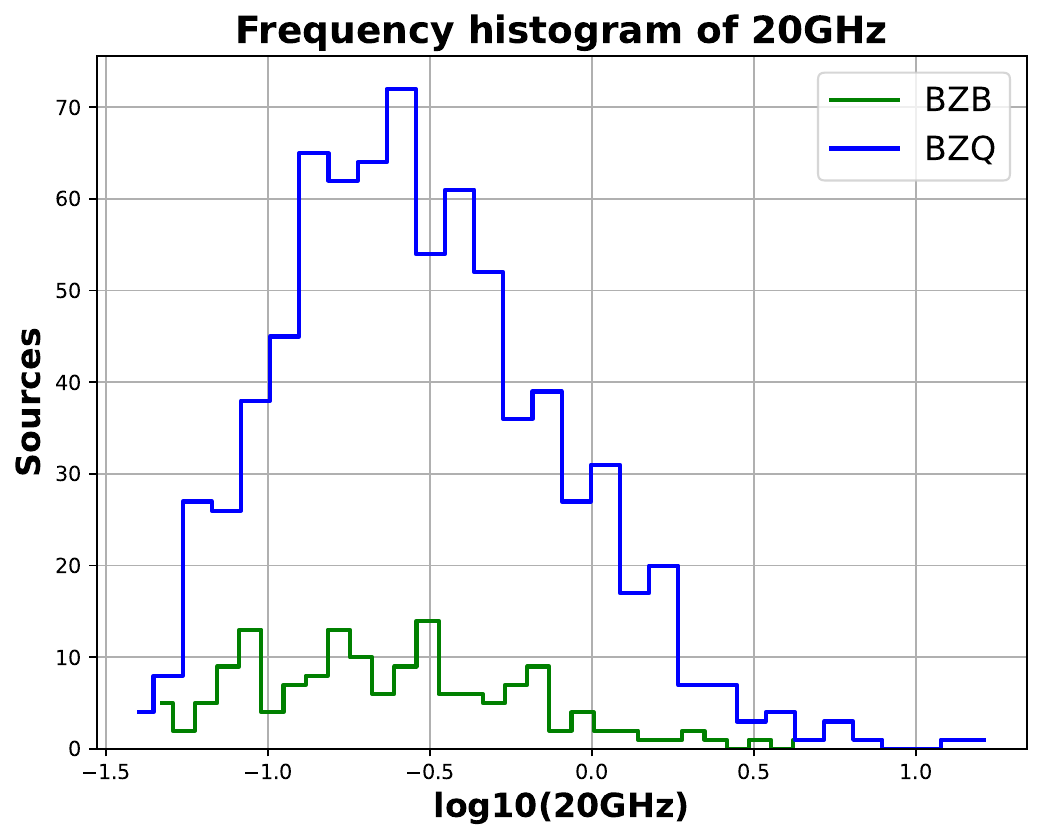}
		\caption{}
		\label{fig:image7}
	\end{subfigure}
	\hfill
	\begin{subfigure}[b]{0.33\textwidth}
		\centering
		\includegraphics[width=\textwidth]{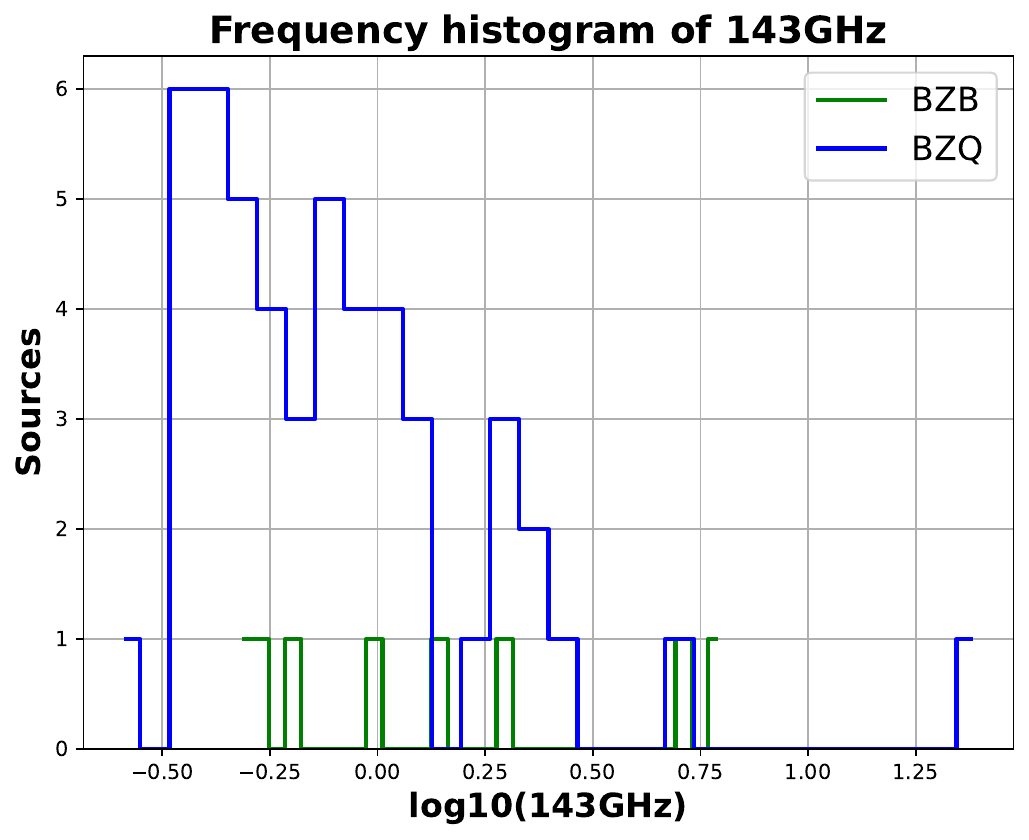}
		\caption{}
		\label{fig:image8}
	\end{subfigure}
	\hfill
	\begin{subfigure}[b]{0.33\textwidth}
		\centering
		\includegraphics[width=\textwidth]{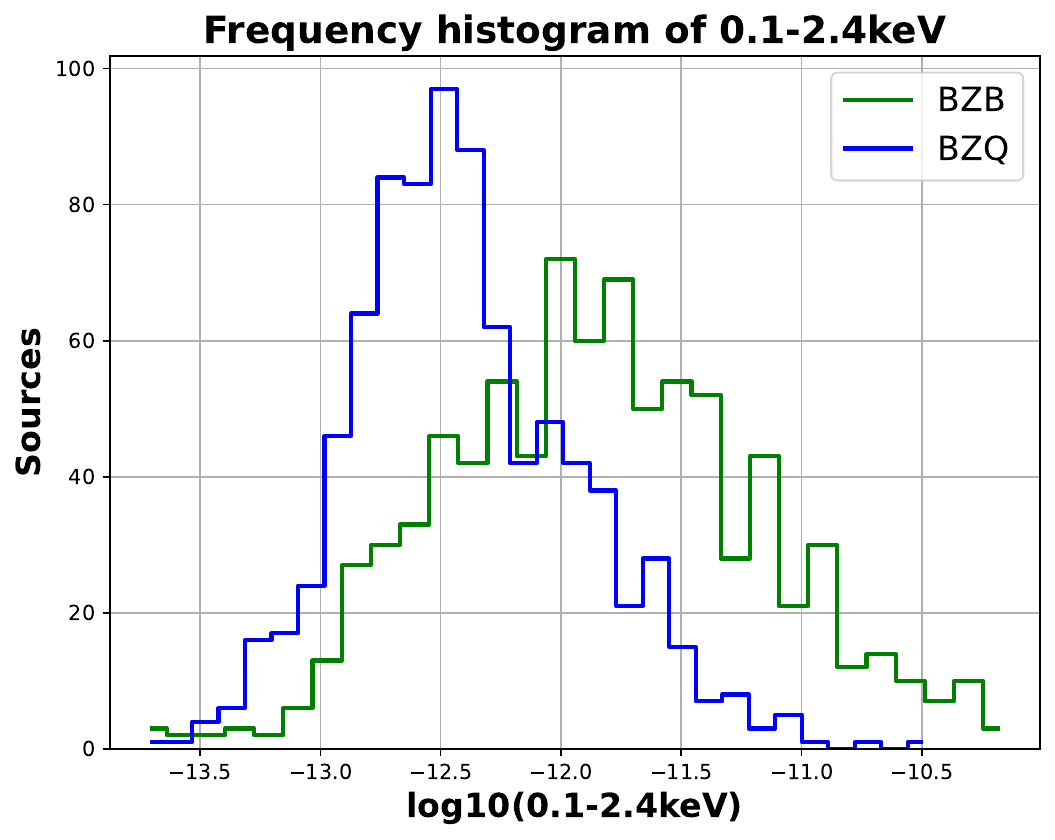}
		\caption{}
		\label{fig:image9}
	\end{subfigure}
	
	\vspace{0.25em}
	
	\begin{subfigure}[b]{0.33\textwidth}
		\centering
		\includegraphics[width=\textwidth]{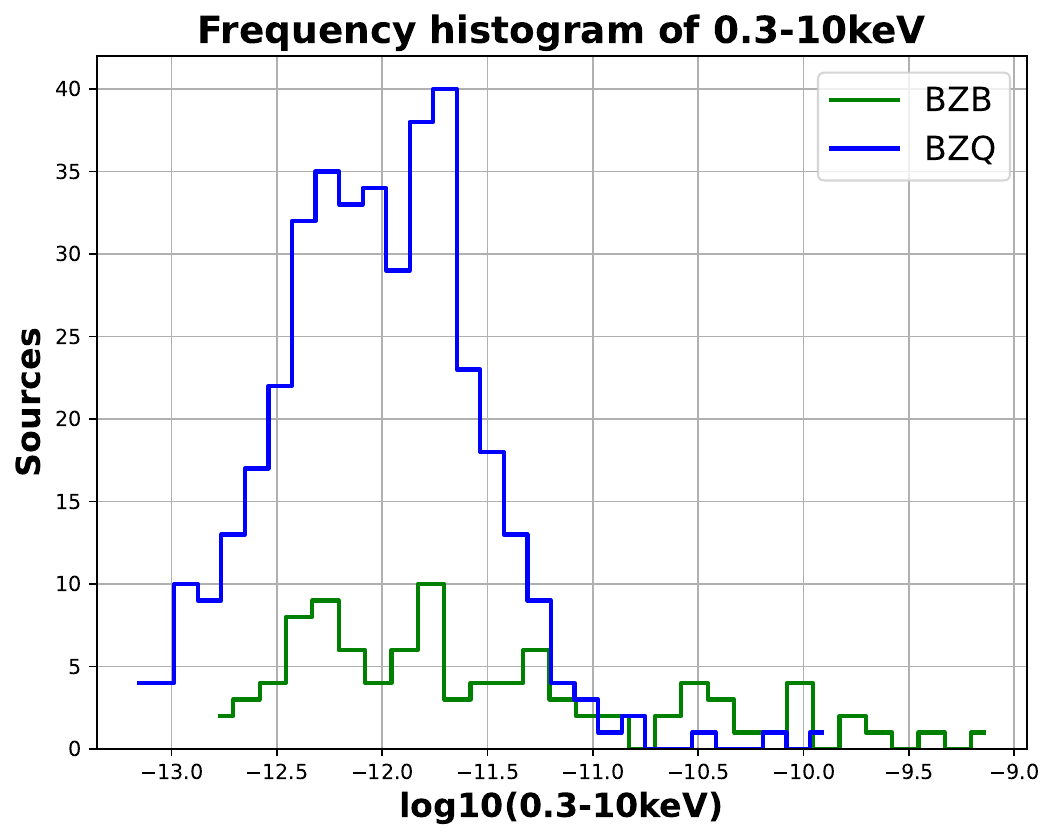}
		\caption{}
		\label{fig:image10}
	\end{subfigure}
	\hfill
	\begin{subfigure}[b]{0.33\textwidth}
		\centering
		\includegraphics[width=\textwidth]{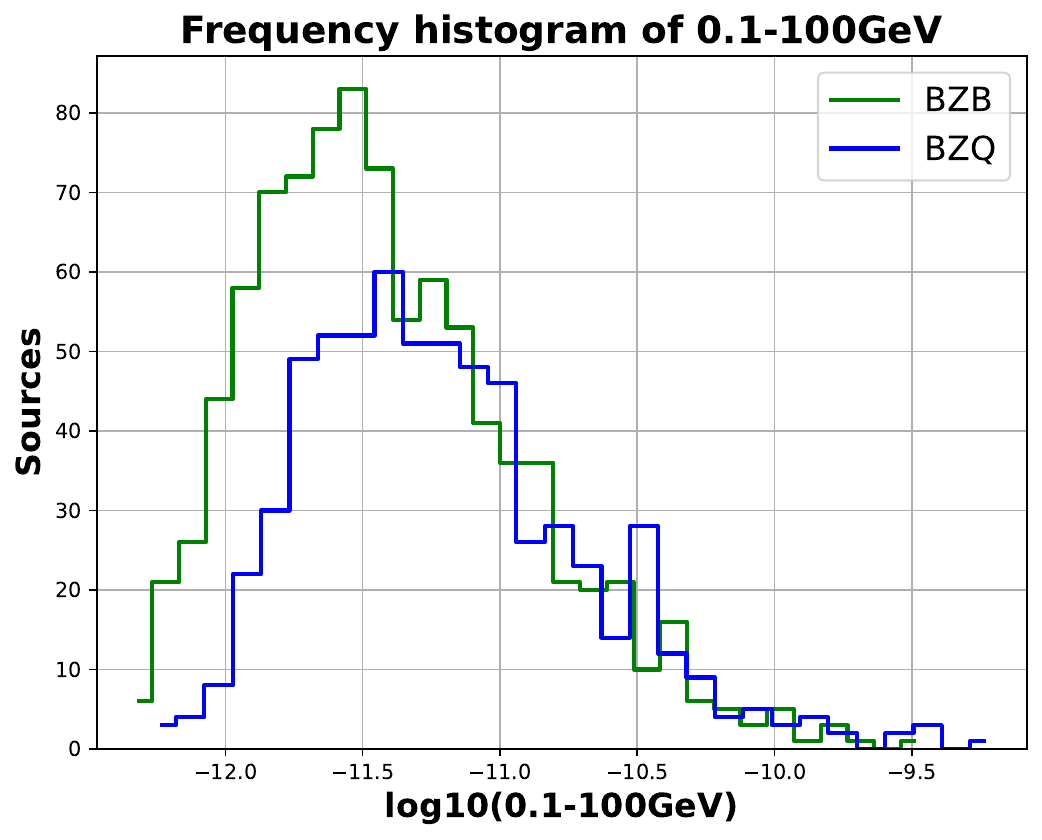}
		\caption{}
		\label{fig:image11}
	\end{subfigure}
	\hfill
	\begin{subfigure}[b]{0.33\textwidth}
		\centering
		\phantom{} 
	\end{subfigure}
	
	\caption{Frequency histograms of flux distributions for blazars in the 5BZCAT\_err across 11 wavebands, showing BZBs in green and BZQs in blue, with flux values on the horizontal axis and frequency on the vertical axis.}
	\label{fig_hist:three_images}
\end{figure*}

\begin{figure*}
	\centering
	\includegraphics[width=1.1\textwidth]{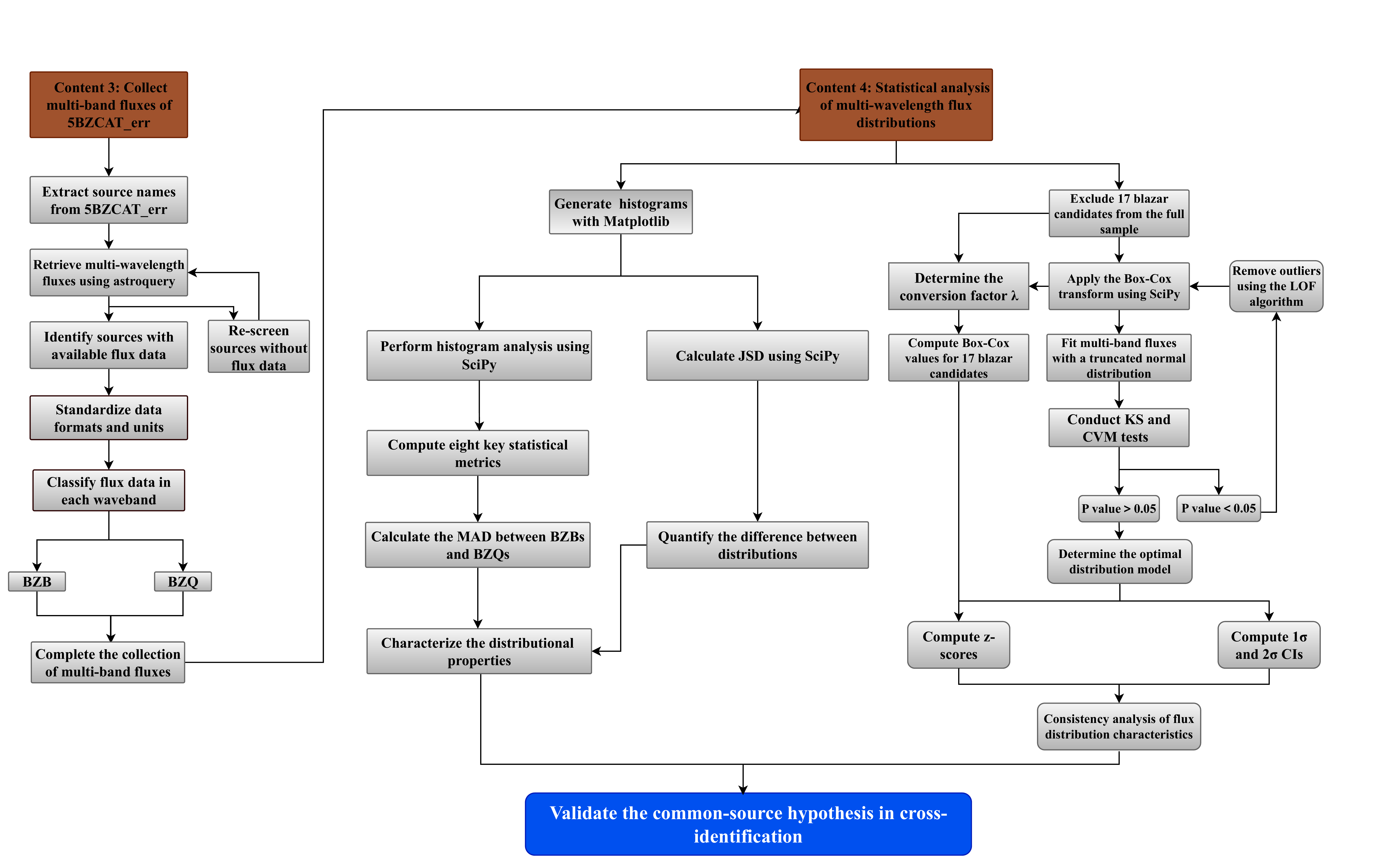}
	\caption{This figure shows the pipeline for the statistical modeling and analysis of multi-band flux distributions in blazars.
	}
	\label{technical-route3}
	
\end{figure*}

A detailed pipeline for this part is presented in Content 4 of Figure \ref{technical-route3}. To facilitate an investigation of the multi-wavelength flux distribution of 5BZCAT\_err blazars, we applied a base-10 logarithmic transformation to the flux values. This approach effectively compresses the wide range of the data, significantly reduces skewness, and adheres to scientific conventions in blazar flux analysis \citep{Ackermann2015}. 
To systematically investigate the differences in the flux distributions of BZBs and BZQs across various wavebands, we employed \textbf{matplotlib.pyplot} to create frequency histograms spanning 11 wavebands (see Figure~\ref{fig_hist:three_images}). Additionally, using \textbf{scipy.stats} \citep{Virtanen2020}, we computed eight commonly used statistical parameters to comprehensively characterize the statistical properties of the distributions from multiple dimensions.

\begin{table*}
	\centering
	\small
	\renewcommand{\arraystretch}{1.2} 
	\begin{threeparttable}
		\caption{Comparative Analysis of Flux Distributions Between BZBs and BZQs}
		\label{tab9}
	\begin{tabular}{lclllllllc}
		\hline\hline
		
		\hline
		Type & Min & Max & Mean & Median & Mode & Std & Skewness & Kurtosis  & JSD(BZB \& BZQ)  \\
		\hline
		\multicolumn{10}{c}{{74\ MHz}} \\
		\hline
		BZB & -0.76 & 1.19 & 0.18 & 0.13 & -0.11 & 0.32 & 0.64 & 0.66 & \multirow{2}{*}{0.11} \\
		BZQ & -0.33 & 2.18 & 0.24 & 0.14 & -0.11 & 0.34 & 1.14 & 1.88 & \\
		\hline
		\multicolumn{10}{c}{{365\ MHz}} \\
		\hline
		BZB & -0.84 & 0.99 & -0.24 & -0.30 & -0.22 & 0.31 & 0.62 & 0.21 & \multirow{2}{*}{0.12} \\
		BZQ & -0.87 & 1.82 & -0.15 & -0.20 & -0.48 & 0.35 & 0.96 & 1.68 & \\
		\hline
		\multicolumn{10}{c}{{843\ MHz}} \\
		\hline
		BZB & -1.94 & 0.55 & -0.90 & -0.88 & -1.55 & 0.54 & 0.06 & -0.48 & \multirow{2}{*}{0.31} \\
		BZQ & -1.89 & 0.79 & -0.50 & -0.46 & -1.07 & 0.46 & -0.12 & 0.59 & \\
		\hline
		\multicolumn{10}{c}{{1.4\ GHz}} \\
		\hline
		BZB & -2.95 & 0.92 & -1.26 & -1.30 & -2.18 & 0.65 & 0.20 & -0.53 & \multirow{2}{*}{0.50} \\
		BZQ & -3.00 & 1.55 & -0.55 & -0.52 & -0.74 & 0.48 & -0.32 & 1.46 & \\
		\hline
		\multicolumn{10}{c}{{5\ GHz}} \\
		\hline
		BZB & -2.55 & 0.59 & -0.91 & -0.89 & -1.70 & 0.63 & -0.04 & -0.57 & \multirow{2}{*}{0.36} \\
		BZQ & -3.22 & 1.64 & -0.46 & -0.46 & 0.08 & 0.48 & -0.15 & 1.73 & \\
		\hline
		\multicolumn{10}{c}{{15\ GHz}} \\
		\hline
		BZB & -1.59 & 0.89 & -0.60 & -0.59 & -1.35 & 0.49 & 0.32 & -0.12 & \multirow{2}{*}{0.25} \\
		BZQ & -1.48 & 1.53 & -0.38 & -0.42 & -0.65 & 0.43 & 0.88 & 1.36 & \\
		\hline
		\multicolumn{10}{c}{{20\ GHz}} \\
		\hline
		BZB & -1.33 & 0.72 & -0.59 & -0.63 & -1.13 & 0.42 & 0.50 & -0.12 & \multirow{2}{*}{0.11} \\
		BZQ & -1.40 & 1.30 & -0.48 & -0.54 & -0.77 & 0.43 & 0.54 & 0.30 & \\
		\hline
		\multicolumn{10}{c}{{143\ GHz}} \\
		\hline
		BZB & -0.31 & 0.82 & 0.16 & 0.07 & -0.31 & 0.41 & 0.50 & -1.20 & \multirow{2}{*}{0.29} \\
		BZQ & -0.59 & 1.45 & -0.06 & -0.11 & -0.29 & 0.35 & 1.75 & 4.80 & \\
		\hline
		\multicolumn{10}{c}{{0.1--2.4 keV}} \\
		\hline
		BZB & -13.70 & -10.06 & -11.78 & -11.80 & -12.77 & 0.65 & 0.08 & -0.19 & \multirow{2}{*}{0.40} \\
		BZQ & -13.70 & -10.39 & -12.34 & -12.40 & -12.82 & 0.47 & 0.49 & 0.34 & \\
		\hline
		\multicolumn{10}{c}{{0.3--10 keV}} \\
		\hline
		BZB & -12.77 & -9.02 & -11.47 & -11.71 & -12.30 & 0.86 & 0.82 & -0.13 & \multirow{2}{*}{0.34} \\
		BZQ & -13.15 & -9.80 & -11.98 & -11.99 & -12.46 & 0.49 & 0.33 & 1.05 & \\
		\hline
		\multicolumn{10}{c}{{0.1--100 GeV}} \\
		\hline
		BZB & -12.31 & -9.39 & -11.35 & -11.43 & -12.31 & 0.50 & 0.71 & 0.34 & \multirow{2}{*}{0.18} \\
		BZQ & -12.23 & -9.13 & -11.15 & -11.22 & -12.23 & 0.50 & 0.77 & 0.70 & \\
		\hline
		MAD & 0.20 & 0.63 & 0.32 & 0.30 & 0.48 & 0.11 & 0.41 & 1.64 & \multirow{1}{*}{-}\\
		\hline
\end{tabular}
\vspace{2mm}
\begin{tablenotes}[para]
\small
\item[] \textbf{Note:}\ 
Type : ... : Source class \\
Min : Jy/erg cm$^{-2}$ s$^{-1}$ : Minimum value of log10(flux) distribution  \\
Max : Jy/erg cm$^{-2}$ s$^{-1}$ : Maximum value of log10(flux) distribution  \\
Mean : Jy/erg cm$^{-2}$ s$^{-1}$ : Mean of log10(flux) distribution  \\
Median : Jy/erg cm$^{-2}$ s$^{-1}$ : Median of log10(flux) distribution  \\
Mode : Jy/erg cm$^{-2}$ s$^{-1}$ : Mode of log10(flux) distribution  \\
Std : Jy/erg cm$^{-2}$ s$^{-1}$ : Standard deviation  \\
Skewness : ... : Measure of skewness  \\
Kurtosis : ... : Measure of kurtosis  \\
JSD(BZB \& BZQ) : ... : Jensen–Shannon distance between the flux distributions of BZBs and BZQs  \\
MAD: Jy/erg cm$^{-2}$ s$^{-1}$ : The Mean Absolute Difference (MAD) between BZBs and BZQs for the eight statistical metrics
\end{tablenotes}
\end{threeparttable}
\end{table*}

\begin{figure*}
	\centering
	\includegraphics[width=0.8\textwidth, height=1.0\textheight, keepaspectratio]{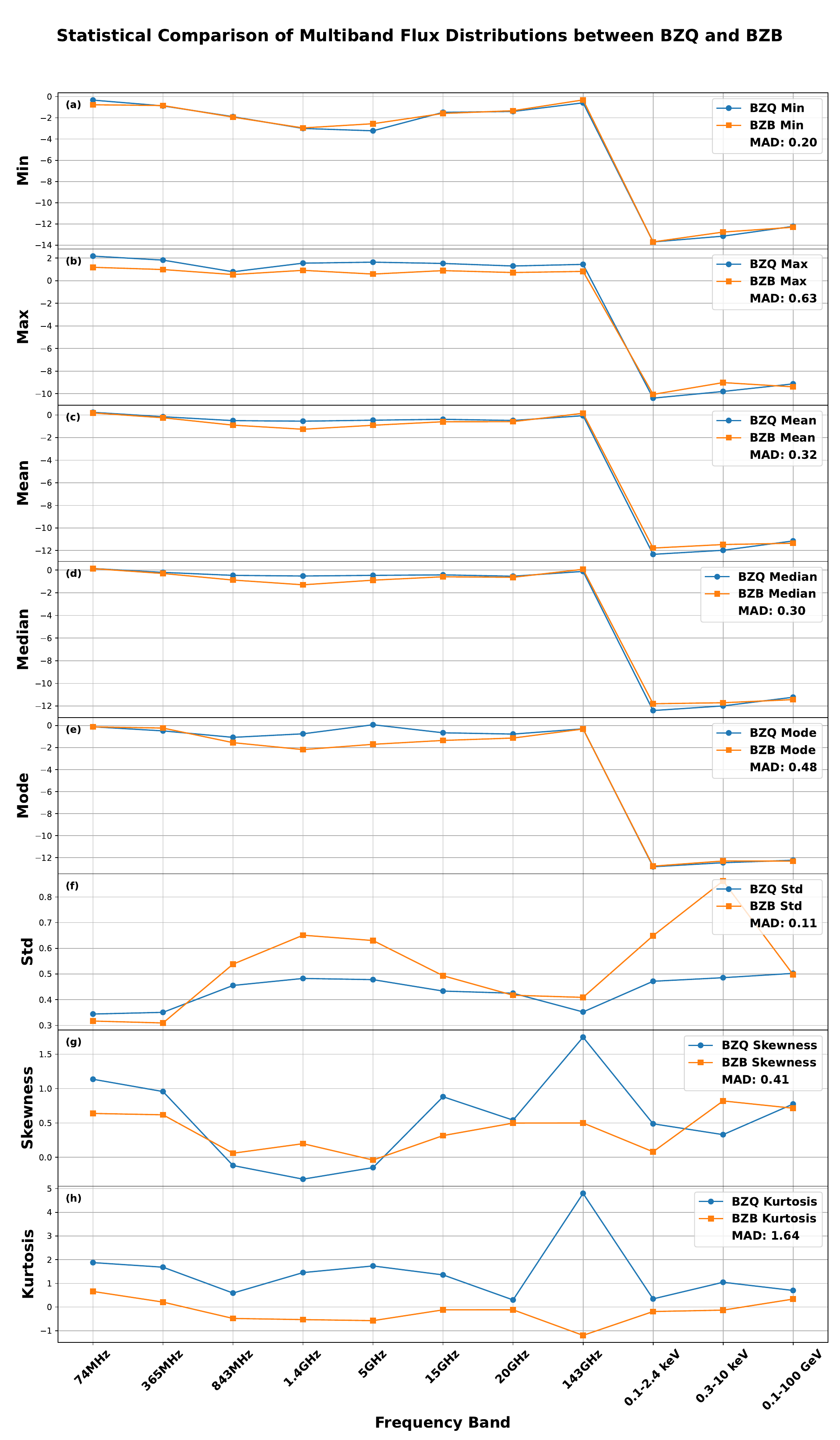}
	\caption{Comparative analysis of multiband flux statistical characteristics between BZQs and BZBs. In each subplot, the blue and orange data points represent the values of eight classical statistical metrics, calculated across 11 wavebands, for BZBs and BZQs, respectively.}
	\label{MAD}
\end{figure*}

These parameters include  the minimum (min), maximum (max), mean, median, mode, standard deviation (std), skewness, and kurtosis. 
The mean represents the average of the logarithmic flux values and reflects the overall level of the distribution. The median denotes the central value of the ordered dataset, indicating the distribution's central tendency and is less sensitive to outliers. 
The mode refers to the most frequently occurring value, representing the most common state within the distribution.
The std reflects the dispersion of the flux data, with larger values indicating greater spread and smaller values indicating tighter concentration. 
Skewness measures the asymmetry of the flux distribution, where positive values indicate right skewness and negative values indicate left skewness.
Kurtosis describes the sharpness of the distribution profile, with higher values indicating a more peaked distribution and lower values indicating a flatter shape. 
Together, these parameters collectively provide a quantitative basis for multi-wavelength flux analysis. 
To clearly investigate the differences in various statistical metrics between BZBs and BZQs, we plotted the relationship between each statistical metric and waveband, as shown in Figure~\ref{MAD}, in conjunction with Table~\ref{tab9}. 

To quantify the overall differences in various statistical metrics between BZBs and BZQs across different wavebands, we computed the MAD for each metric, using the following formula:

\begin{equation}
\mathrm{MAD} = \frac{1}{n} \sum_{i=1}^{n} \left| S_{\mathrm{BZQ},i} - S_{\mathrm{BZB},i} \right|
\label{equ1}
\end{equation}
where \( S_{\mathrm{BZQ},i} \) and \( S_{\mathrm{BZB},i} \) denote the values of a given statistical metric (e.g., min, max, mean, etc.) for BZQs and BZBs, respectively, in the \( i \)-th frequency band. \( n \) represents the total number of frequency bands.

The MAD value quantifies the average difference in statistical metrics between BZQs and BZBs, with smaller values indicating closer similarity and larger values implying greater disparity.
Subsequently, we calculated the MAD values of eight statistical metrics for BZBs and BZQs across different wavebands, as presented in the last column of Table~\ref{tab9}.
From subfigures (a) to (e) in Figure~\ref{MAD}, a pronounced downward trend is observed in the curves from 143 GHz to 0.1–2.4 keV.
This trend is primarily attributed to the inconsistency in flux units: fluxes from 74 MHz to 143 GHz are measured in Jy, whereas those from 0.1 keV to 100 GeV are measured in erg cm$^{-2}$ s$^{-1}$. 
However, this unit discrepancy has no substantial impact on the MAD-based analysis of the relevant metrics, as changes in units do not affect the relative distribution profiles of the two object classes within the same energy band. 
Based on the MAD values and the observed differences across statistical metrics, we draw the following conclusions:

(1) Panel (a) of Figure~\ref{MAD} shows that the difference in the min flux values between BZBs and BZQs is relatively small, as reflected by a MAD value of 0.2. This suggests that the lower limits of the fluxes produced by their respective radiation mechanisms are not significantly different.

(2) Panel (b) of Figure~\ref{MAD} reveals a substantial difference in the max flux values between BZBs and BZQs, with a corresponding MAD of 0.63. Across the 74 MHz to 143 GHz range, BZQs exhibit max flux values approximately one order of magnitude higher than those of BZBs.
In the 0.3–10 keV band, the max flux value of BZBs is slightly higher than that of BZQs. However, in the 0.1–2.4 keV and 0.1–100 GeV bands, the overall difference is negligible, with the max values of BZBs and BZQs nearly overlapping.

(3) As shown in panels (c), (d), and (e) of Figure~\ref{MAD}, the trends of the mean, median, and mode for BZQs and BZBs exhibit remarkable similarity. 
This similarity stems primarily from their shared mathematical properties, with all three serving as measures of central tendency in a distribution. 
In the radio band, particularly within the frequency range of 843 MHz to 15 GHz, the mean, median, and mode values for BZQs are consistently higher than those for BZBs. This suggests that BZQs generally exhibit stronger radio emission in these bands. 
In the 0.1–10 keV energy range, the mean and median values for BZBs are slightly higher than those for BZQs. Across the 0.1–100 GeV range, the values of all three metrics are similar between BZBs and BZQs, showing no significant differences.

(4) 
Panel (f) of Figure \ref{MAD} shows the variation trends of the std for BZQs and BZBs across different frequency bands. As indicated by a MAD value of 0.11, the overall average difference in std between BZBs and BZQs is not significant. 
However, across the frequency range from 843 MHz to 100 GeV, the flux distributions of BZBs generally exhibit greater dispersion than those of BZQs. 
The std values for BZQs are mostly confined to a narrower range of 0.34–0.50, whereas those for BZBs span a broader range of 0.30–0.90. This suggests that BZB flux distributions are more widely spread, while those of BZQs are relatively more concentrated.

(5) Panel (g) of Figure \ref{MAD} shows the variation of skewness for BZQs and BZBs across different wavebands. 
Overall, both source types exhibit positive skewness in most frequency bands, indicating that their flux distributions are right-skewed. 
Furthermore, the MAD of skewness between BZQs and BZBs is 0.41, suggesting notable differences in the asymmetry of their flux distributions across wavebands. 
This discrepancy is especially evident at 143 GHz, where the skewness of BZQ surpasses that of BZB by approximately 1.3. 
The skewness of the BZQ sample ranges from -0.3 to 1.8, suggesting that the asymmetry of its flux distribution varies noticeably across different energy ranges. 
In contrast, the skewness of the BZB sample shows a relatively smooth trend, with values primarily concentrated between 0 and 0.8, suggesting a more stable distribution structure across different wavebands.

(6) Panel (h) of Figure \ref{MAD} shows that the MAD of kurtosis between BZQs and BZBs  is 1.64, indicating a significant difference in the peakedness of their flux distributions. 
Overall, the kurtosis values for the BZQ sample are consistently higher than those for BZBs across all frequency bands, with the most significant difference observed at 143 GHz. At this frequency, the kurtosis of BZQs rises sharply to 4.8, while that of BZBs remains around -1.2. 
The resulting kurtosis difference of nearly 6 clearly reflects a marked distinction in their distribution characteristics.

The above MAD analysis reveals varying degrees of statistical differences in the multi-band flux distributions between BZBs and BZQs, reflecting disparities across different distributional characteristics.  
Among all considered metrics, kurtosis, as a statistical indicator of distributional peakedness and tail weight, exhibits the highest MAD value of 1.64. This suggests that kurtosis is likely the most effective metric for distinguishing the differences in multi-band flux distributions between BZBs and BZQs.

To more comprehensively assess the differences in the multi-band flux distributions between BZBs and BZQs, this study not only employs MAD to analyze point-to-point differences of various statistical metrics, but also introduces the Jensen–Shannon Distance (JSD) as a complementary measure.
The JSD evaluates the overall difference between the two types of objects based on their probability density distributions (PDDs) and is particularly effective in capturing differences in distributional  characteristics. 
Compared to unidimensional statistical metrics, the JSD provides a unique advantage in revealing distributional structure differences through a comprehensive assessment. When the logarithm base is 2, the JSD ranges from 0 to 1, where 0 indicates identical distributions and 1 indicates completely different ones \citep{Endres2003,Connor2013}. 
In general, a value of $\mathrm{JSD} > 0.3$ indicates a significant difference between the two distributions \citep{Carton2018}. 
Due to its boundedness, it is well-suited for the comparative analysis of probability distributions in this study. 

In the analysis process, we first applied a base-10 logarithmic transformation to the flux data in each waveband.
We then used the \textbf{gaussian\_kde} method from the \textbf{scipy.stats} module to perform kernel density estimation (KDE),
which constructs a normalized probability density function that approximates the underlying distribution of the input data via Gaussian kernel smoothing \citep{Virtanen2020}. 
Next, we used the \textbf{scipy.spatial.distance.jensenshannon} function\footnote{\url{https://docs.scipy.org/doc/scipy/reference/generated/scipy.spatial.distance.jensenshannon.html}} to calculate the JSD values between the flux PDDs of BZBs and BZQs across 11 wavebands. 
The corresponding JSD results are summarized in the last column of Table~\ref{tab9}. The JSD quantifies the divergence between two probability distributions and is derived from the Jensen–Shannon divergence, $\mathrm{JSD}'(p, q)$, which is defined as follows:

\begin{equation}
\mathrm{JSD}'(p, q) = \frac{1}{2} D(p \parallel m) + \frac{1}{2} D(q \parallel m)
\label{equ2}
\end{equation}
Let \( p \) and \( q \) be two probability distributions, and let \( m = \frac{1}{2}(p + q) \) denote their symmetric mixture distribution. The function \( D(\cdot \parallel \cdot) \) represents the Kullback–Leibler(KL)  divergence. The JSD is then defined as the square root of the $\mathrm{JSD}'(p, q)$: 

\begin{equation}
\text{JSD}(p, q) = \sqrt{\mathrm{JSD}'( p, q)} 
\label{equ3}
\end{equation}
The $\mathrm{JSD}'(p, q)$ is a symmetrized measure of distributional difference based on the  KL  divergence, as introduced by \cite{Lin2002}. The KL divergence quantifies the information loss incurred when one probability distribution is used to approximate another, and is defined as follows \citep{hershey2007approximating}:

\begin{equation}
D(p \parallel m) = \int p(x) \log \frac{p(x)}{m(x)} \, dx, \quad
D(q \parallel m) = \int q(x) \log \frac{q(x)}{m(x)} \, dx
\label{equ4}
\end{equation}

\begin{figure*}
	\centering
	\includegraphics[width=0.7\textwidth]{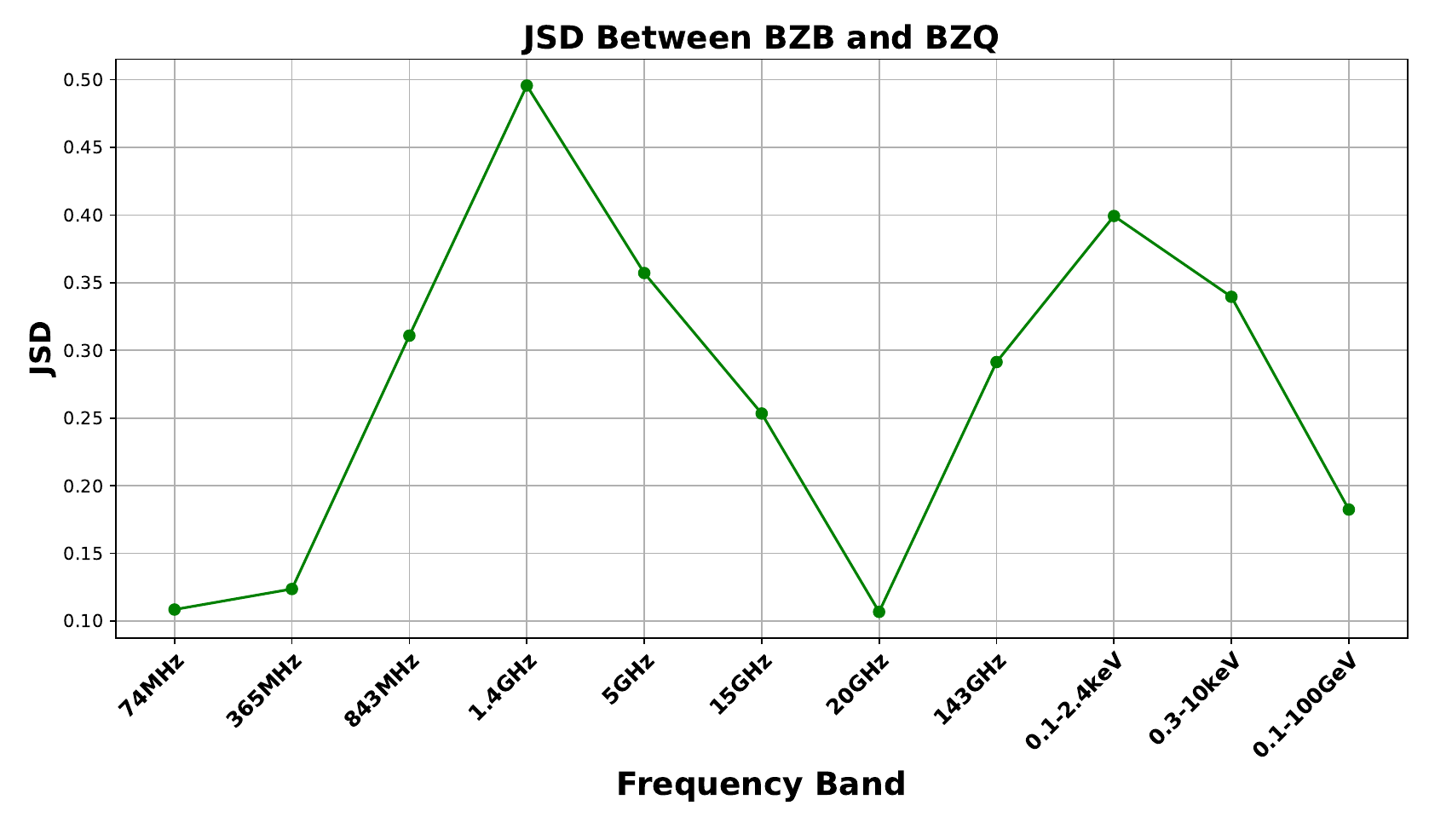}
	\caption{This figure shows  the trend of JSD variation of the flux distributions between BZBs and BZQs across 11 wavebands.}
	\label{jsd5}
\end{figure*}

To intuitively analyze the variation trend of JSD across different wavebands, we plotted the JSD data from Table~\ref{tab9} in Figure ~\ref{jsd5}.  
Figure~\ref{jsd5} illustrates notable differences in the probability distributions of BZBs and BZQs across various wavebands, especially at 1.4 GHz and 0.1–2.4 keV, where the JSD values exceed 0.3.
In contrast, the JSD values at 74 MHz and 20 GHz are below 0.15, indicating relatively small differences in flux distributions between the two source types. 
This may suggest that BZBs and BZQs share similar radiative characteristics in these low-energy bands. 
Overall, the trend of JSD variation shown in Figure~\ref{jsd5} highlights the varying degrees of distributional divergence between BZBs and BZQs across different wavebands.

\begin{figure*}
	\centering
	\begin{subfigure}[b]{0.48\textwidth}
		\centering
		\includegraphics[width=\linewidth]{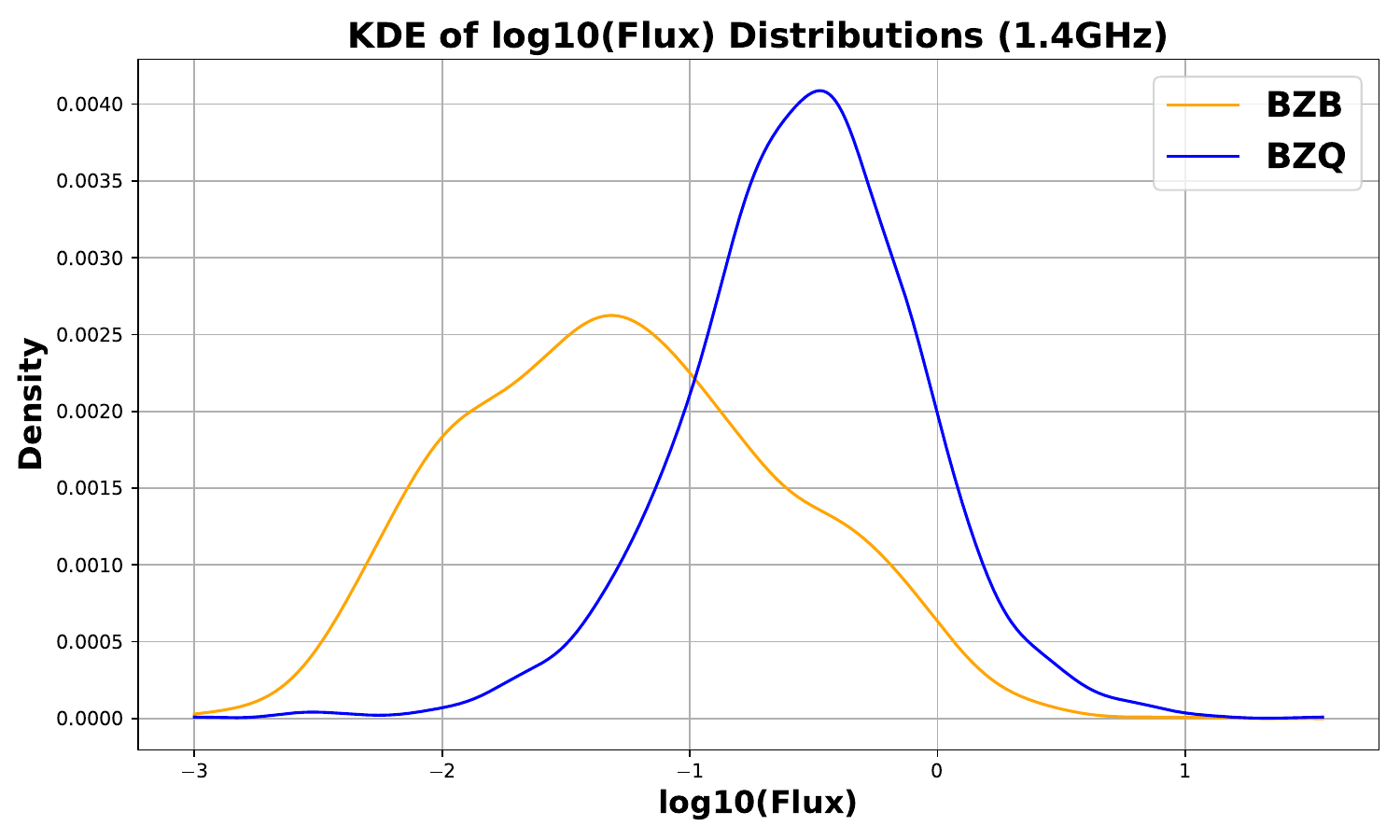}
		\caption{}
		\label{fig:flux_kde_xray}
	\end{subfigure}
	\hfill
	\begin{subfigure}[b]{0.48\textwidth}
		\centering
		\includegraphics[width=\linewidth]{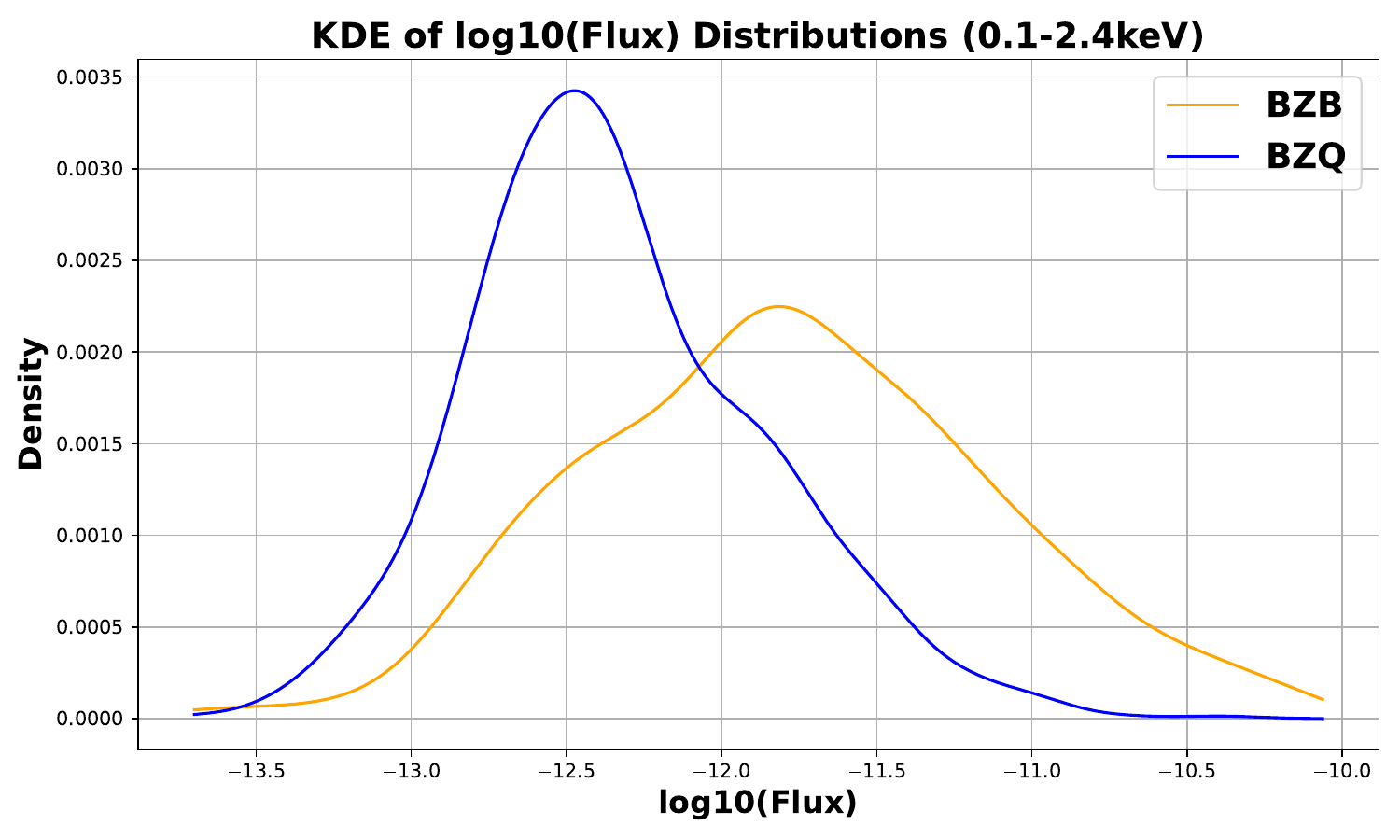}
		\caption{}
		\label{fig:flux_kde_radio}
	\end{subfigure}
	\caption{Kernel density estimations of the log$_{10}$(Flux) distributions at 1.4  GHz and 0.1-2.4 keV for BZBs and BZQs.}
	\label{fig:flux_kde_compare}
\end{figure*}

The above analysis shows that the MAD values of the eight classical statistical measures and the JSD values across different wavebands are all greater than zero, 
indicating systematic differences in the flux distributions between BZBs and BZQs. 
A univariate analysis using MAD as the difference metric reveals that the two source types differ significantly in max, mode, skewness, and kurtosis, with corresponding MAD values exceeding 0.4. Among these, kurtosis shows the largest discrepancy, with a MAD value of 1.64. 
Using JSD as a metric, we identified substantial differences in the probability distributions of the two source types at 1.4 GHz, 843 MHz, 5 GHz, 0.1–2.4 keV, and 0.3–10 keV, where the JSD values exceed 0.3 and are markedly higher than those in other wavebands. 
These distributional differences are also clearly visible in the probability density function (PDF) curves generated by KDE, such as those for 0.1–2.4 keV and 1.4 GHz, as shown in Figure~\ref{fig:flux_kde_compare}. 
The distribution differences in other wavebands, whether analyzed using MAD or JSD, are relatively small but still objectively present. 

In conclusion, we found that the two types of blazars exhibited systematic differences in their radiation behaviors across multiple wavebands. These differences are reflected not only in the central tendencies of the distributions but also in their skewness and kurtosis.

\subsection{Statistical Modeling Scheme} \label{appendixC.3}

A detailed description of the pipeline is presented in Content 4 of Figure \ref{technical-route3}, and the specific modeling steps are outlined as follows:

First, the 17 new GeV blazar candidates are excluded from the overall sample and designated as test objects for subsequent validation. Then, the flux data in each waveband are transformed using the Box–Cox transformation, implemented via the \textbf{scipy.stats.boxcox()} function.  
This transformation, originally proposed by \cite{Box1964}, aims to convert non-normally distributed data into an approximately normal form, thereby better satisfying the normality assumption required for subsequent statistical modeling. Its mathematical form is given as follows: 

\begin{equation}
y = 
\begin{cases}
\frac{x^{\lambda} - 1}{\lambda}, & \lambda \neq 0 \\
\ln(x), & \lambda = 0  
\end{cases}
\label{equ5}
\end{equation}
Here, $\lambda$ denotes the transformation parameter, with the optimal value determined through max likelihood estimation (MLE). 
For each flux dataset, the \textbf{boxcox()} function automatically determines a unique transformation parameter $\lambda$ via MLE, yielding a strictly monotonic and invertible mapping between the original and transformed data. 
For each of the 11 observational bands, the Box–Cox transformation was  applied separately to the flux distributions of BZBs and BZQs, with the   optimal $\lambda$ value independently estimated for each waveband. 
Next, we employed a TND model, implemented via \textbf{scipy.stats.truncnorm}, to fit the Box–Cox transformed flux data.
The TND is constructed by constraining and renormalizing the standard normal distribution (SND) within a finite interval [a,b], making it well-suited for modeling flux data that are naturally bounded due to physical constraints or observational limitations.
Let the observation interval be $[a,b]$, with its PDF given by: 

\begin{equation}
f(x \mid \mu, \sigma, a, b) = \frac{1}{\sigma} \frac{\varphi \left( \frac{x - \mu}{\sigma} \right)}{\Phi \left( \frac{b - \mu}{\sigma} \right) - \Phi \left( \frac{a - \mu}{\sigma} \right)}, \quad a \leq x \leq b,
\label{equ6}
\end{equation} 
where \( \mu \) and \( \sigma \) represent the mean and std of the original SND. The parameters \( a \) and \( b \) denote the lower and upper truncation bounds. The PDF of the SND is given by
$ \varphi(z) = \frac{1}{\sqrt{2\pi}} e^{-z^2/2},
$ 
and its cumulative distribution function (CDF) is expressed as 
$
\Phi(z) = \int_{-\infty}^{z} \varphi(t) \, dt.
$ 
The term \( \Phi\left(\frac{b - \mu}{\sigma}\right) - \Phi\left(\frac{a - \mu}{\sigma}\right) \) serves as a normalization factor, ensuring that the \( f(x \mid \mu, \sigma, a, b) \) integrates to one over the interval \([a, b]\). 
Here the parameters $\mu$ and $\sigma$ of the TND  were estimated via MLE, a procedure conveniently implemented using the \texttt{scipy.stats.truncnorm()} function to fit the flux data.

In this study, we employed the KS and CVM tests to evaluate the GoF of the transformed data under the TND. The KS test is a nonparametric method that assesses whether a sample follows a specified distribution by comparing the empirical distribution function (EDF) of the transformed data with the CDF of the TND and using the maximum absolute difference as the test statistic. 
The advantage of the KS test lies in its simplicity and intuitive interpretation: it effectively captures localized discrepancies in a distribution and is particularly sensitive to large departures, thereby enabling precise identification of significant deviations \citep{Massey1951}. 
The CVM test is a global GoF procedure based on the integrated squared difference. It evaluates overall fit by comparing the sample's EDF with the hypothesized CDF.
The advantage of the CVM test is its sensitivity to global GoF: by aggregating deviations over the entire distribution, it more accurately reflects overall differences and is therefore particularly effective for detecting small-to-moderate departures \citep{cramer1928composition}.

Using the KS test in tandem with the CVM test provides a more comprehensive assessment of GoF. The KS test is sensitive to local deviations and helps pinpoint potential points of departure, whereas the CVM test evaluates fit globally by aggregating discrepancies over the entire distribution, thereby reducing the risk of misjudgment based solely on local evidence. Taken together, the two tests enhance the accuracy and robustness of the GoF evaluation.

To implement the aforementioned GoF tests, we directly leverage the \textbf{SciPy} library \citep{Virtanen2020}, specifically calling the \textbf{kstest} and \textbf{cramervonmises} functions to execute the KS and CVM tests, respectively. These utilities enable a rapid evaluation of the p-values produced by both methods. 
Here we assess GoF using the $p$-values associated with the KS and CVM test statistics. 
In astronomy, $p = 0.05$ is commonly regarded as an acceptance threshold \citep{hovatta2016optical, salvestrini2022molecular, pavlik2020primordial, wang2023comprehensive}. When $p > 0.05$, we fail to reject the null hypothesis $H_0$ that the sample follows the TND, indicating that the fit is adequate. 
These tests and the $p$-value criterion are applied separately to the flux data of BZBs and BZQs at each waveband.

In addition to the above statistical tests, we employ model fit plots and Q--Q plots as auxiliary diagnostics. Model fit plots make discrepancies between the observed and theoretical distributions visually apparent, enabling rapid identification of potential departures and outliers; 
Q–Q plots compare the sample quantiles and theoretical quantiles, revealing agreement or departures across the entire distribution. To quantify the relationship between the sample quantiles and theoretical quantiles in the Q--Q plot, we employ the Pearson correlation coefficient $r$ to measure linear concordance; the closer $r$ is to 1, the stronger the agreement, whereas smaller values indicate pronounced departures.

\begin{figure*}
	\centering
	\begin{subfigure}[b]{0.52\textwidth}
		\centering
		\includegraphics[width=1\linewidth,height=7cm]{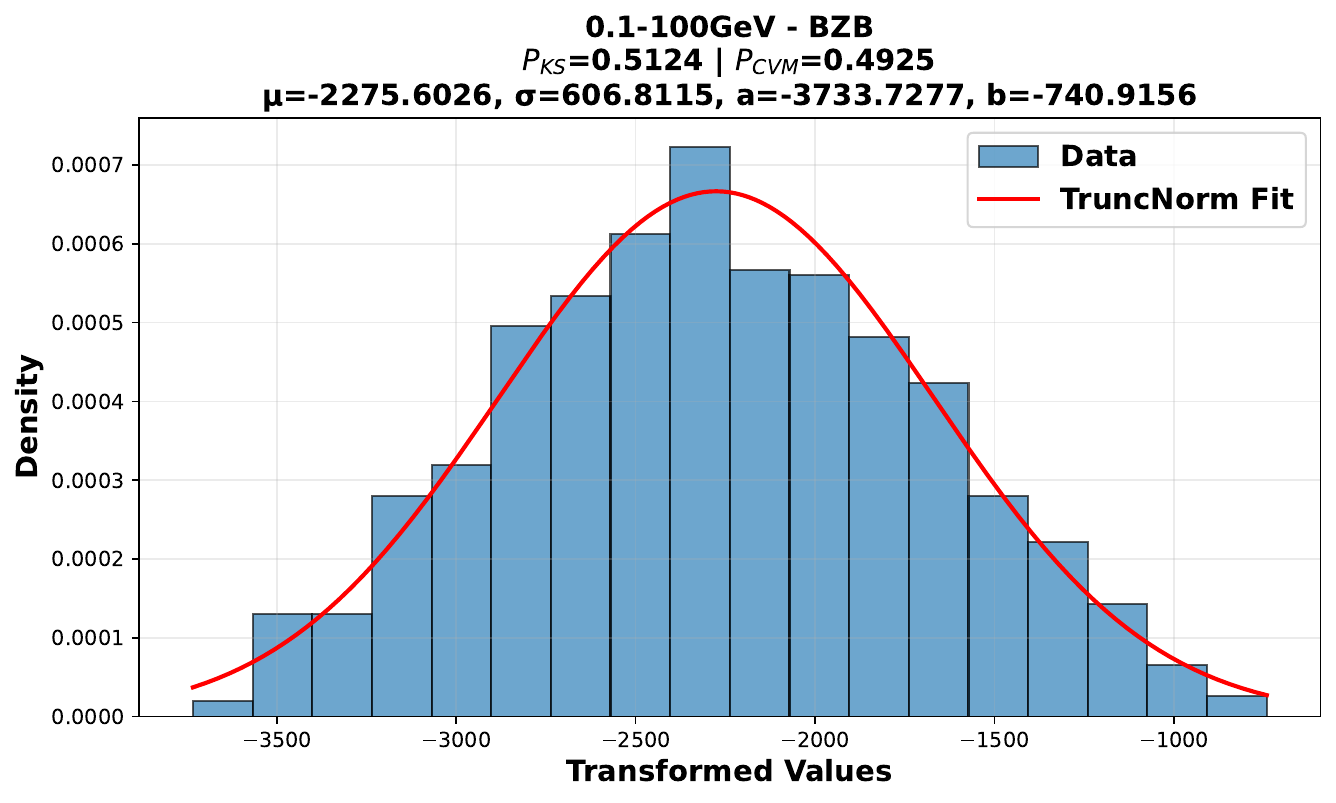}  
		\caption{}
		\label{fig:model-fit}
	\end{subfigure}
	\hfill
	\begin{subfigure}[b]{0.45\textwidth}
		\centering
		\includegraphics[width=\linewidth]{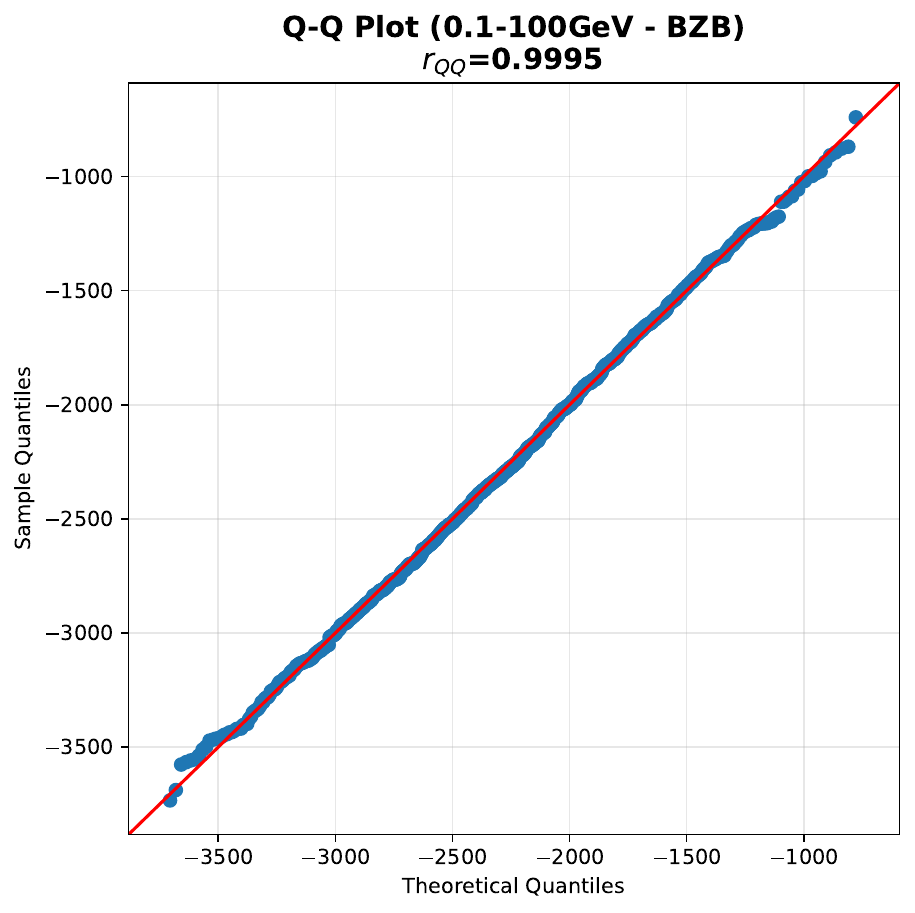}
		\caption{}
		\label{fig:qq}
	\end{subfigure}
	\caption{(a) Left plot: Fitted TND for the 0.1-100 GeV band, with the relevant fitting parameters and the p-values from the KS and CVM tests displayed at the top of the plot; (b) Right plot: Q-Q plot used to assess the fitting results, with the Pearson correlation coefficient \( r_{\rm QQ} = 0.9995 \).}
	\label{fig:fit_qq_plots}
\end{figure*}

To visually demonstrate the effectiveness of the above methods, we use the 0.1–100 GeV band dataset of BZBs from 5BZCAT as an example, with the model fit plot (\ref{fig:model-fit}) and the Q–Q plot (\ref{fig:qq}) shown in Figure~\ref{fig:fit_qq_plots}.
From panel (\ref{fig:model-fit}), a clear agreement between the observed and theoretical distributions can be seen, with no evident deviations. Meanwhile, panel (\ref{fig:qq}) shows that the sample quantiles and theoretical quantiles lie almost perfectly on a straight line, with a Pearson correlation coefficient of 
$r$=0.9995 further quantifying this near-perfect linear concordance. 
These auxiliary diagnostic results, together with the previous statistical tests, jointly indicate that the proposed model can accurately and robustly describe the flux distribution in this band.

By integrating visual diagnostics and classical statistical tests, it is possible to achieve mutual cross-validation at global and local levels, rapidly identify deviations, and mitigate the risk of misjudgment from any single method.
This leads to a more accurate and robust evaluation of model fit and a deeper understanding of the distributional features.

During the preliminary fitting stage, the fits for certain BZB and BZQ subsamples from the 5BZCAT\_err, spanning the 74 MHz, 843 MHz, 1.4 GHz, 5 GHz, 0.1--2.4 keV, and 0.3--10,keV bands, failed to pass the KS or CVM tests, yielding p-values ($\boldsymbol{p_{\text{KS}}}$ and $\boldsymbol{p_{\text{CVM}}}$) consistently below 0.05 (as detailed in Table \ref{tab10}).

\begin{table*}[h]
	\centering
	\small
	\renewcommand{\arraystretch}{1.2}
	
	\begin{threeparttable}
		\caption{LOF-Based Outlier Filtering and Goodness-of-Fit Reevaluation}
		\label{tab10}
		\begin{tabular}{llccccccccc}
			
			\toprule
			\textbf{Waveband} & \textbf{Type} & $\boldsymbol{k}$ & $\boldsymbol{\alpha}$ &
			$\boldsymbol{N_{0}}$ & $\boldsymbol{N_{\mathrm{proc}}}$ & $\boldsymbol{N_{\mathrm{out}}}$ &
			$\boldsymbol{p'_{\mathrm{KS}}}$ & $\boldsymbol{p'_{\mathrm{CVM}}}$ &
			$\boldsymbol{p_{\mathrm{KS}}}$ & $\boldsymbol{p_{\mathrm{CVM}}}$ \\
			\midrule
			74\,MHz   & BZB & 5 & 0.01 & 149  & 147  & 2  & 0.055 & 0.025 & 0.638 & 0.644 \\
			74\,MHz   & BZQ & 6 & 0.03 & 551  & 538  & 13 & 0.005 & 0.005 & 0.104 & 0.121 \\
			843\,MHz  & BZQ & 5 & 0.01 & 365  & 361  & 4  & 0.070 & 0.015 & 0.447 & 0.363 \\
			1.4\,GHz  & BZB & 5 & 0.01 & 1152 & 1140 & 12 & 0.015 & 0.015 & 0.418 & 0.438 \\
			1.4\,GHz  & BZQ & 5 & 0.01 & 1632 & 1615 & 17 & 0.005 & 0.005 & 0.177 & 0.181 \\
			5\,GHz    & BZQ & 9 & 0.01 & 1141 & 1129 & 12 & 0.005 & 0.005 & 0.209 & 0.140 \\
			0.1--2.4\,keV & BZQ & 5 & 0.01 & 849  & 840  & 9  & 0.015 & 0.005 & 0.103 & 0.202 \\
			0.3--10\,keV  & BZQ & 5 & 0.01 & 395  & 391  & 4  & 0.104 & 0.050 & 0.658 & 0.823 \\
			\bottomrule
		\end{tabular}%
		
		\vspace{2mm}
		\begin{tablenotes}[para]
			
			\small
			\item[] \textbf{Note:}\;
			Waveband   : ... : Observational waveband \\
			Type : ... : Source class \\
			$k$ : ... : Parameter (number of neighbors) for the LOF algorithm \\
			$\alpha$ : ... : Contamination parameter (expected outlier ratio) for the LOF algorithm \\
			$N_{0}$ : ... : Original sample size (before outlier removal) \\
			$N_{\text{proc}}$ : ... : Processed sample size (after outlier removal) \\
			$N_{\text{out}}$ : ... : Number of outliers detected and removed ($N_{0} = N_{\text{proc}} + N_{\text{out}}$) \\
			$p'_{\text{KS}}$ : ... : P-value of the KS test on the original data (before removal) \\
			$p'_{\text{CVM}}$ : ... : P-value of the CVM test on the original data (before removal) \\
			$p_{\text{KS}}$ : ... : P-value of the KS test on the processed data (after removal) \\
			$p_{\text{CVM}}$ : ... : P-value of the CVM test on the processed data (after removal)\\
		\end{tablenotes}
		
	\end{threeparttable}
\end{table*}

 This result may be attributed to the presence of outliers that deviate substantially from the majority of observations—possibly arising from measurement errors or rare physical phenomena. 
These outliers may distort the data distribution and undermine the validity of the assumed PDF model.

To systematically identify and remove potential outliers, we applied the LOF algorithm \citep{Breunig2000} to the raw flux data in Table~\ref{tab10}  prior to the Box–Cox transformation. 
The LOF method evaluates the degree of abnormality of a data point by comparing its local reachability density with that of its neighbors, thereby detecting local outliers in the data distribution. This algorithm can be conveniently implemented using the \textbf{sklearn.neighbors.LocalOutlierFactor} class\footnote{\url{https://scikit-learn.org/stable/modules/generated/sklearn.neighbors.LocalOutlierFactor.html}}. 

To determine the optimal hyperparameters for LOF, we conducted an exhaustive grid search over two key parameters: the neighborhood size ($\boldsymbol{k}$) and the contamination proportion ($\boldsymbol{\alpha}$). The neighborhood size was varied from 5 to 50 in increments of one, while the contamination proportion ranged from 0.01 to 0.20 in steps of 0.01. 
A total of \( 46 \times 20 = 920 \) parameter combinations were evaluated. For each configuration, outliers were identified among the samples of the specified waveband(s) in Table~\ref{tab10}.   
The resulting cleaned dataset was then fitted using the TND, and the GoF was assessed via the KS and CVM tests. Only configurations yielding p-values greater than 0.05 were retained. 
Among all parameter combinations that passed the KS and CVM tests, we selected the configuration with the lowest contamination level. 
This selection effectively balances outlier removal with maximal preservation of the original dataset for subsequent analysis. 

During this process, all identified outliers were packaged into \texttt{outliers.tar.xz}; see Section \ref{sec4}  for details. The hyperparameters obtained using LOF are shown in Table \ref{tab10}. 
Subsequently, after outlier removal, the Box–Cox transformation and TND fitting were re-applied to the flux dataset specified in Table \ref{tab10}, followed by the KS and CVM tests. 
The results confirmed consistency with the assumed distribution model, with the corresponding p-values for $P_{\text{KS}}$ and $P_{\text{CVM}}$ both being $>$ 0.05. 
The optimal fitting results across all bands are summarized in Table~\ref{tab11}, showing p-values of KS and CVM tests consistently greater than 0.05. 
These results demonstrate the broad applicability of the analysis pipeline shown in Figure \ref{technical-route3}, offering a robust approach to modeling the flux distribution in all wavebands.

\begin{table*}
	\small
	\renewcommand{\arraystretch}{1.15}
	\begin{threeparttable}
		\caption{Distribution Fitting Results Across 11 Wavebands}
		\label{tab11}
		\begin{tabular}{ccccccccccc} 
		
		\toprule
		\textbf{Waveband} & \textbf{Type} & \multicolumn{1}{c}{\textbf{$\lambda$}} & \multicolumn{1}{c}{\textbf{a}} & \multicolumn{1}{c}{\textbf{b}} & \multicolumn{1}{c}{\textbf{$\mu$}} & \multicolumn{1}{c}{\textbf{$\sigma$}} & \multicolumn{1}{c}{\textbf{$P_{\text{KS}}$}} & \multicolumn{1}{c}{\textbf{$P_{\text{CVM}}$}} & \multicolumn{1}{c}{\textbf{$r_{\text{QQ}}$}} & \multicolumn{1}{c}{\textbf{TruncNorm}} \\ 
		\midrule
		\multirow{2}{*}{74MHz} & BZB & -0.482 & -0.974 & 1.519 & 0.304 & 0.594 & 0.638 & 0.644 & 0.985& \ding{51} \\
		& BZQ & -0.518 & -0.924 & 1.787 & 0.367 & 0.575 & 0.104 & 0.121 & 0.996 & \ding{51} \\
		\midrule
		\multirow{2}{*}{365MHz} & BZB & -0.334 & -2.727 & 1.594 & -0.723 & 0.873 & 0.493 & 0.254 & 0.997& \ding{51} \\
		& BZQ & -0.376 & -3.003 & 2.111 & -0.509 & 0.878 & 0.333 & 0.164 & 0.999& \ding{51} \\
		\midrule
		\multirow{2}{*}{843MHz} & BZB & -0.020 & -4.659 & 1.244 & -2.261 & 1.507 & 0.085 & 0.249 & 0.996& \ding{51} \\
		& BZQ & 0.031 & -4.075 & 1.874 & -1.107 & 1.037 & 0.447 & 0.363 & 0.995& \ding{51} \\
		\midrule
		\multirow{2}{*}{1.4GHz} & BZB & -0.053 & -8.179 & 1.032 & -3.189 & 1.812 & 0.418 & 0.438 & 0.998& \ding{51} \\
		& BZQ & 0.088 & -5.169 & 2.228 & -1.155 & 0.977 & 0.177 & 0.181 & 0.995& \ding{51} \\
		\midrule
		\multirow{2}{*}{5GHz} & BZB & 0.015 & -5.632 & 1.367 & -2.022 & 1.509 & 0.527 & 0.582 & 0.999& \ding{51} \\
		& BZQ & 0.073 & -5.725 & 2.504 & -1.003 & 0.984 & 0.209 & 0.141 & 0.994& \ding{51} \\
		\midrule
		\multirow{2}{*}{15GHz} & BZB & -0.105 & -4.448 & 1.851 & -1.600 & 1.422 & 0.219 & 0.124 & 0.997& \ding{51} \\
		& BZQ & -0.264 & -5.537 & 2.291 & -1.133 & 1.205 & 0.333 & 0.169 & 0.999& \ding{51} \\
		\midrule
		\multirow{2}{*}{20GHz} & BZB & -0.213 & -4.307 & 1.403 & -1.762 & 1.429 & 0.512 & 0.547 & 0.998& \ding{51} \\
		& BZQ & -0.199 & -4.512 & 2.257 & -1.376 & 1.226 & 0.448 & 0.443 & 0.999& \ding{51} \\
		\midrule
		\multirow{2}{*}{143GHz} & BZB & -0.397 & -0.825 & 1.333 & -270.591 & 39.714 & 0.547 & 0.522 & 0.992& \ding{51} \\
		& BZQ & -0.593 & -2.062 & 1.453 & -0.327 & 0.812 & 0.836 & 0.677 & 0.993& \ding{51} \\
		\midrule
		\multirow{2}{*}{0.1-2.4keV} & BZB & -0.019 & -43.269 & -29.143 & -35.536 & 2.590 & 0.438 & 0.323 & 0.999& \ding{51} \\
		& BZQ & -0.119 & -347.722 & -159.090 & -239.513 & 30.888 & 0.103 & 0.202 & 0.999& \ding{51} \\
		\midrule
		\multirow{2}{*}{0.3-10keV} & BZB & -0.208 & -2189.835 & -357.956 & -1264.802 & 589.898 & 0.537 & 0.358 & 0.996& \ding{51} \\
		& BZQ & 0.040 & -17.655 & -15.793 & -16.804 & 0.368 & 0.658 & 0.823 & 0.995& \ding{51} \\
		\midrule
		\multirow{2}{*}{0.1-100GeV} & BZB & -0.240 & -3733.728 & -740.916 & -2275.603 & 606.812 & 0.512 & 0.493 & 1.000& \ding{51} \\
		& BZQ & -0.236 & -3258.043 & -601.364 & -1874.681 & 483.885 & 0.786 & 0.567 & 0.999& \ding{51} \\
		\bottomrule
  \end{tabular}

\vspace{2mm}
\begin{tablenotes}[para]
\small
\item[] \textbf{Note:}\ Waveband   : ... : Observational waveband \\
Type       : ... : Source class  \\
$\lambda$  : ... : Box–Cox transformation factor  \\
a          : ... : Lower bound of the TND  \\
b          : ... : Upper bound of the TND  \\
$\mu$      : ... : Mean of the Box–Cox transformed fluxes  \\
$\sigma$   : ... : Standard deviation of the Box–Cox transformed fluxes  \\
$P_{\text{KS}}$ : ... : P-value corresponding to the KS statistic  \\
$P_{\text{CVM}}$ : ...: P-value corresponding to the CVM statistic  \\
$r_{\text{QQ}}$  : ...: Pearson correlation coefficient of the QQ plot  \\
TruncNorm  : ... : Flag indicating if the TND model fit is valid \\
\end{tablenotes}
\end{threeparttable}
\end{table*}


\bsp	
\label{lastpage}
\end{document}